\documentclass[preprint,showkeys,showpacs]{revtex4-1}

\usepackage{graphicx}
\usepackage{graphics}
\usepackage{hyperref}
\usepackage{amssymb}
\usepackage{amsmath}
\usepackage{hyperref}
\usepackage{braket}
\usepackage{paralist}
\usepackage{theorem}
\usepackage{multirow}
\usepackage{mathtools}
\usepackage{upgreek}
\theoremheaderfont{\normalsize \bfseries}
\theorembodyfont{\normalsize \rmfamily}
\newtheorem{prop}{Property}

\usepackage{type1cm,eso-pic,color}

\begin{document}
\author{Ivan V. Bazarov}
\affiliation{Physics Department, Cornell University, Ithaca, NY 14853, USA}

\title{Synchrotron radiation representation in phase space}
\begin{abstract}
The notion of brightness is efficiently conveyed in geometric optics as density of rays in phase space. Wigner has introduced his famous distribution in quantum mechanics as a quasi-probability density of a quantum system in phase space. Naturally, the same formalism can be used to represent light including all the wave phenomena. It provides a natural framework for radiation propagation and optics matching by transferring the familiar `baggage' of accelerator physics ($\beta$-function, emittance, phase space transforms, etc.) to synchrotron radiation. This paper details many of the properties of the Wigner distribution and provides examples of how its use enables physically insightful description of partially coherent synchrotron radiation in phase space.
\end{abstract}
\date{\today}

\maketitle

\section{Introduction}
The concept of phase space plays an important role in accelerator physics. Useful tools such as Twiss parameters, emittance, phase space propagation have been in long use in the accelerator community. The extension of classical phase space concept to synchrotron radiation is straightforward for geometric optics, applicable for incoherent radiation. The Wigner distribution, or Wigner distribution function (WDF), was recognized to be a general framework to represent quantum \cite{PhysRev.40.749} and therefore wave phenomena in phase space \cite{WALTHER:68, KwangJe198671}. The approach allows light characterization of arbitrary degree of coherence \cite{Bastiaans:86} and polarization \cite{2007JEOS....2E7030L} in phase space, though its application by the accelerator community has so far been mostly limited to simplest cases of Gaussian or Gauss-Schell beams \cite{Kim89, Coisson:mb0007}. This provides a set of useful analytical expressions for quick estimates of performance of modern x-ray sources with improved coherence properties. In particular, the concepts of the diffraction limit and brightness have been extended to cover partially coherent radiation cases following the notion of Gaussian distributions in the phase space for both the synchrotron radiation of the undulator central cone and density of electrons. More detailed approach to coherent or partially coherent sources inevitably calls on physically rigorous wave description of the radiation, either through using cross-spectral density \cite{Geloni2008463} or the Wigner distribution. In particular, since neither undulator radiation, nor electron distribution in the phase need to be Gaussian, the general framework becomes essential to be able to describe the performance of x-ray sources with improved coherence.

The Wigner distribution function provides a natural and elegant description of the properties of light, and can serve as a useful tool in accelerator and x-ray beamline design including electron to x-ray beam matching, light propagation, fully accounting for arbitrary polarization and coherence properties of radiation. The intuitive picture provided by the WDF, being the phase space density of light or generalized brightness, is particularly appealing to the accelerator community trained to view many aspects of the beam dynamics in phase space. Not only can the WDF be readily computed from the first principles, the first measurement of x-ray Wigner distribution has been reported in the literature \cite{PhysRevLett.98.224801}. The knowledge of the Wigner distribution represents the entirety of what can be known about the radiation and its importance will only increase with advent of more coherent x-ray sources.

The purpose of this paper is to review many of the useful properties of the Wigner distribution and demonstrate that the WDF can be used with physical insight to describe partially coherent synchrotron radiation. In what follows, the Wigner distribution properties are first reviewed in Section \ref{sec:wig_qm} using the language of quantum mechanics. Various examples illustrate the physical meaning of the WDF for both pure and mixed quantum states. The case of the synchrotron radiation as discussed in Section \ref{sec:wig_opt} is then viewed as a natural extension of the quantum mechanical treatment. Coherence and dispersion properties of light as conveniently conveyed by the WDF are emphasized. A special attention is given to light polarization, being an important characteristic of synchrotron radiation. Practical matters of computing the WDF are covered in Section \ref{sec:syn_rad}, which outlines the general procedure for obtaining the Wigner distribution first for a single electron, and then extending the result to include electron bunches of Energy Recovery Linac as an example. Since neither synchrotron radiation nor electron beam in this case have Gaussian phase space density, some consideration is given to generalizing the concepts of emittance and brightness to describe non-Gaussian distributions.

\section{Wigner distribution in quantum mechanics\label{sec:wig_qm}}
The Wigner distribution, initially introduced to account for quantum phenomena in statistical mechanics \cite{PhysRev.40.749}, provides a convenient description of a quantum mechanical system in phase space. The Wigner distribution itself does not possess any new information not already contained in quantum state itself, which is fully described (together with its complete time evolution through Hamiltonian $
\hat{\mathcal{H}}$) either by a pure state $\psi$ or more generally for a mixed state by its density matrix $\hat{\rho} = \sum\limits_{j} p_j \ket{\psi_j}\!\bra{\psi_j}$, with state weights $\sum\limits_{j} p_j = 1$. The utility of the Wigner distribution is in convenient and visual representation of the quantum system (and by extension wave optics phenomena) in terms of quasi-probability of having both phase space quantities (e.g. $x, p$). Such characterization, being very familiar to accelerator physicists, is a natural framework of description for a unified phenomena of both classical and wave nature reusing many of the concepts from the accelerator field (emittance, $\beta$-function, phase space propagation, brightness, etc.). Quasi-probability refers to the fact that while the Wigner distribution is normalized to 1 and is used to compute averages of various quanitites as expected for a probability density function, the function can take on local negative values. This deviation from non-negativity is essential for general quantum or wave phenomena where position and momentum operators do not commute and the uncertainty principle must hold contrary to the classical description. Nevertheless, this non-positivity does not preclude measurement of Wigner distribution using tomography techniques \cite{PhysRevLett.98.224801, PhysRevLett.70.1244}.

The properties of the Wigner distribution have been studied extensively in the context of quantum mechanics \cite{Hillery1984121, Tatarskii83}, wave optics \cite{Bastiaans:86, 2007JEOS....2E7030L, testorf2009phase} and signal processing \cite{Dragoman_2005, Sheng2000, Faye2000}. To provide a suitable context, the properties of the WDF are reviewed in this section. For simplicity, we limit our consideration here to a 1D scalar wavefunction, $\psi(x)$. Extension to higher dimensions and polarization as required for synchrotron radiation is detailed in Section \ref{sec:syn_rad}.

\subsection{Pure quantum state}
First, we consider a pure quantum state $\psi$. Particularly insightful definition of the Wigner distribution can be given in Dirac notation:
\begin{equation}\label{eq:wig1}
W(x,p) = \int \braket{\psi|x+\tfrac{x'}{2}}\!\braket{x+\tfrac{x'}{2}|p}\!\braket{p|x-\tfrac{x'}{2}}\!\braket{x-\tfrac{x'}{2}|\psi} dx'.
\end{equation}
(The integration here and elsewhere in this paper is taken over the entire range $-\infty$ to $+\infty$ unless stated otherwise.) The integrand is the quantum equivalent of a classical phase-space trajectory as seen by reading Dirac brackets from right to left:  \begin{inparaenum}[(1)]
\item the probability amplitude for a particle in state $\psi$ to have a position $(x-\tfrac{x'}{2})$;
\item the amplitude for a particle with position $(x-\tfrac{x'}{2})$ to have momentum $p$;
\item the amplitude for a particle with momentum $p$ to have position $(x+\tfrac{x'}{2})$;
and finally \item the amplitude for a particle with position $(x+\tfrac{x'}{2})$ to (still) be in the state $\psi$.
\end{inparaenum}
The integration over the entire space $x'$ therefore creates a superposition of all possible quantum trajectories of state $\psi$, which interfere constructively and destructively, providing a quasi-probability distribution in phase space \cite{2009arXiv0912.2333R}. Using a well known identity (with $h = 2\pi\hbar$ the Planck constant)
\begin{equation*}
\braket{x|p} = \tfrac{1}{\sqrt{h}} e^\frac{ipx}{\hbar},
\end{equation*}
we rewrite
\begin{equation*}
\braket{x+\tfrac{x'}{2}|p}\!\braket{p|x-\tfrac{x'}{2}} = \tfrac{1}{h} e^\frac{ipx'}{\hbar}.
\end{equation*}
Eq.~\ref{eq:wig1} then assumes its most frequently quoted form
\begin{equation}\label{eq:wigx}
W(x,p) = \frac{1}{h} \int \psi^*(x+\tfrac{x'}{2}) \psi(x-\tfrac{x'}{2}) e^\frac{ipx'}{\hbar} dx',
\end{equation}
where $\psi^*(x+\tfrac{x'}{2}) \equiv \braket{\psi|x+\tfrac{x'}{2}}$ and $\psi(x-\tfrac{x'}{2}) \equiv \braket{x-\tfrac{x'}{2}|\psi}$.

In the same spirit, Eq.~\ref{eq:wig1} can be rewritten in terms of integration over the entire momentum space,
\begin{equation}\label{eq:wig2}
W(x,p) = \int \braket{\psi|p+\tfrac{p'}{2}}\!\braket{p+\tfrac{p'}{2}|x}\!\braket{x|p-\tfrac{p'}{2}}\!\braket{p-\tfrac{p'}{2}|\psi} dp',
\end{equation}
leading to an equivalent definition of the Wigner distribution function now in terms of momentum representation of the state $\varPsi(p) \equiv \braket{p|\psi}$:
\begin{equation}\label{eq:wigp}
W(x,p) = \frac{1}{h} \int \varPsi^*(p+\tfrac{p'}{2}) \varPsi(p-\tfrac{p'}{2}) e^{-\frac{ipx'}{\hbar}} dp'.
\end{equation}
The momentum and position representations, $\psi(x)$ and $\varPsi(p)$, are related via the Fourier transform
\begin{equation}
\begin{aligned}
\varPsi(p) & = \tfrac{1}{\sqrt{h}} \int \psi(x) e^{-\frac{ipx}{\hbar}} dx,\\
\psi(x) & = \tfrac{1}{\sqrt{h}} \int \varPsi(p) e^{\frac{ipx}{\hbar}} dp.
\end{aligned}
\end{equation}

A summary of the main properties of the Wigner distribution function is given below. Properties that are revisited later for a more general case of a mixed state are denoted by an asterisk (*).

\begin{prop}[Realness]\label{prop:real}
\begin{equation}
W(x,p) \in \mathbb{R}.
\end{equation}
\end{prop}
This property follows from $W^*(x, p) = W(x,p)$.

\begin{prop}[Normalization and Marginals*]\label{prop:norm}
The WDF is normalized to $1$ with its projections (or marginals) corresponding to nonnegative probability densities in either position or momentum
\begin{equation}\label{eq:norm}
\begin{aligned}
& \iint W(x,p)\,dx \, dp = 1, \\
& \int W(x,p)\,dp = |\psi(x)|^2, \\
& \int W(x,p)\,dx = |\varPsi(p)|^2.
\end{aligned}
\end{equation}
\end{prop}
The proof is by substitution of the Wigner definition into Eqs.~\ref{eq:norm} and then using the identify
\begin{equation}\label{eq:delta}
\int e^\frac{i a b}{\hbar} da = h \,\delta(b).
\end{equation}

\begin{prop}[Boundness]\label{prop:bound}
\begin{equation}\label{eq:bound}
|W(x,p)| \leq \frac{2}{h} = \frac{1}{\pi\hbar}.
\end{equation}
\end{prop}
This property can be proven using the Cauchy-Schwarz inequality on the definition of the Wigner function.

It is illustrative to consider when the WDF assumes $\pm\tfrac{2}{h}$ extrema. An arbitrary wavefunction $\psi(x)$ can be written in terms of even $\psi_e(-x) = \psi_e(x)$ and odd $\psi_o(-x) = -\psi_o(x)$ parts: $\psi(x) = \psi_e(x) + \psi_o(x)$. Then, the WDF at the origin becomes
\begin{align*}
W(0,0) & = \frac{1}{h} \int \psi^*(\tfrac{x'}{2})\psi(-\tfrac{x'}{2})\,dx',\\
& = \frac{2}{h} \int \psi^*(x)\psi(-x)\,dx.
\end{align*}
This can be written in terms of the wavefunction's even and odd parts:
\begin{equation}\label{eq:maxmin}
W(0,0) = \frac{2}{h} \int \left( |\psi_e(x)|^2 - |\psi_o(x)|^2 \right)\,dx.
\end{equation}
As can be seen from Eq.~\ref{eq:maxmin} and the wavefunction normalization, $W(0,0) = \tfrac{2}{h}$ if $\psi(x)$ is even and $W(0,0) = -\tfrac{2}{h}$ if $\psi(x)$ is odd and vice versa \cite{Tatarskii83}.

\begin{prop}[Expectation values]\label{prop:expect}
The expectation value of an operator $\hat{A}$ can be found from its phase-space representation function $A(x,p)$ according to
\begin{equation}
\braket{\hat{A}} = \iint A(x,p) W(x,p)\,dx \, dp,
\end{equation}
where $W(x,p)$ acts as a phase-space probability density. The function $A(x,p)$ and operator $\hat{A}$ satisfy the following relationships \cite{Hillery1984121}
\begin{align}
& A(x,p) = \int \braket{x-\tfrac{x'}{2}|\hat{A}|x+\tfrac{x'}{2}} e^\frac{ipx'}{\hbar} dx',\label{eq:wig-trans}\\
& \braket{x_1|\hat{A}|x_2} = \frac{1}{h} \int A(\tfrac{x_1+x_2}{2}, p) e^\frac{ip(x_1-x_2)}{\hbar} dp.\label{eq:weyl-trans}
\end{align}
The pair of Eqs.~\ref{eq:wig-trans} and \ref{eq:weyl-trans} is referred to as the Wigner-Weyl transformation \cite{leaf:65}.
\end{prop}

Refer to \cite{Hillery1984121} for proof. A practical significance of this property is that any linear combination of functions of only position operators or
only momentum operators correspond to the classical phase-space representation given by Eq.~\ref{eq:wig-trans} where one replaces $\hat{x} \rightarrow x$ and $\hat{p} \rightarrow p$. In particular, the $n$-th moments of the distribution are readily obtained using
\begin{align*}
& \braket{p^n} = \int p^n W(x,p)\,dx \, dp,\\
& \braket{x^n} = \int x^n W(x,p)\,dx \, dp.
\end{align*}
Quantum mechanical correspondence to the classical correlation expectation of position-momentum is more involved due to non-commuting nature of the operators. Indeed, the operator $\hat{x}\hat{p}$ is not Hermitian, i.e. $(\hat{x}\hat{p})^\dagger = \hat{p}^\dagger\hat{x}^\dagger = \hat{p}\hat{x} \neq \hat{x}\hat{p}$ since $[\hat{x}, \hat{p}] = i\hbar \neq 0$. As a result, the expectation value $\braket{\hat{x}\hat{p}}$ is generally complex and $\braket{\hat{x}\hat{p}} \neq \braket{\hat{p}\hat{x}}$. A solution is to write $\braket{\tfrac{1}{2}(\hat{x}\hat{p}+\hat{p}\hat{x})} = \braket{xp} = \braket{px}$, where the symmetric operator is now Hermitian and its corresponding phase space function is found from Eq.~\ref{eq:wig-trans} to be $\tfrac{1}{2}(\hat{x}\hat{p}+\hat{p}\hat{x}) \rightarrow xp$. Therefore,
\begin{equation*}
\braket{xp} = \braket{px} = \iint xp \, W(x,p)\,dx \, dp.
\end{equation*}

The above equations allows us to compute the $\boldsymbol\Sigma$-matrix of the quantum phase-space distribution familiar to accelerator physicists
\begin{equation}
\boldsymbol\Sigma = \left(
\begin{array}{cc}
\left< x^2 \right>        & \left< x p \right> \\
\left<p x \right>  & \left< p^2 \right>
\end{array}
\right) = \left(
\begin{array}{cc}
\epsilon\beta   & -\epsilon\alpha\\
-\epsilon\alpha & \epsilon\gamma
\end{array}
\right) = \epsilon \left(
\begin{array}{cc}
\beta   & -\alpha\\
-\alpha & \gamma
\end{array}
\right) = \epsilon \mathbf{T},\label{eq:sig-mat}
\end{equation}
with the usual meaning of emittance $\epsilon = \sqrt{\det\boldsymbol\Sigma}$ and Twiss parameters satisfying $\det\mathbf{T} = 1$. The Heisenberg uncertainty principle can then be written as
\begin{equation}
\epsilon \geq \frac{\hbar}{2}.
\end{equation}

\begin{prop}[Time evolution]\label{prop:evol}
For a time-independent Hamiltonian $\hat{\mathcal{H}} = \hat{p}^2/2m+V(\hat{x})$, the time evolution for the Wigner distribution $W$ is governed by
\begin{equation}\label{eq:time_evol}
\frac{\partial W}{\partial t} = -\frac{p}{m}\frac{\partial W}{\partial x} + \frac{1}{i\hbar}\left[ V\left(x+\tfrac{i\hbar}{2}\tfrac{\partial}{\partial p} \right)-V \left(x-\tfrac{i\hbar}{2}\tfrac{\partial}{\partial p} \right) \right] W.
\end{equation}
\end{prop}
The proof is straightforward using time-dependant Schr\"{o}dinger equation. Refer to \cite{Tatarskii83} for details. In particular, for a linear force $F(x) = F_0 - k x$ with a potential energy $V(x) = V_0 - F_0 x + \tfrac{1}{2} k x^2$ ($F_0$, $k$, and $V_0$ are arbitrary constants), $\hbar$ drops out from the Eq.~\ref{eq:time_evol} and we recover the classical Liouville's evolution of the phase-space distribution
\begin{equation}
\frac{\partial W}{\partial t} + \frac{p}{m}\frac{\partial W}{\partial x} + F \, \frac{\partial W}{\partial p} = 0.
\end{equation}
This property further illustrates the connection to the classical concept of phase space. In particular, classical invariants and transformation rules directly carry over to the quantum phase space density in case of no or linear forces.

\begin{prop}[State cross-correlation*]\label{prop:cross}
Cross-correlation of the wavefunction can be recovered from the WDF of a pure state via a Fourier transform
\begin{equation}
\begin{aligned}\label{eq:cross_pure}
\psi(x_1)\psi^*(x_2) & = \int W\!\left(\tfrac{x_1+x_2}{2}, p \right) e^\frac{i(x_1-x_2)p}{\hbar} dp,\\
\varPsi(p_1)\varPsi^*(p_2) & = \int W\!\left(x, \tfrac{p_1+p_2}{2}\right) e^{-\frac{ix(p_1-p_2)}{\hbar}} dx.
\end{aligned}
\end{equation}
\end{prop}

This property is proven by substituting the WDF definition and using the identity (\ref{eq:delta}). Note the similarity to Weyl's relationship, the Eq.~\ref{eq:weyl-trans}.

It should be noted that the cross-correlation function of Eq.~\ref{eq:cross_pure} is just a density matrix of a pure state $\psi$ in either position or momentum basis
\begin{equation}
\begin{aligned}
\psi(x_1)\psi^*(x_2) & = \braket{x_1|\psi}\!\braket{\psi|x_2},\\
\varPsi(p_1)\varPsi^*(p_2) & = \braket{p_1|\psi}\!\braket{\psi|p_2}.
\end{aligned}
\end{equation}
This connection of the Wigner distribution to the density operator matrix will continue for mixed states as discussed later.

\begin{prop}[State recovery*]\label{prop:invert}
Property \ref{prop:cross} allows to recover the wavefunction from the WDF modulo a complex constant
\begin{equation}
\begin{aligned}
& \psi(x)\psi^*(0) = \int W\!\left(\tfrac{x}{2}, p \right) e^\frac{ixp}{\hbar} dp,\\
& \varPsi(p)\varPsi^*(0) = \int W\!\left(x, \tfrac{p}{2}\right) e^{-\frac{ixp}{\hbar}} dx.
\end{aligned}
\end{equation}
\end{prop}

\begin{prop}[Integrated product]\label{prop:scalar} For two WDFs corresponding to pure states $\psi$ and $\chi$
\begin{align*}
W_{(\psi)}(x,p) & = \frac{1}{h} \int \psi^*(x+\tfrac{x'}{2}) \psi(x-\tfrac{x'}{2}) e^\frac{ipx'}{\hbar} dx',\\
W_{(\chi)}(x,p) & = \frac{1}{h} \int \chi^*(x+\tfrac{x'}{2}) \chi(x-\tfrac{x'}{2}) e^\frac{ipx'}{\hbar} dx',
\end{align*}
the integrated (overlapping) product is related to scalar state product according to
\begin{equation}\label{eq:scalar_product}
\iint W_{(\psi)}(x,p) W_{(\chi)}(x,p)\,dx \, dp = \frac{1}{h} \Bigl|\int \psi^*(x) \chi(x)\,dx \Bigr|^2 = \frac{1}{h} \left|\braket{\psi|\chi}\right|^2.
\end{equation}
\end{prop}
The proof of this property again involves a substitution of the WDF definition into Eq.~\ref{eq:scalar_product}, the use of identity (\ref{eq:delta}) and a change of integration variables.

\begin{prop}[Generalized integrated product]\label{prop:gscalar}
To account for Wigner distribution of a superposition of quantum states, we introduce a generalized Wigner distribution (now generally complex) using
\begin{equation}\label{eq:wiggen}
W_{(\psi_1,\psi_2)}(x,p) = \frac{1}{h} \int \psi_1^*(x+\tfrac{x'}{2}) \psi_2(x-\tfrac{x'}{2}) e^\frac{ipx'}{\hbar} dx'.
\end{equation}
Note that $W_{(\psi_1,\psi_2)}^* = W_{(\psi_2,\psi_1)}$. The standard Wigner distribution $W_{(\psi)} \equiv W_{(\psi,\psi)}$ is obtained by setting $\psi_1(x) = \psi_2(x) = \psi(x)$ in Eq.~\ref{eq:wiggen}.

Then the integrated overlapping product of two generalized WDFs becomes
\begin{equation}\label{eq:gen_scalar}
\iint W_{(\psi_1, \psi_2)} (x,p) W_{(\chi_1, \chi_2)} (x,p)\,dx \, dp = \frac{1}{h} \braket{\psi_1|\chi_2}\!\braket{\chi_1|\psi_2}
\end{equation}
\end{prop}

\begin{prop}[Superposition of states]\label{prop:super}
Consider a superposition of states $\ket{\psi} = \sum\limits_{n} \alpha_n \ket{\phi_n}$ (either finite or infinite sum). States $\phi_n$ need not be orthogonal, but all states are assumed normalized: $\braket{\psi|\psi} = 1$ and $\braket{\phi_n|\phi_n} = 1$. The Wigner distribution can then be written as
\begin{equation}\label{eq:sup}
W_{(\psi)} = \sum\limits_{n}\sum\limits_{m} \alpha_n^* \alpha_m W_{(\phi_n,\phi_m)}.
\end{equation}
Note that the superposition of states generally leads to appearance of cross-terms $W_{(\phi_n,\phi_m)}$ in the Wigner distribution.
Realness of $W_{(\psi)}$ is readily verified by noting that the off-diagonal terms in Eq.~\ref{eq:sup} are complex conjugates of each other $(\alpha_n^* \alpha_m W_{(\phi_n,\phi_m)})^* = \alpha_m^* \alpha_n W_{(\phi_m,\phi_n)} = \alpha_m^* \alpha_n W_{(\phi_n,\phi_m)}^*$ and therefore their sum must be real.

For example, consider a stationary Hamiltonian $\hat{\mathcal{H}}$ producing a complete orthogonal basis $\ket{n}$ with corresponding energy eigenvalues $E_n$
\begin{equation}
\hat{\mathcal{H}} \ket{n} = E_n \ket{n}.
\end{equation}
Then, the time evolution of an arbitrary state characterized by the initial vector $\ket{\psi_0} = \ket{\psi(t = 0)}$ adopts the familiar form
\begin{equation}\label{eq:state-evolve}
\ket{\psi(t)} = \sum\limits_n a_n \ket{n} e^{-\frac{i E_n t}{\hbar}},
\end{equation}
where the expansion coefficients are found in terms of projections of the initial state on the eigenbasis $a_n = \braket{n|\psi_0}$ and must satisfy normalization requirement $\sum_n |a_n|^2 = 1$.

The time evolution of the WDF for $\ket{\psi(t)}$ is given by
\begin{equation}\label{eq:wig_eigenbasis_decomp}
W_{(\psi)}(x,p;t) = \sum\limits_n \sum\limits_m a_n^* a_m e^{\frac{i(E_n-E_m)t}{\hbar}} W_{(n, m)}(x,p).
\end{equation}
Eq.~\ref{eq:wig_eigenbasis_decomp} can be also rewritten using (\ref{eq:gen_scalar}) in terms of integrated products of the Wigner functions
\begin{equation}
W_{(\psi)}(x,p;t) = h \sum\limits_n \sum\limits_m \left[\iint  W_{(\psi_0)}(x',p') W_{(m, n)}(x',p')\,dx' dp' \right] W_{(n, m)}(x,p) e^{\frac{i(E_n-E_m)t}{\hbar}}.
\end{equation}
Note that the decomposition of the Wigner distribution for a pure state into an othronormal basis generally requires the presence of non-vanishing interference cross-terms. We shall revisit this subject when considering incoherent addition of states.
\end{prop}

\begin{prop}[Gaussian state]\label{prop:gauss}
A positive WDF can only be realized for a wavefunction of the form
\begin{equation}
\psi(x) = e^{-(a x^2 + b x + c)},~\Re\{a\} > 0,
\end{equation}
leading to a joint Gaussian WDF in position and momentum \cite{Tatarskii83}. We also note in passing a well-known fact that a Gaussian state yields the phase-space probability density with the smallest rms spread (quantum emittance) $\epsilon = \hbar/2$.
\end{prop}

\begin{prop}[Convolution]\label{prop:conv}
The distribution function obtained by convolving two WDFs each corresponding to an arbitrary pure state is everywhere positive.
\end{prop}
This property is introduced in \cite{1983PhLA...94..415B} and is mentioned here for completeness. It should be noted that the distribution function so obtained is no longer the quasi-probability suitable for finding the expectation values of the original state. For further implications of this property, including it being a possible mathematical tool of the concept of measurement in quantum mechanics, the reader is directed to the discussion in \cite{1983PhLA...94..415B}.

\begin{figure}[htb]
\begin{center}
	\includegraphics[width=0.8\columnwidth]{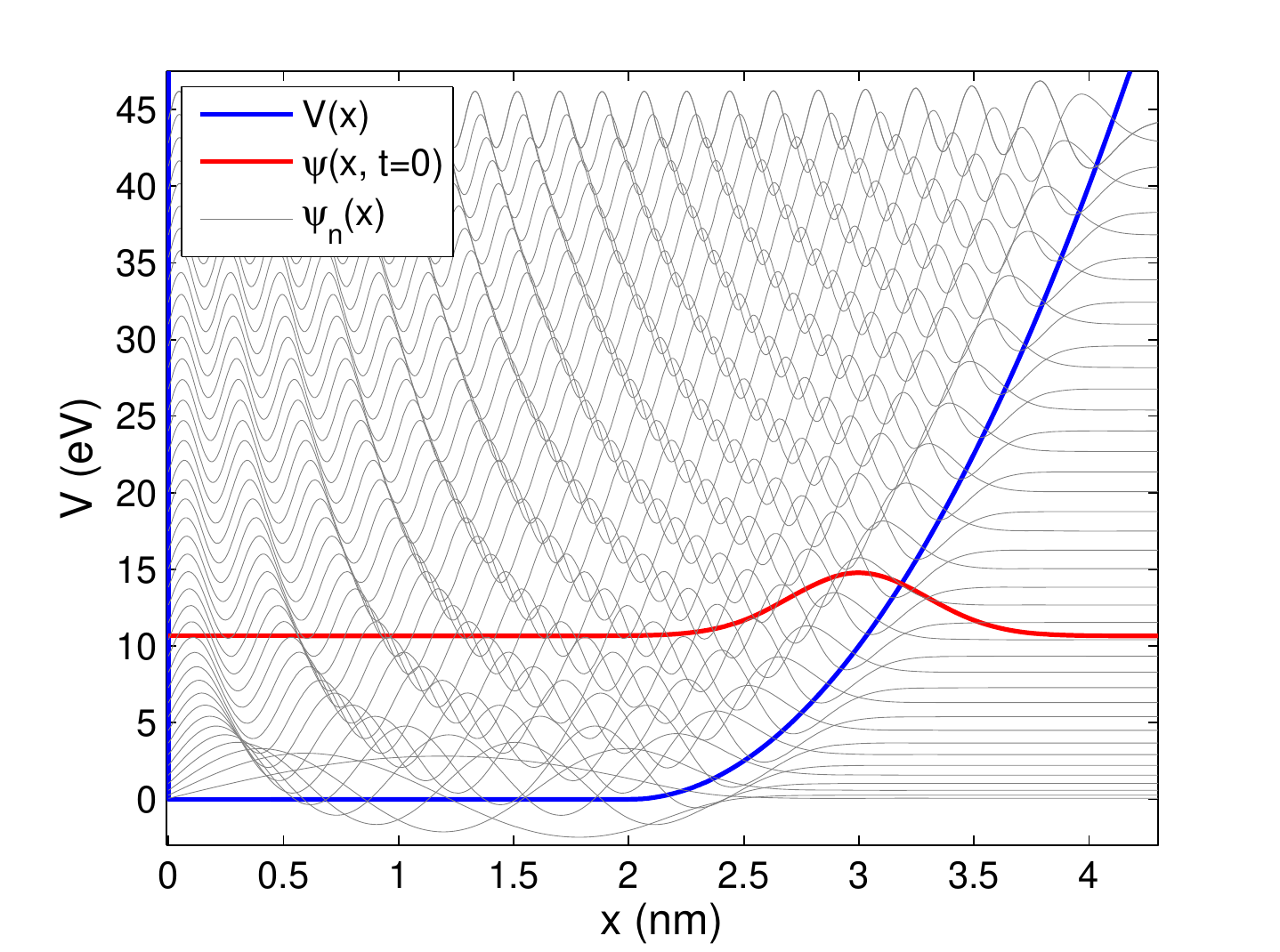}
	\caption{Potential for an electron $V(x)$. The initial state $\psi(x,t=0)$ and the first 40 energy eigenstates $\psi_n(x)$ offset by their eigenvalues are shown.\label{fig:potential_initialstate_eigenstate}}
\end{center}
\end{figure}

\subsubsection{Example: wave packet time evolution in 1D potential}
Next, we consider several examples that illustrate the concept of the Wigner distribution. The first example shows the phase-space motion of an electron in 1D potential depicted in Fig.~\ref{fig:potential_initialstate_eigenstate}. The potential consists of a perfectly reflecting barrier on the left and a simple harmonic oscillator (SHO) potential on the right. In units of eV the potential is given by (the position $x$ is in nm):
\begin{equation}
V(x) = \begin{cases} \infty, & x \leq 0 \\
0, & 0 < x \leq 2 \\
10(x-2)^2, & x > 2 \end{cases}
\end{equation}

\begin{figure}[thb]
\begin{center}
	\includegraphics[width=0.8\columnwidth]{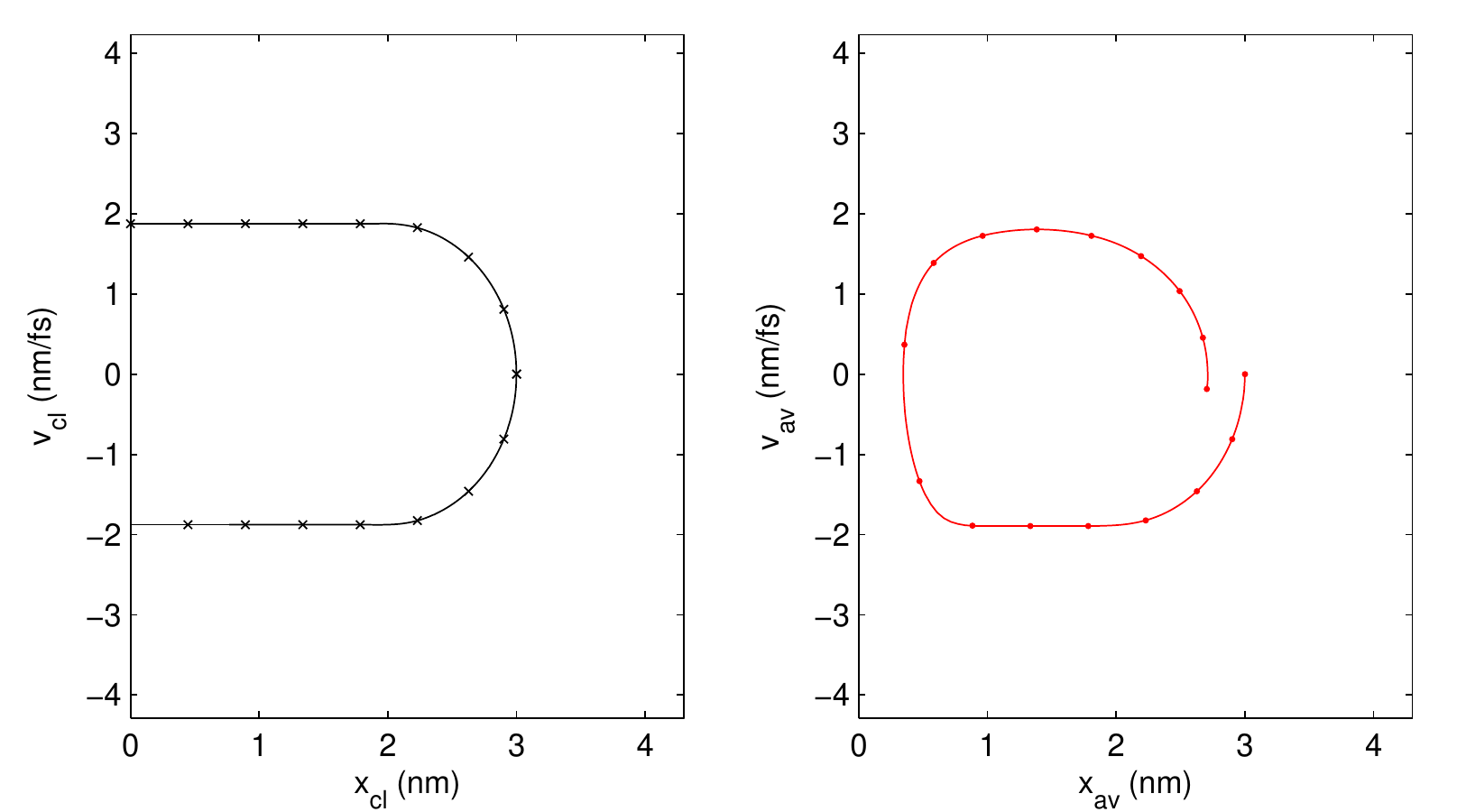}
	\caption{Motion of a classical (left) and quantum (right) electron in phase space. In both cases the initial position is 3\,nm with no velocity. Points show time steps from $0$ to $T$ with steps of $T/16$, where $T=3.8$\,fs is the natural period of the motion.\label{fig:centroid_phase_space}}
\end{center}
\end{figure}

The initial wave packet is described by a Gaussian $\psi(x, t=0) \propto e^{-\frac{(x-x_0)^2}{2 \sigma_x^2}}$ with $x_0 = 3$ and $\sigma_x = 0.3$. Fig.~\ref{fig:potential_initialstate_eigenstate} shows $\psi(x, t=0)$ and first 40 energy eigenstates $\psi_n(x)$. The intial quantum state is then evolved according to Eq.~\ref{eq:state-evolve}. The motion in the phase space of a classical particle with an initial position $x=3$ and a zero velocity is shown in Fig.~\ref{fig:centroid_phase_space} along with expectation values for position $x_{av} = \braket{x}$ and velocity $v_{av}=\braket{p}\!/m$ for the quantum case. Points on the plot correspond to $T/16$ time steps where $T$ is the natural period of motion of the system, equal to about 3.8\,fs in this case. The apparent damping in the quantum case corresponds to delocalization of the initial Gaussian wave packet.

\begin{figure}[thb]
\begin{center}
	\includegraphics[width=0.9\columnwidth]{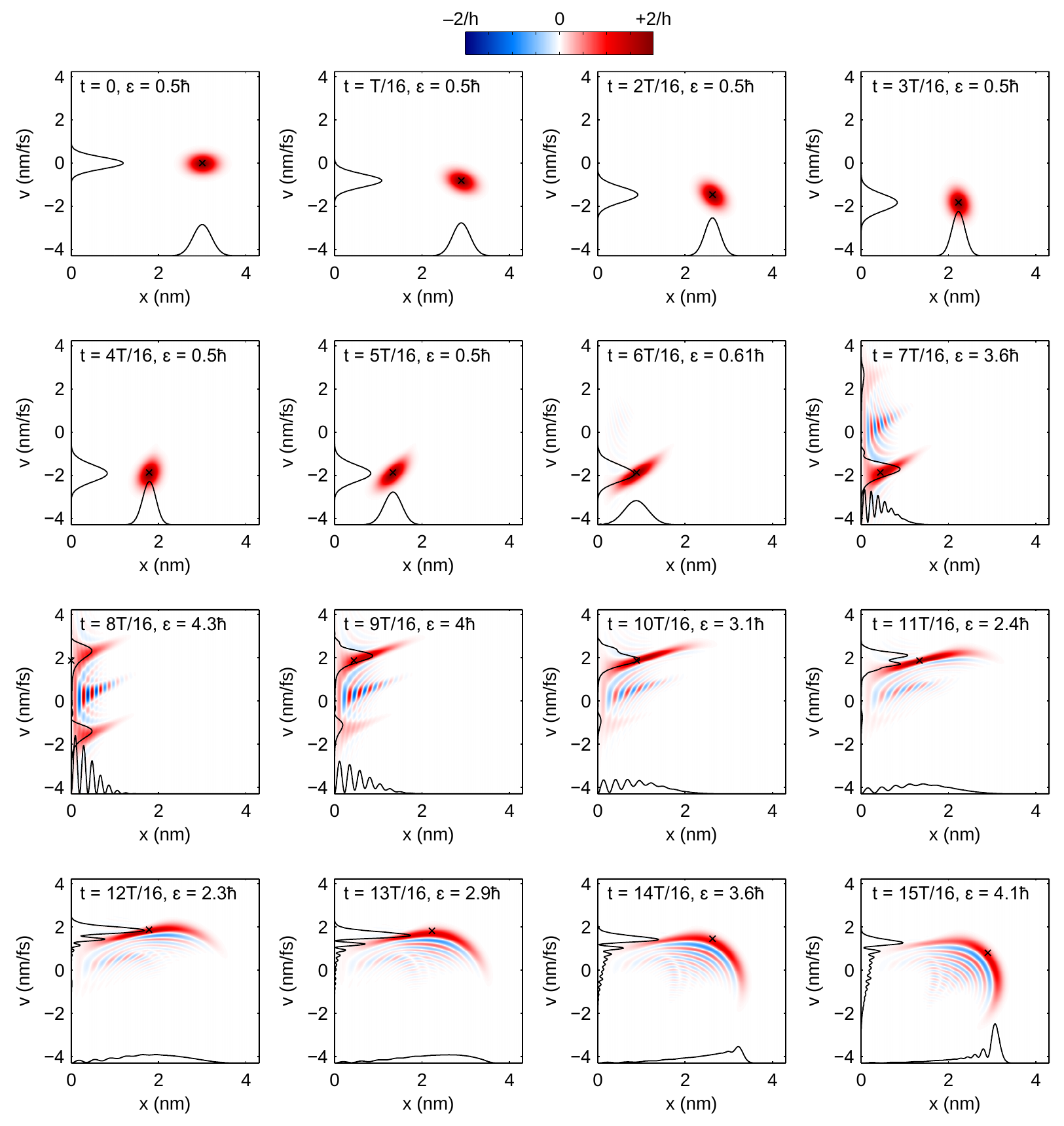}
	\caption{Time evolution of the Wigner distribution. Solid curves show projections of the WDF, probability densities in position and momentum (velocity) respectively. The electron's classical counterpart is depicted by $\times$.\label{fig:wigner_period_over_16_steps}}
\end{center}
\end{figure}

Fig.~\ref{fig:wigner_period_over_16_steps} shows the Wigner distribution of the quantum motion along with its classical counterpart (marked by $\times$) at the same times as depicted on Fig.~\ref{fig:centroid_phase_space}. The Wigner projections $\left|\braket{x|\psi}\right|^2$ and $\left|\braket{p|\psi}\right|^2$ are also shown. In addition, the rms emittance of the quantum phase space distribution is shown. As can be seen from Fig.~\ref{fig:wigner_period_over_16_steps}, the emittance of the initial wave packet is $\hbar/2$, increasing when the wave packet reaches the hard reflective potential boundary (and more generally when discontinuities in the potential are encountered). Here and in all subsequent plots of the Wigner distribution we use the same color map: blue and red colors correspond to negative and positive values respectively and the white corresponds to zero. It should be noted that despite increase in the phase space area, the mode obviously remains pure at all times.

\subsubsection{Example: eigenstates of a simple harmonic oscillator}

\begin{figure}[bth]
\begin{center}
	\includegraphics[width=1.0\columnwidth]{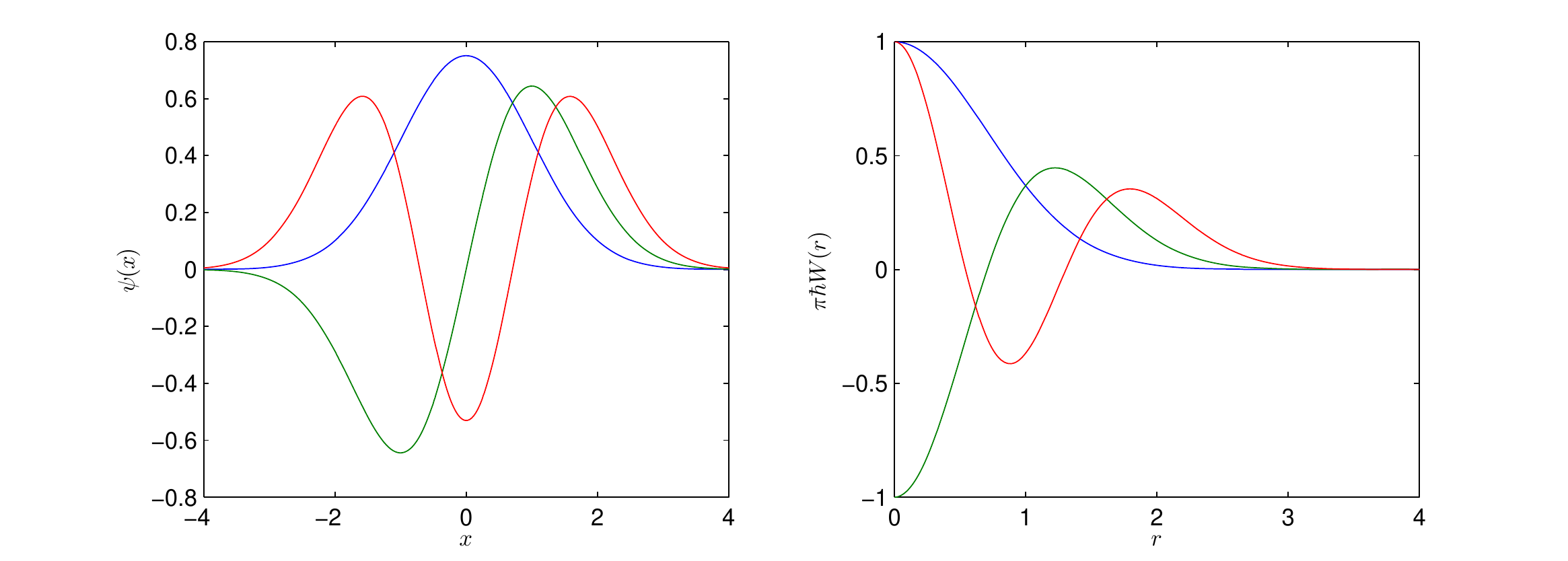}
	\caption{First three states (left) of a simple harmonic oscillator along with the radial Wigner distribution (right).\label{fig:sho}}
\end{center}
\end{figure}

Another example we consider is the WDF of energy eigenstates for a SHO \cite{Schleich2001}, which adopts the same math as Hermite-Gaussian modes in optics (at a waist and a single coordinate). The well-known eigenstates of the Hamiltonian $\hat{\mathcal{H}} = -\tfrac{1}{2}\tfrac{d^2}{dx^2}+\tfrac{1}{2} x^2$ are of the form
\begin{equation}
\psi_n(x) = \frac{\pi^{-\frac{1}{4}}}{\sqrt{2^n n!}} e^{-\frac{x^2}{2}} H_n(x),\mbox{ with }n = 0, 1, 2, \ldots.
\end{equation}
$H_n$ are Hermite polynomials. For this example we set the usual constants $\hbar, m, \omega \rightarrow 1$ to minimize clutter in the expressions. The Wigner distribution is then a radial function in phase space with $r^2 = x^2 + p^2$
\begin{equation}\label{eq:wig_sho}
W = \frac{(-1)^n}{\pi} e^{-r^2} L_n(2 r^2).
\end{equation}
Here $L_n$ are Laguerre polynomials. The first three states along with $W(r)$ are shown in Fig.~\ref{fig:sho}. We make a couple of observations regarding Eq.~\ref{eq:wig_sho}:
\begin{itemize}
\item The WDF is the maximum possible value at the origin for even $n$ and the minimum possible for odd $n$ in accordance with Eq.~\ref{eq:maxmin}: $W(0) = (-1)^n/\pi$ or in the regular units
\begin{equation}
W(x = 0, p = 0) = \frac{(-1)^n}{\hbar\pi}.
\end{equation}
\item The uncertainty in position and momentum or emittance for mode $n$ is $2n+1$ times $1/2$ in the natural units or $\hbar/2$ in regular units. In optics, $\epsilon/(\hbar/2)$ quantity is known as the $M^2$-parameter, i.e.
\begin{equation}
M^2 = 2n+1.
\end{equation}
\end{itemize}
In natural units used in this example we have
\begin{equation}\label{eq:M2}
\braket{x^2} = \braket{p^2} = \epsilon = \frac{M^2}{2}.
\end{equation}
Thus, the ground Gaussian state has the smallest possible uncertainty of $1/2$ (or $\hbar/2$), whereas each subsequent excitation adds an additional node to the wavefunction and the radial Wigner distribution, and increases the emittance by 1 (or $\hbar$).

\subsection{Mixed quantum state}
Generally real quantum systems cannot be described as a superposition of pure modes (which is itself a pure mode) instead adopting a statistical language to describe an \textit{incoherent} mixture of pure states characterized via the density operator
\begin{equation}
\hat{\rho} = \sum\limits_{j} p_j \ket{\psi_j}\!\bra{\psi_j},
\end{equation}
with state probabilities $p_j$ adding up to one. When only one coefficient $p_j = 1$ for some $j$ is present the formalism is reduced to that of a pure state. Recall that an expectation value of an operator $\hat{A}$ is given in terms of a trace
\begin{equation}
\braket{\hat{A}} = \sum\limits_{j} p_j \braket{\psi_j|\hat{A}|\psi_j} = \mathrm{Tr}(\hat{\rho} \hat{A}).
\end{equation}
Other standard properties are
\begin{align}
& \mathrm{Tr}(\hat{\rho}) = 1, \\
& \mathrm{Tr}(\hat{\rho}^2) \leq 1, \label{eq:tr2}
\end{align}
where the equal sign in Eq.~\ref{eq:tr2} is for a pure state case and less than 1 otherwise.

The definitions Eq.~\ref{eq:wigx} and \ref{eq:wigp} are now replaced with
\begin{equation}\label{eq:wig_den_states}
\begin{aligned}
W(x,p) & = \frac{1}{h} \int \braket{x-\tfrac{x'}{2} | \hat{\rho} | x+\tfrac{x'}{2}} e^\frac{ipx'}{\hbar} dx',\\
& = \frac{1}{h} \int \braket{p-\tfrac{p'}{2} | \hat{\rho} | p+\tfrac{p'}{2}} e^{-\frac{ip'x}{\hbar}} dp'.
\end{aligned}
\end{equation}
In other words, the WDF of a mixed state is a weighted sum of the WDFs corresponding to individual pure states of the density matrix
\begin{equation}\label{eq:wig_incoh}
W(x,p) = \sum\limits_{j} p_j W_{(\psi_j)}(x,p).
\end{equation}
Since, the Wigner distribution is a quadratic (intensity-like) function of the state, addition of individual WDFs corresponds to an incoherent addition, which is to be contrasted with coherent superposition of Eq.~\ref{eq:sup}. This language of a pure state vs.~a mixed state, incoherent addition vs.~coherent superposition carries over directly to optics and allows characterization of partially coherent sources.

Next, we present a new property that relates the WDF to modal purity. We then revisit some of the properties introduced earlier to extend them for the mixed state case. Most of the properties discussed previously, namely Properties \ref{prop:real} through \ref{prop:invert} directly carry over to the mixed state case after the necessary modifications to Properties \ref{prop:norm}, \ref{prop:cross}, and \ref{prop:invert}.

\begin{prop}[Measure of modal purity]
The integrated WDF squared is a measure of modal purity
\begin{equation}\label{eq:purity}
\mathrm{Tr}(\hat{\rho}^2) = h \iint W^2(x,p)\,dx\,dp \leq 1.
\end{equation}
The equal sign in Eq.~\ref{eq:purity} is for a pure state.
\end{prop}
The proof of this property follows directly from Eq.~\ref{eq:wig_incoh} and Property \ref{prop:scalar}.

\begin{prop}[Marginals --- Property \ref{prop:norm} revisited]
\label{prop:norm_rev}
\begin{equation}\label{eq:marginals}
\begin{aligned}
& \int W(x,p)\,dp = \braket{x|\hat{\rho}|x} = \sum\limits_j p_j |\psi_j(x)|^2,\\
& \int W(x,p)\,dx = \braket{p|\hat{\rho}|p} = \sum\limits_j p_j |\varPsi_j(p)|^2.
\end{aligned}
\end{equation}
\end{prop}
Again, this property reinforces the notion of simply adding the intensities (here probability densities) for mixed states.

\begin{prop}[Density matrix --- Property \ref{prop:cross} revisited]
\label{prop:cross_rev}
The density matrix is related to the Wigner distribution via a Fourier transform
\begin{equation}\label{eq:density_matrix}
\begin{aligned}
& \rho(x_1,x_2) \equiv \braket{x_1|\hat{\rho}|x_2} = \int W\!\left(\tfrac{x_1+x_2}{2}, p \right) e^\frac{i(x_1-x_2)p}{\hbar} dp,\\
& \varrho(p_1,p_2) \equiv \braket{p_1|\hat{\rho}|p_2} = \int W\!\left(x, \tfrac{p_1+p_2}{2} \right) e^{-\frac{ix(p_1-p_2)}{\hbar}} dx,
\end{aligned}
\end{equation}
which are simply the inverse of the Wigner function definitions
\begin{equation}\label{eq:wig_def_again}
\begin{aligned}
W(x,p) & = \int \rho(x-\tfrac{x'}{2}, x+\tfrac{x'}{2}) e^\frac{ipx'}{\hbar} dx',\\
& = \int \varrho(p-\tfrac{p'}{2}, p+\tfrac{p'}{2}) e^{-\frac{i p' x}{\hbar}} dp'.
\end{aligned}
\end{equation}

\end{prop}

\begin{prop}[Mode decomposition --- Property \ref{prop:invert} revisited]
\label{prop:invert_rev}
Property \ref{prop:invert} to invert the wavefunction from its WDF is only applicable for a pure state returning a meaningless ``wavefunction'' otherwise. Since the density matrix is a positive-semidefinite Hermitian operator with unit trace, it has an orthonormal basis of eigenstates $\phi_n$ whose corresponding real eigenvalues $\lambda_n \geq 0$ and $\sum_n \lambda_n = 1$ \cite{menskii­2000quantum}:
\begin{equation}\label{eq:decomp}
\hat{\rho} = \sum\limits_{n = 1}^N \lambda_n \ket{\phi_n}\!\bra{\phi_n},~\lambda_n\geq0,~\sum\limits_{n}\lambda_n = 1.
\end{equation}
Therefore, from Eq.~\ref{eq:wig_def_again}, the WDF can also be written as an incoherent sum of orthogonal pure modes
\begin{equation}\label{eq:wig_decomp}
W(x,p) = \sum\limits_{n=1}^N \lambda_n W_{(\phi_n)}(x,p).
\end{equation}
\end{prop}

The knowledge of the Wigner function or the corresponding density matrix $\hat{\rho}$ allows finding the modes and their weights via the standard eigenvector and eigenvalue problem, i.e.~as seen from multiplying both sides of Eq.~\ref{eq:decomp} by $\ket{\phi_m}$ and using orthonormality condition $\braket{\phi_n|\phi_m} = \delta_{nm}$
\begin{equation}\label{eq:eigen}
\hat{\rho} \ket{\phi_m} = \lambda_m \ket{\phi_m}.
\end{equation}
Any convenient complete orthonormal basis can be chosen to represent the density matrix $\hat{\rho}$ and its eigenstates $\phi_n$ \cite{Flewett:09}.

Some additional comments about this property are in order. The decomposition given by Eq.~\ref{eq:wig_decomp} is distinct from Eq.~\ref{eq:wig_incoh} in that the decomposition yields orthogonal states, which is generally not the case in how the mixed state has been originally set up. As a result, the modes of the decomposition, Eq.~\ref{eq:decomp}, may bear little resemblance to the original preparation states of the density matrix (i.e.~different mixtures may correspond to the same density operator). For the case of a mixed state with orthogonal preparation states, Eq.~\ref{eq:wig_incoh}, the decomposition recovers the modes and their weights exactly. Small negative eigenvalues $\lambda_n$ usually indicate an experimental error in arriving at the density matrix (or the Wigner distribution) \cite{Flewett:09} and can serve as a diagnostics and a self-consistency check. Finally, as is generally the case for pure states, the modes $\phi_n$ need not be simple in the sense that they don't necessarily have a small momentum-position uncertainty (i.e.~$M^2$ can be $M^2 \gg 1$). As an example, consider a relatively complicated mode of Fig.~\ref{fig:wigner_period_over_16_steps} after one or more oscillation periods, which, despite having a large dispersion, is still a pure (fully coherent) mode with $\mathrm{Tr}(\hat{\rho}^2) = 1$. The use of Property \ref{prop:invert_rev} recovers just that mode, which itself may have a very rich spectrum in some other basis.

\subsubsection{Example: superposition of two states}

In this example we demonstrate the difference between a coherent superposition and an incoherent mixture of two Gaussian states. A numerical example of a mixed mode decomposition further demonstrates the use of Property \ref{prop:invert_rev}.

First, let us consider another property useful in this example.
\begin{prop}[Quadratic phase]\label{prop:phase}
The effect of multiplying an arbitrary wavefunction $\psi(x)$ by a pure phase factor of the form $e^{i(c_0 + c_1 x + \tfrac{1}{2} c_2 x^2)}$ for arbitrary real coefficients $c_0$, $c_1$, and $c_2$
\begin{equation}
\widetilde{\psi}(x) = \psi(x) e^{i(c_0 + c_1 x + \tfrac{1}{2} c_2 x^2)},
\end{equation}
leads to the WDF $W_{(\widetilde{\psi})}$ related to the original $W_{(\psi)}$ via the linear momentum transformation
\begin{equation}
W_{(\widetilde{\psi})}(x,p) = W_{(\psi)}(x,\widetilde{p}),\mbox{ with }\widetilde{p} = p - \hbar(c_1 + c_2 x).
\end{equation}
\end{prop}
In other words, multiplying the wavefunction by a linear phase factor amounts to a shift in momentum, whereas the quadratic phase shift adds a linear correlation (chirp) to the momentum vs.~position.

\begin{figure}[hbt]
\begin{center}
	\includegraphics[width=0.9\columnwidth]{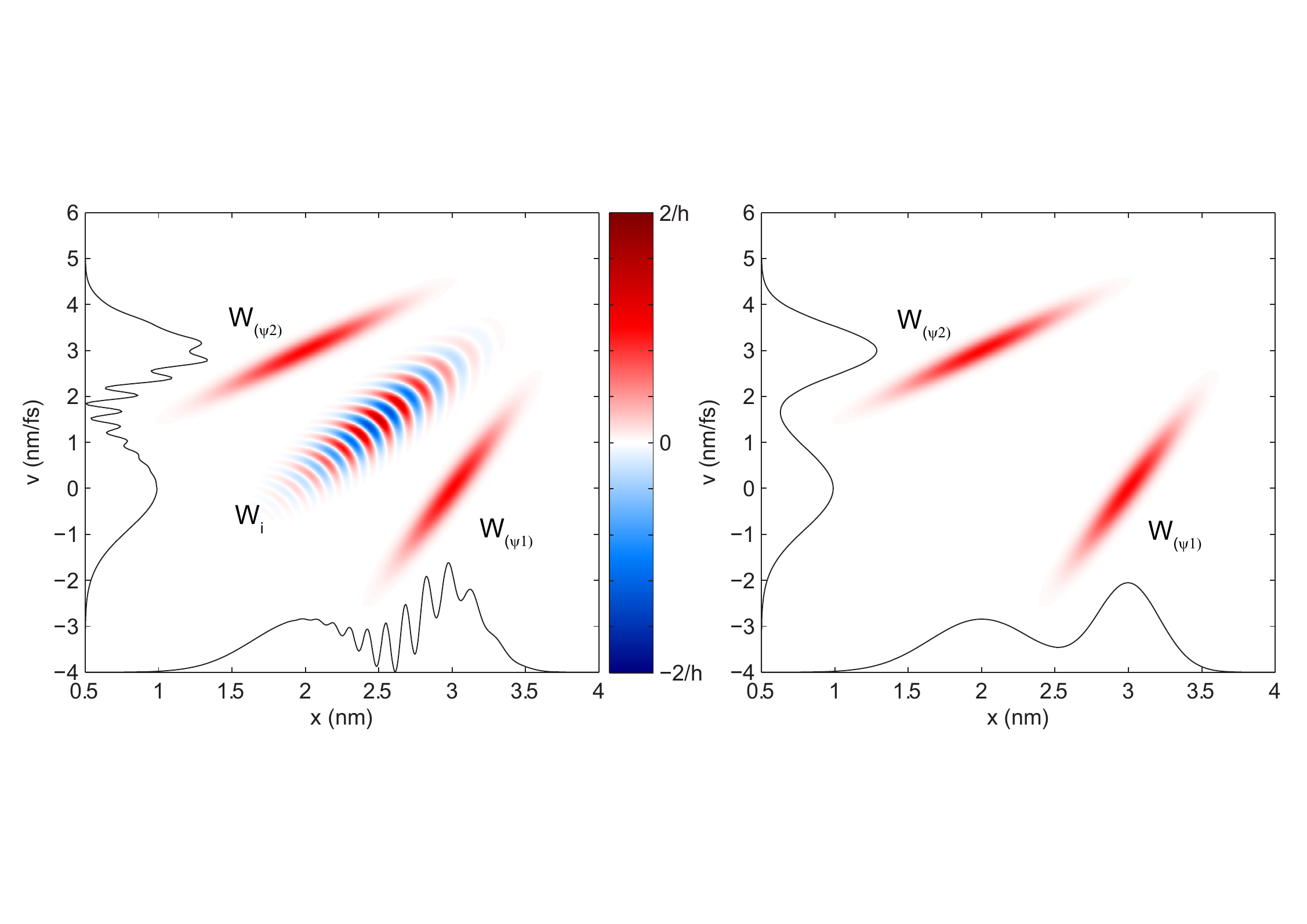}
	\caption{The Wigner distribution for coherent (left) and incoherent (right) superposition of two quantum states with equal weights. Solid curves show projections of the WDF, probability densities in position and velocity respectively. The states are of the form $\psi(x) \propto e^{-\Delta x^2/2\sigma_x^2}e^{i\tfrac{m}{\hbar}(v_0 \Delta x + \tfrac{1}{2}\tfrac{dv}{dx} \Delta x^2)}$, where $\Delta x = x-x_0$. The parameters $\{x_0, \sigma_x, v_0, dv/dx\}$ are equal to \{3\,nm, 0.3\,nm, 0\,nm/fs, 4\,fs$^{-1}$\} and \{2\,nm, 0.5\,nm, 3\,nm/fs, 1.5\,fs$^{-1}$\} for the two Gaussian states depicted.\label{fig:2gauss}}
\end{center}
\end{figure}

Fig.~\ref{fig:2gauss} shows a superposition and an incoherent addition of two Gaussian wave packets. In this example, the position is in nm and velocity is in nm/fs (an electron is assumed). Coherent superposition is for two Gaussian wave packets with equal weights (the Gaussians are nearly orthogonal as seen from the fact that their Wigner distributions don't overlap and the Property \ref{prop:scalar}).

\begin{figure}[thb]
\begin{center}
	\includegraphics[width=0.9\columnwidth]{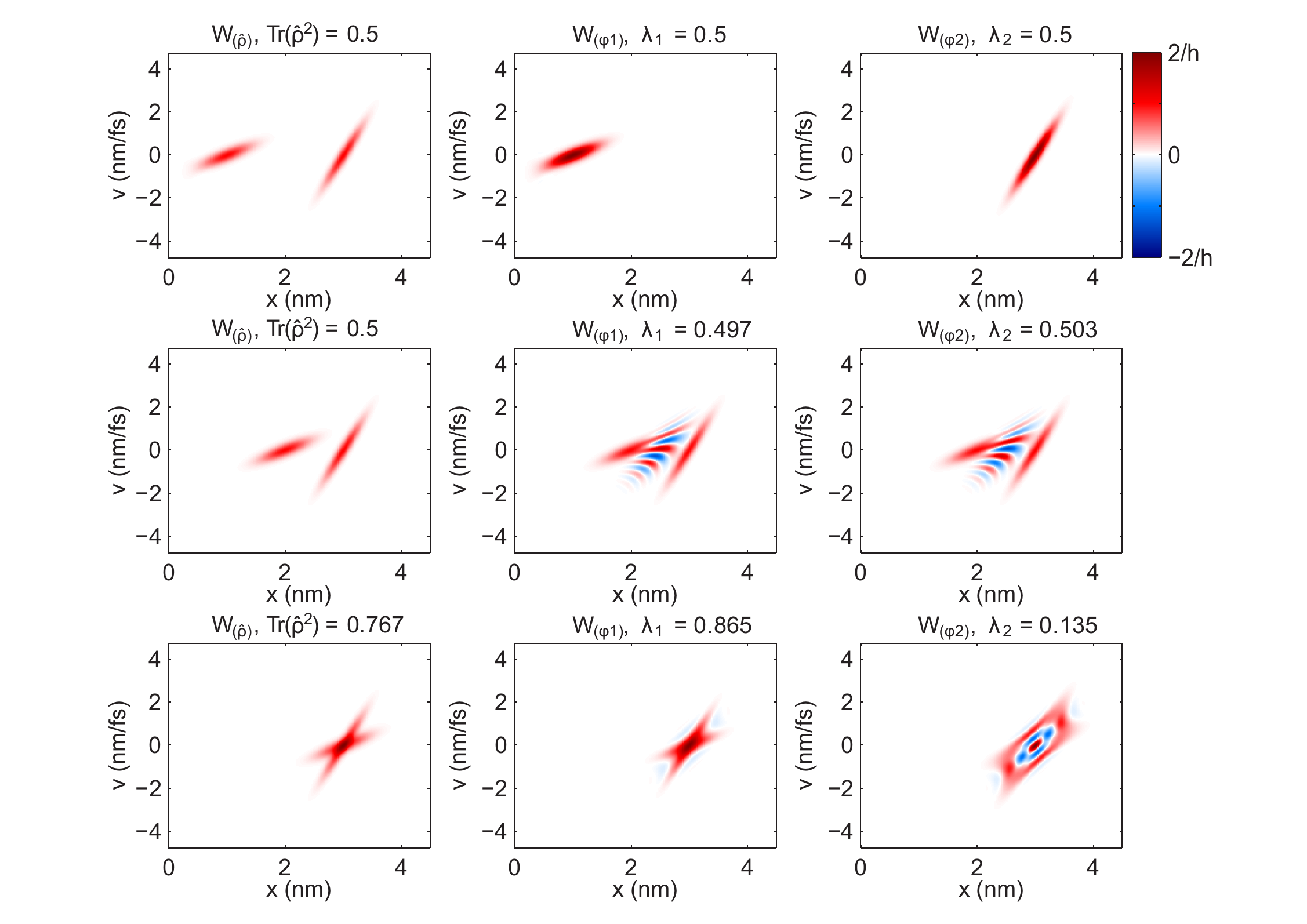}
	\caption{The Wigner distribution for a mixed state of two Gaussians of equal weight (left column) and corresponding orthogonal mode decomposition (middle and right columns). The preparation states are of the form $\psi(x) \propto e^{-\Delta x^2/2\sigma_x^2}e^{i\tfrac{m}{2\hbar}\tfrac{dv}{dx} \Delta x^2}$, where $\Delta x = x-x_0$. The parameters $\{x_0, \sigma_x, dv/dx\}$ are equal to \{3\,nm, 0.3\,nm, 4\,fs$^{-1}$\} for fixed and \{1-2-3\,nm, 0.4\,nm, 1\,fs$^{-1}$\} for displaced Gaussians. The reconstructed WDF is recovered through $W_{(\hat{\rho})} = \lambda_1 W_{(\phi_1)} + \lambda_2 W_{(\phi_2)}$\label{fig:decomp}}
\end{center}
\end{figure}

Consider a superposition of two states $\ket{\psi} \propto \ket{\psi_1} + \ket{\psi_2}$. Using Property \ref{prop:super}, the Wigner distribution has 3 terms
\begin{equation}\label{eq:two_states_superposition}
W_{(\psi)} \propto W_{(\psi_1)} + W_{(\psi_2)} + W_i,
\end{equation}
where the interference term $W_i = W_{(\psi_1,\psi_2)}+W_{(\psi_1,\psi_2)}^*$. The interference term is responsible for oscillations seen in Fig.~\ref{fig:2gauss}, and it is easy to show that $W_i$ for two orthogonal states carries no energy
\begin{equation}
\iint W_i(x,p)\,dx \, dp = 0,\mbox{ if }\braket{\psi_1|\psi_2} = 0.
\end{equation}
On the other hand, a mixed state $\hat{\rho} \propto \ket{\psi_1}\!\bra{\psi_1} + \ket{\psi_2}\!\bra{\psi_2}$ has only 2 terms from each individual states in its Wigner distribution
\begin{equation}\label{eq:two_states_incoherent}
W_{(\hat{\rho})} \propto W_{(\psi_1)} + W_{(\psi_2)},
\end{equation}
without the interference term.

Fig.~\ref{fig:decomp} demonstrates orthogonal mode decomposition using Property \ref{prop:invert_rev} for a mixed state with two Gaussians of equal weights. As seen from the Fig.~\ref{fig:decomp}, recovered modes generally reflect the shape of the mixed state and can have large dispersion (emittance) if the mixed state itself has had a large dispersion.

\subsubsection{Example: Gauss-Schell model}

Now we consider a quantum mechanical analog of what is known as a Guass-Schell model in optics \cite{Gori80}. Using same natural units of SHO with $\hbar, m, \omega \rightarrow 1$, we rewrite the Wigner distribution similar to Eq.~\ref{eq:wig_sho} in a generalized Gaussian form
\begin{equation}\label{eq:wig_gs}
W(r) = \frac{1}{M^2 \pi} e^{-\frac{r^2}{M^2}},~M^2 \geq 1,
\end{equation}
where as previously $r^2 = x^2 + p^2$. Setting $M^2 \rightarrow 1$ recovers a pure Gaussian ground state, whereas $M^2 > 1$ corresponds to a mixed state. As previously, Eq.~\ref{eq:M2} applies for our choice of units
\[
\braket{x^2} = \braket{p^2} = \epsilon = \frac{M^2}{2}.
\]
Next, we use Eq.~\ref{eq:density_matrix} to recover the density matrix
\begin{equation}\label{eq:den_gs}
\rho(x_1,x_2) = \frac{1}{\sqrt{\pi} M} e^{-\frac{1}{4}[M^2(x_1-x_2)^2+M^{-2}(x_1+x_2)^2]}.
\end{equation}
A more common form of presenting the density matrix is as a Schell-model source
\begin{equation}
\rho(x_1,x_2) \equiv \sqrt{I(x_1)} \sqrt{I(x_2)} \mu(x_1-x_2),
\end{equation}
where the probability density
\begin{equation}
I(x) = \rho(x,x) = \frac{1}{\sqrt{2\pi}\sigma_x} e^{-\frac{x^2}{2\sigma_x^2}},\mbox{ with } \sigma_x = \frac{M}{\sqrt{2}},
\end{equation}
and the \textit{degree of spatial coherence}
\begin{equation}
\mu(x_1-x_2) = e^{-\frac{(x_1-x_2)^2}{2 \sigma_\mu^2}},\mbox{ with } \sigma_\mu = \frac{\sqrt{2}M}{\sqrt{M^4-1}}.
\end{equation}

The function $\mu(\Delta x)$ is bound $0 \leq |\mu(\Delta x)| \leq 1$ with 1 or 0 corresponding to a perfect or no phase correlation respectively. $\sigma_\mu$ is known as a coherence length in optics. E.g.~$M^2 \rightarrow 1$ yields $\sigma_\mu \rightarrow \infty$ (a perfect phase correlation or pure state) whereas $M^2 \rightarrow \infty$ gives $\sigma_\mu \rightarrow 0$ (no phase correlation in the state).

Decomposition eigen problem (\ref{eq:eigen}) can be rewritten as
\begin{align*}
\braket{x|\hat{\rho}|\phi_m} & = \lambda_m \braket{x|\phi_m},\\
\int \braket{x|\hat{\rho}|x'}\!\braket{x'|\phi_m} dx' & = \lambda_m \braket{x|\phi_m},\\
\int \rho(x,x') \phi_m(x')\,dx' & = \lambda_m \phi_m(x).
\end{align*}
This Fredholm integral equation yields the following spectrum of eigenvalues and eigenfunctions for the density matrix of Eq.~\ref{eq:den_gs} \cite{Gori80,Starikov:82}
\begin{align}
& \phi_m(x) = \frac{\pi^{-\frac{1}{4}}}{\sqrt{2^m m!}} e^{-\frac{x^2}{2}} H_m(x),\\
& \lambda_m = \lambda_0 q^m,\mbox{ with }m = 0, 1, 2, \ldots,
\end{align}
where
\begin{align}
& \lambda_0 = \frac{2}{M^2+1},\\
& q = \frac{M^2-1}{M^2+1}.
\end{align}
Thus, the Gauss-Schell model adopts a particularly simple mode decomposition, which are the pure states of a simple harmonic oscillator. Additionally, it can be checked that
\begin{align}
\mathrm{Tr}(\hat{\rho}) & = \sum\limits_{m=0}^{\infty} \lambda_m = 1,\\
\mathrm{Tr}(\hat{\rho}^2) & = \sum\limits_{m=0}^{\infty} \lambda_m^2 = \frac{1}{M^2}.\label{eq:M2confusing}
\end{align}

Eq.~\ref{eq:M2confusing} can be a source of confusion in that one may be tempted to equate $M^2$ (phase-space area, or emittance, or dispersion) directly to spectral purity $\mathrm{Tr}(\hat{\rho}^2)$ for an \textit{arbitrary} mixed state. This temptation should be resisted since quantum-mechanically the two concepts, the phase-space uncertainty $M^2$ and the mode purity $\mathrm{Tr}(\hat{\rho}^2)$, are distinct as argued previously. To the extent that the Gauss-Schell model is applicable to a quantum or optical system, such blurred interpretation of $M^2$ simultaneously being a measure of dispersion and coherence may be justified. A notable exception is when $M^2 \rightarrow 1$, which corresponds to both perfect coherence and minimum uncertainty of a pure Gaussian state. We shall see later, however, that the synchrotron radiation from an undulator source by a \textit{single} electron is far from a Gaussian and, therefore, such dual interpretation of $M^2$ needs to be rejected for the diffraction-limited electron beams. Similarly, the use of the Gauss-Schell model on a distinctly non-Gaussian phase space distribution function has little merit.

\section{Wigner distribution for synchrotron radiation\label{sec:wig_opt}}

The connection of the Wigner distribution to describing partially coherent sources is usually made through the cross-spectral density function $\Gamma(\mathbf{r_1}, \mathbf{r_2}, \omega)$ \cite{Bastiaans:86}
\begin{equation}\label{eq:csd}
\Gamma(\mathbf{r_1}, \mathbf{r_2}; \omega) = \braket{E(\mathbf{r_1}; \omega) E^*(\mathbf{r_2}; \omega)}.
\end{equation}
Here $E(\mathbf{r_1},\omega)$ is frequency representation of the electric field, which assumed for now to be a scalar function (e.g.~linearly polarized light) of 2D transverse coordinates $\mathbf{r} = (x,y)$ (e.g.~the detector plane) and $\braket{\ldots}$ means ensemble average (e.g.~over electron bunches for synchrotron radiation). For Eq.~\ref{eq:csd} to fully describe coherence properties, the source needs to be stationary in that all ensemble averages do not vary with respect to time (or at least first and second moments are time-independent, which is a requirement for wide-sense stationary processes). The synchrotron radiation with its pulsed bunch structure is generally non-stationary. However, as argued in \cite{Geloni2008463}, one can use the cross-spectral density in the form of Eq.~\ref{eq:csd} if individual synchrotron radiation pulses last much longer than their coherence time (the time scale of short-term field fluctuations, inversely related to the source bandwidth), or $\sigma_t \gg N_u \omega_0$ for an undulator source with $N_u$ undulator periods and resonant (radiation) frequency $\omega_0$ and electron bunches of $\sigma_t$ duration. This condition is usually well satisfied (though an extension of the formalism can be straightforwardly made to describe nearly transform-limited sources in time-frequency domains). One also typically defines the spectral degree of coherence \cite{mandel1995optical}
\begin{equation}\label{eq:sdc}
\mu(\mathbf{r_1}, \mathbf{r_2}; \omega) = \frac{\Gamma(\mathbf{r_1}, \mathbf{r_2}; \omega)}{\sqrt{\Gamma(\mathbf{r_1}, \mathbf{r_1}; \omega)}\sqrt{\Gamma(\mathbf{r_2}, \mathbf{r_2}; \omega)}}.
\end{equation}
The modulus of the spectral degree of coherence ranges from 0 to 1 for incoherent to fully coherent sources, $0 \leq |\mu| \leq 1$. For a fully coherent radiation, $|\mu| = 1$ everywhere. This quantity is directly related to the fringe visibility in interference experiments.

A fully equivalent characterization can of course be made in time domain \cite{papoulis1968systems}. In what follows, we restrict our treatment to frequency domain, being a more natural choice for x-rays. Therefore, the frequency dependence for the functions will be understood while the symbol itself will usually be omitted from the expressions, e.g. $E(\mathbf{r}) \equiv E(\mathbf{r};\omega)$.

The Wigner distribution for optics is then given by \cite{WALTHER:68}
\begin{align}
W(\mathbf{r},\boldsymbol\uptheta) & = \left(\frac{1}{\lambda}\right)^{\!2} \int \Gamma(\mathbf{r}-\tfrac{\mathbf{r'}}{2}, \mathbf{r}+\tfrac{\mathbf{r'}}{2}) e^{i k \mathbf{r'}\cdot\boldsymbol\uptheta} d^2\mathbf{r'},\label{eq:wig_opt_pos}\\
& = \left(\frac{1}{\lambda}\right)^{\!2} \int \varGamma(\boldsymbol\uptheta-\tfrac{\boldsymbol\uptheta\mathbf{'}}{2}, \boldsymbol\uptheta+\tfrac{\boldsymbol\uptheta\mathbf{'}}{2}) e^{-i k \mathbf{r}\cdot\boldsymbol\uptheta\mathbf{'}} d^2\boldsymbol\uptheta\mathbf{'},\label{eq:wig_opt_angle}
\end{align}
where transverse position $\mathbf{r}$ and angle $\boldsymbol\uptheta=(\theta_x,\theta_y)$ form a conjugate pair similar to position-momentum in quantum mechanics (small angle approximation is used throughout). Cross-spectral density in position and angular representations are defined according to
\begin{align}
\Gamma(\mathbf{r_1}, \mathbf{r_2}) & = \braket{E(\mathbf{r_1}) E^*(\mathbf{r_2})},\\
\varGamma(\boldsymbol\uptheta_\mathbf{1}, \boldsymbol\uptheta_\mathbf{2}) & = \braket{\mathcal{E}(\boldsymbol\uptheta_\mathbf{1}) \mathcal{E}^*(\boldsymbol\uptheta_\mathbf{2})},
\end{align}
where the angular representation $\mathcal{E}(\boldsymbol\uptheta)$ of radiation (far field) is related to its spatial representation $E(\mathbf{r})$ via the Fourier transform pair
\begin{equation}
\begin{aligned}
\mathcal{E}(\boldsymbol\uptheta) & = \frac{1}{\lambda} \int E(\mathbf{r}) e^{-ik\mathbf{r}\cdot\boldsymbol\uptheta} d^2\mathbf{r},\\
E(\mathbf{r}) & = \frac{1}{\lambda} \int \mathcal{E}(\boldsymbol\uptheta) e^{ik\mathbf{r}\cdot\boldsymbol\uptheta} d^2\boldsymbol\uptheta.
\end{aligned}
\end{equation}
The radiation wavenumber $k$ above is given by $k = 2\pi/\lambda = \omega/c$ in terms of wavelength $\lambda$, frequency $\omega$, and the speed of light $c$.

The connection to quantum mechanics now becomes obvious. Refer to Table~\ref{tab:correspondence}. All the properties introduced in the previous section have their counterparts in optics. In particular, we note that the radiation wavelength $\lambda$ in wave optics plays a role of Planck constant $h$ in quantum mechanics. E.g.~geometric optics is recovered in the limit $\lambda \rightarrow 0$ just as the classical behavior can be obtained through $h \rightarrow 0$.
\begin{table}[h!]
\caption{Correspondence between quantum and optical formalisms.}
\label{tab:correspondence}
\begin{center}
\begin{tabular}{|l|c||c|r|}
\hline
\multicolumn{2}{|c||}{Quantum mechanics}&\multicolumn{2}{|c|}{Wave optics}\\
\hline\hline
State & $\psi(x)$ & $E(\mathbf{r})$ & Field \\
Planck constant & $h$ & $\lambda$ & Wavelength \\
Uncertainty principle & $\epsilon \geq h/4\pi$ & $\epsilon_{x,y} \geq \lambda/4\pi $ & Diffraction limit \\
Conjugate pair & $x \xleftrightarrow[]{~\mathcal{FT}~} p $ & $\mathbf{r} \xleftrightarrow[]{~\mathcal{FT}~} \boldsymbol\uptheta $ & Conjugate pair \\
Density matrix & $\rho(x_1, x_2)$ & $\Gamma(\mathbf{r_1}, \mathbf{r_2})$ & Cross-spectral density \\
Classical mechanics & $h \rightarrow 0 $ & $\lambda \rightarrow 0$ & Geometric optics \\
\hline\hline
Phase space density & $W(x,p)$ & $W(\mathbf{r},\boldsymbol\uptheta)$ & Spectral brightness \\
Normalized to & 1 & spectral flux & Normalized to \\
Measure of state purity & $h\iint W^2 dx\,dp$ & $\lambda^2\frac{\iint W^2 d^2\mathbf{r}\,d^2\boldsymbol\uptheta}{(\iint W d^2\mathbf{r}\,d^2\boldsymbol\uptheta)^2}$ & Measure of coherence \\
\hline
\end{tabular}
\end{center}
\end{table}

Also, the overall degree of coherence $\mu_g^2$, which is directly equivalent to $\mathrm{Tr}(\hat{\rho}^2)$ of density matrix $\hat{\rho}$ in quantum mechanics, can be expressed in terms of the Wigner distribution function
\begin{equation}
\mu_g^2 = \lambda^2\frac{\iint W^2(\mathbf{r}, \boldsymbol\uptheta)\, d^2\mathbf{r}\,d^2\boldsymbol\uptheta}{(\iint W(\mathbf{r}, \boldsymbol\uptheta)\,d^2\mathbf{r}\,d^2\boldsymbol\uptheta)^2},
\end{equation}
where the denominator, the total flux squared, plays a normalization role so that $0 \leq \mu_g^2 \leq 1$.

\subsection{Polarized light}

Treatment of polarized light \cite{Luis:07} is directly analogous to WDF of spin-$\tfrac{1}{2}$ quantum particle \cite{PhysRevA.30.2613}, which require a 2-component spinor to characterize a state. The reason that a spin-1 particle (photon) can be described by a 2-component spinor (as opposed to 3) is well known in that only $\pm\hbar$ spin projections along the direction of propagation (helicity) are realized for a massless particle.

The Wigner distribution now becomes a $2\times2$ matrix $\mathbf{W}(\mathbf{r}, \boldsymbol\uptheta)$ (complex for off-diagonal elements) with components defined according to
\begin{equation}\label{eq:wigmat}
W_{kl}(\mathbf{r}, \boldsymbol\uptheta) = \left( \frac{1}{\lambda} \right)^{\!2} \int \braket{E_k^*(\mathbf{r}+\tfrac{\mathbf{r'}}{2}) E_l(\mathbf{r}-\tfrac{\mathbf{r'}}{2})} e^{ik\mathbf{r'}\cdot\boldsymbol\uptheta} d^2\mathbf{r'}, \mbox{ with } k,l = x,y.
\end{equation}
Generalizing the formalism of polarized light \cite{goldstein2003polarized, Luis:07}, $\mathbf{W}(\mathbf{r}, \boldsymbol\uptheta)$ can be represented as a scalar function on the Poincar\'{e} sphere using
\begin{equation}
W(\mathbf{r}, \boldsymbol\uptheta, \Omega) = \boldsymbol\Omega\cdot \mathbf{S}(\mathbf{r}, \boldsymbol\uptheta),
\end{equation}
where generalized Stokes parameters are found from
\begin{equation}
S_j(\mathbf{r}, \boldsymbol\uptheta) = \mathrm{Tr}[\sigma_j \mathbf{W}(\mathbf{r}, \boldsymbol\uptheta)],\mbox{ with }j = 0, 1, 2, 3.
\end{equation}
Here $\sigma_j$ are $2\times2$ Pauli matrices with $\sigma_0$ being an identity matrix, and $\boldsymbol\Omega$ is a vector mapping Stokes parameters onto the Poincar\'{e} sphere with polar $\chi$ and azimuthal $\phi$ angles
\begin{equation}
\Omega = \tfrac{1}{2}\left(
\begin{array}{c}
1\\
\sqrt{3} \sin\chi \cos\phi\\
\sqrt{3} \sin\chi \sin\phi\\
\sqrt{3} \cos\chi
\end{array}
\right).
\end{equation}

The generalized Stokes parameters, which now play a role of a 4-component phase space distribution, can we written explicitly in terms of the WDF components
\begin{equation}\label{eq:gstokes}
\begin{aligned}
S_0(\mathbf{r}, \boldsymbol\uptheta) & = W_{xx}(\mathbf{r}, \boldsymbol\uptheta) + W_{yy}(\mathbf{r}, \boldsymbol\uptheta) \\
S_1(\mathbf{r}, \boldsymbol\uptheta) & = W_{xy}(\mathbf{r}, \boldsymbol\uptheta) + W_{yx}(\mathbf{r}, \boldsymbol\uptheta) \\
S_2(\mathbf{r}, \boldsymbol\uptheta) & = i[W_{xy}(\mathbf{r}, \boldsymbol\uptheta) - W_{yx}(\mathbf{r}, \boldsymbol\uptheta)] \\
S_3(\mathbf{r}, \boldsymbol\uptheta) & = W_{xx}(\mathbf{r}, \boldsymbol\uptheta) - W_{yy}(\mathbf{r}, \boldsymbol\uptheta).
\end{aligned}
\end{equation}
In what follows we may occasionally refer to the generalized Stokes parameters as simply Wigner distribution functions. These four functions completely characterize radiation of arbitrary degree of coherence and polarization in phase space. They can be propagated in linear optics just like the scalar WDF. The usual Stokes parameters are found from those in Eqs.~\ref{eq:gstokes} by integrating away angles
\begin{equation}
s_j(\mathbf{r}) = \int S_j(\mathbf{r}, \boldsymbol\uptheta)\,d^2\boldsymbol\uptheta,\mbox{ with }j = 0, 1, 2, 3.
\end{equation}
Thus, the generalized Stokes parameters have their usual meaning for polarized light (the exact ordering of the Stokes components 1 through 3 may differ in literature): $S_0(\mathbf{r}, \boldsymbol\uptheta)$ represents total intensity in phase space, $S_1(\mathbf{r}, \boldsymbol\uptheta)$ represents $+45^\circ/-45^\circ$ linearly polarized light (for $+/-$ respectively), $S_2(\mathbf{r}, \boldsymbol\uptheta)$ corresponds to right/left-hand circular polarization, and $S_3(\mathbf{r}, \boldsymbol\uptheta)$ to $x/y$-linear polarization. We note that exact signs here apply only to intensity projections $s_j(\mathbf{r})$ since the WDF ($S_0$) is allowed to take on negative values while its projections (marginals) are guaranteed to be positive. For example, a Gaussian mode with $x$-polarization will have $S_0(\mathbf{r}, \boldsymbol\uptheta) = S_3(\mathbf{r}, \boldsymbol\uptheta) > 0$ with other Stokes parameters being 0, or for left-hand circular polarization $S_2(\mathbf{r}, \boldsymbol\uptheta) = -S_0 (\mathbf{r}, \boldsymbol\uptheta)$. Whereas fully polarized light satisfies $s_0 = \sqrt{s_1^2 + s_2^2 + s_3^2}$ and partial polarization manifests itself as $s_0 > \sqrt{s_1^2 + s_2^2 + s_3^2}$, the generalized Stokes parameter $S_0$ can take on local negative values and deviate from these expressions.

As shown in \cite{Luis:07}, the overall degree of coherence for vectorial waves can be written as
\begin{equation}
\mu_g^2 = 2\pi \lambda^2\frac{\iint d^2\mathbf{r}\,d^2\boldsymbol\uptheta \int_{4\pi} d^2 \Omega \, W^2(\mathbf{r}, \boldsymbol\uptheta, \Omega)}{[\iint d^2\mathbf{r}\,d^2\boldsymbol\uptheta \int_{4\pi} d^2 \Omega \, W(\mathbf{r}, \boldsymbol\uptheta, \Omega)]^2},
\end{equation}
where $d^2 \Omega = \sin\chi \, d\chi \, d\phi$. Or equivalently in terms of generalized Stokes parameters
\begin{equation}
\mu_g^2 = \tfrac{1}{2} \lambda^2\frac{\iint \mathbf{S}^2(\mathbf{r}, \boldsymbol\uptheta)\,d^2\mathbf{r}\,d^2\boldsymbol\uptheta}{[\iint S_0(\mathbf{r}, \boldsymbol\uptheta)\,d^2\mathbf{r}\,d^2\boldsymbol\uptheta]^2},
\end{equation}
and explicitly in terms of the WDF components as
\begin{equation}
\mu_g^2 = \lambda^2\frac{\iint (W_{xx}^2+ 2 W_{xy}W_{yx} + W_{yy}^2)\,d^2\mathbf{r}\,d^2\boldsymbol\uptheta}{[\iint (W_{xx} + W_{yy})\,d^2\mathbf{r}\,d^2\boldsymbol\uptheta]^2}.
\end{equation}

\subsection{Wigner distribution projections}

One of practical limitations of the Wigner distribution is that generally one needs to employ four-dimensional arrays as a function of light frequency (and possibly time if the temporal structure inside an individual synchrotron pulse is important) times 4 for Stokes parameters to represent the radiation fully. In addition to large memory requirements, one typically prefers to visualize two-dimensional projections rather than the entire phase space, much as it is done in accelerator physics for particle tracking. Here, we mention some of the properties of such projected WDFs, limiting our discussion to linearly polarized light for simplicity. Important 2D projections of the Wigner distribution are intensity $I(x,y)$, the far field (angular) intensity $\mathcal{I}(\theta_x, \theta_y)$, $x-\theta_x$ and $y-\theta_y$ phase space projections, $\mathcal{B}_x(x,\theta_x)$ and $\mathcal{B}_y(y,\theta_y)$.
\begin{align}
I(x,y) & \equiv \iint W(x,y,\theta_x,\theta_y)\,d\theta_x\,d\theta_y,\\
\mathcal{I}(\theta_x,\theta_y) & \equiv \iint W(x,y,\theta_x,\theta_y)\,dx\,dy,\\
\mathcal{B}_x(x,\theta_x) & \equiv \iint W(x,y,\theta_x,\theta_y)\,dy\,d\theta_y,\\
\mathcal{B}_y(y,\theta_y) & \equiv \iint W(x,y,\theta_x,\theta_y)\,dx\,d\theta_x.
\end{align}
Carrying out the integration and using the identity $\int e^{ikab} da = 2\pi \delta(b)/k$, one obtains
\begin{align}
I(x,y) & = \braket{E^*(x,y)E(x,y)},\\
\mathcal{B}_x(x,\theta_x) & = \frac{1}{\lambda} \iint \Gamma_x(x-\tfrac{x'}{2},x+\tfrac{x'}{2}) e^{ikx'\theta_x} dx',\label{eq:bx}\\
\mathcal{B}_y(y,\theta_y) & = \frac{1}{\lambda} \iint \Gamma_y(y-\tfrac{y'}{2},y+\tfrac{y'}{2}) e^{iky'\theta_y} dy',\label{eq:by}\\
\mathcal{I}(\theta_x,\theta_y) & = \braket{\mathcal{E}^*(\theta_x,\theta_y) \mathcal{E}(\theta_x,\theta_y)},
\end{align}
where $\Gamma_x(x_1, x_2) \equiv \int \braket{E(x_1,y)E^*(x_2,y)} dy$ and $\Gamma_y(y_1, y_2) \equiv \int \braket{E(x,y_1)E^*(x,y_2)} dx$.

If the radiation modes are separable, i.e.~can be written in the form $E(x,y) = \phi_x(x) \phi_y(y)$ (for example Hermite-Gaussian modes), then all the properties discussed in Section \ref{sec:wig_qm} for two-dimensional WDF in quantum mechanics apply to the Wigner 2D projections after normalization $W_{x,y} = \mathcal{B}_{x,y}/\mathcal{F}$, where $\mathcal{F} = \iint I(x,y)\,dx\,dy$ is the total (spectral) flux. It includes the interpretation of the $\lambda \iint W_{x}^2(x,\theta_x)\,dx\,d\theta_x$ and a similar expression for $y$-plane to be a measure of coherence $\mu_{gx,y}^2$ (the analog of the $\mathrm{Tr}(\hat{\rho}^2)$ in quantum mechanics). On the other hand, for non-separable radiation fields (e.g.~general radially symmetric modes), the same interpretation of $\lambda \iint W_{x}^2(x,\theta_x)\,dx\,d\theta_x = \mu_{gx}^2$ cannot be made.

Nevertheless, for simple linear optics without coupling of $x,y$-planes (drifts and lenses), the WDF projections can be propagated in the same way as the full four-dimensional Wigner distribution. We also note that a pure mode with symmetric fields $E(-x,-y) = E(x,y)$, which are of a practical importance to synchrotron radiation, the on-axis 2D brightness takes on the possible maximum value
\begin{equation}
\max\{\mathcal{B}_{x,y}(0,0)\} = \frac{2}{\lambda} \mathcal{F},
\end{equation}
where $\mathcal{F}$ is the total spectral flux contained in the mode.

\subsection{Light propagation}

One of the strong appeals of the Wigner distribution function is in its natural propagation for linear optics, which is entirely similar to the classical phase space evlolution. As a result, the formalism developed in the accelerator physics for classical phase space distributions can be directly carried over to (partially) coherent synchrotron radiation. In analogy to Property \ref{prop:evol}, the local values of WDF stay constant on phase space trajectories subject to classical transformation in drifts and lenses along the longitudinal position $z$
\begin{equation}
W(\mathbf{r}(z_2),\boldsymbol\uptheta(z_2)) = W(\mathbf{r}(z_1),\boldsymbol\uptheta( z_1)),
\end{equation}
where
\begin{equation}
\left(
\begin{array}{c}
x\\
\theta_x\\
y\\
\theta_y
\end{array}
\right)_{\!z2} = \mathbf{M}(z_1 \rightarrow z_2) \left(
\begin{array}{c}
x\\
\theta_x\\
y\\
\theta_y
\end{array}
\right)_{\!z1},\mbox{ with }\det\mathbf{M} = 1.
\end{equation}
Similarly, for a decoupled in $x,y$-plane transport, the 2D projections of the WDF follow the classical transformation
\begin{align}
\mathcal{B}_x(x(z_2),\theta_x(z_2)) & = \mathcal{B}_x(x(z_1),\theta_x(z_1)),\\
\mathcal{B}_y(y(z_2),\theta_y(z_2)) & = \mathcal{B}_y(y(z_1),\theta_y(y_1)).
\end{align}

Drift or lens transformations lead to a rotated or sheered WDF, and since the projections of the WDF are accessible for measurement, this allows a reconstruction of the Wigner distribution through tomography, similar to the use of tomography in phase space reconstruction in accelerators.

The introduction of spatial filters (e.g.~a pinhole or a slit) naturally leads to diffraction phenomena and the Wigner distribution gets altered in a non-trivial way. Whereas, the electric field after an aperture with transmission $t(\mathbf{r})$ is simply $E(\mathbf{r}) \rightarrow E(\mathbf{r}) t(\mathbf{r})$, the Wigner distribution is given by the convolution of the angular variables $\boldsymbol\uptheta$ of the input Wigner function with that of the spatial filter \cite{2007JEOS....2E7030L}
\begin{equation}
W(\mathbf{r},\boldsymbol\uptheta) \rightarrow \int W(\mathbf{r},\boldsymbol\uptheta) W_t(\mathbf{r},\boldsymbol\uptheta-\boldsymbol\uptheta\mathbf{'})\, d^2\boldsymbol\uptheta\mathbf{'},
\end{equation}
where
\[
W_t(\mathbf{r},\boldsymbol\uptheta) = \left(\frac{1}{\lambda}\right)^{\!2} \int t^*(\mathbf{r}+\tfrac{\mathbf{r'}}{2}) t(\mathbf{r}-\tfrac{\mathbf{r'}}{2}) e^{i k \mathbf{r'}\cdot\boldsymbol\uptheta} d^2\mathbf{r'}.
\]

\section{Numerical evaluation\label{sec:syn_rad}}

In this section we discuss practical matters pertaining to computing the Wigner distribution for undulator radiation. As we shall see, as long as the effects of $\cosh$-dependence of the undulator fields can be ignored, the synchrotron radiation in phase space can be obtained by a convolution of the WDF from a single electron with that of the entire electron beam phase space. This is a consequence of the well-known fact that the electrons in a single bunch do not interfere with each other unless there is a microbunching structure on the wavelength scale. In which case, the computation of radiation fields proceeds differently. In what follows, we limit our examples to the ``electron only interferes with itself'' scenario, as applicable for non-free-electron-laser (non-FEL) emission regimes.

\subsection{Radiation field generation}

Calculation of radiation fields is well established, e.g.~see \cite{chubar:1995, landau1975classical}. The frequency representation of the electric field is given by
\begin{equation}\label{eq:chubar}
\mathbf{E}(\mathbf{r};\omega) = \frac{i e \omega}{4\pi\epsilon_0 c} \int \frac{1}{R}\left[\boldsymbol\upbeta-\mathbf{n}\left(1+\frac{i c}{\omega R}\right) \right] e^{i\omega(\tau+R/c)} d\tau,
\end{equation}
for an observer at $\mathbf{r}$, the position vector from the observer to the electron $\mathbf{R} = \mathbf{r} - \mathbf{r_e}$, with $\mathbf{r_e}(\tau)$ being the electron's trajectory as a function of time $\tau$, the velocity $\boldsymbol\upbeta = c^{-1} d\mathbf{r_e}/d\tau$, and the unit vector $\mathbf{n} = \mathbf{R}/R$ with $R = |\mathbf{R}|$. The expression \ref{eq:chubar} is exact and is convenient for numerical evaluation in that once the trajectory $\mathbf{r_e}(\tau)$ is found, the integral evaluation is direct. We assume transversality of the field, i.e.~$\mathbf{E} \approx (E_x, E_y, 0)$. An expression with paraxial approximation for the field can be obtained \cite{Geloni2007167}, however, it represents little advantage over the exact expression for numerical work. A simulation tool has been developed that solves for the electron trajectory in arbitrary field configuration and evaluates the radiation integral, Eq.~\ref{eq:chubar}.

The undulator magnetic fields are taken of the usual form
\begin{equation}\label{eq:und_bfield}
\begin{aligned}
B_x & =  B_{0x} \sin(k_u z) \cosh(k_u x), \\
B_y & =  B_{0y} \cos(k_u z) \cosh(k_u y), \\
B_z & =  B_{0x} \cos(k_u z) \sinh(k_u x) - B_{0y} \sin(k_u z) \sinh(k_u y).
\end{aligned}
\end{equation}
Here $\lambda_u = 2\pi/k_u$ is the undulator period, $B_{0x,y}$ are the maximum magnetic fields in both planes with $B_{0x} = 0$ for a conventionally oriented planar undulator. The total undulator length is taken to be $L_u = N_u \lambda_u$, and the relation to the undulator $K$ parameter is via the usual $K_{x,y} = e B_{0x,y} \lambda_u/2\pi m_e c$. To ensure on-axis orbits with no net deflection, the undulator fields are $1/4$ and $3/4$ of their nominal values for the first and second period halves on either undulator end.

For numerical evaluation of the Wigner distribution function, the Fourier transform of Eqs.~\ref{eq:bx}, \ref{eq:by} is replaced with its discrete analog. A detector is placed at an arbitrary position $z$ downstream of the undulator, and the electric field is evaluated on a transverse grid of positions $\mathbf{r}_{kl} = (x_k,y_l,z)$. The phase space distribution is then typically back-propagated to the undulator center using the usual transforms. It should be noted that the discrete Fourier transform can suffer from aliasing problems and, in order to avoid this problem, the maximum angular extent of the radiation must be within $\pi/k \Delta_{x,y}$, where $\Delta_{x,y}$ is the grid size of the radiation field sampling. To avoid very small grid sizes, it is convenient to use Property \ref{prop:phase} to first remove the quadratic phase present in the radiation pattern. This is equivalent to introduction of a perfect thin lens, which is subsequently removed after the WDF is evaluated but prior to phase-space propagation to a point of interest.

\subsection{Electron bunch effect}

The effect of adding radiation from many electrons is equivalent to an earlier considered example of superposition from two quantum states. For any two electrons in the bunch, the electric field will differ by a phase factor $e^{i\omega t_j}$, where $t_j$ represents time of the electron inside the bunch. It is easy to see that the interference term of Eq.~\ref{eq:two_states_superposition} averages out to 0 since it contains essentially random phase factors $e^{\pm i\omega(t_j - t_k)}$ inside the averaging brackets. In other words, the uncertain phase relationship between the two electrons on the optical scale leads to a density matrix case analogue where the interference term drops out and the WDF is simply an incoherent sum over all the electrons.

Therefore, if the Wigner distribution function of a single electron does not change its shape, but simply shifts in position and angle, as one would expect for an undulator in which the trajectories remains linear with small position and angle offsets, the overall radiation pattern is just a convolution (summation) of the electron distribution in phase space with the WDF of a \textit{single} electron. Additional effects arise for either segmented undulator (with focusing between the segments), due to the vertical focusing of a planar (horizontally deflecting) undulator, or due to the effect of a larger off-axis field which has $\cosh$-like dependence in the vertical plane for the planar undulator or in both planes for a helical undulator.

In other words, the most general form of the Wigner distribution obtained from incoherent addition of radiation from individual electrons is of the form
\begin{equation}\label{eq:wig_add_gen}
W(\mathbf{r},\boldsymbol\uptheta) = N_e \int W_0(\mathbf{r},\boldsymbol\uptheta ; \mathcal{V}_e) P( \mathcal{V}_e )\,d^4 \mathcal{V}_e,
\end{equation}
Here $N_e$ is the total number of electrons inside a bunch described by the probability density function $P(\mathcal{V}_e)$ such as $\int P(\mathcal{V}_e) d^4 \mathcal{V}_e = 1$ with $\mathcal{V}_e$ representing 4 phase-space variables of electrons: two transverse components of position $(x_e, y_e)$ and angle $(\theta_{ex}, \theta_{ey})$. If, on the other hand, the different electron trajectories simply lead to an offset in position and angle of the WDF of a single electron
\begin{equation}\label{eq:wig_offset}
W_0(\mathbf{r}, \boldsymbol\uptheta; \mathbf{r_e}, \boldsymbol\uptheta_\mathbf{e}) = W_0(\mathbf{r}-\mathbf{r_e},\boldsymbol\uptheta-\boldsymbol\uptheta_\mathbf{e}), \end{equation}
the integral of Eq.~\ref{eq:wig_add_gen} is then replaced with a convolution integral
\begin{equation}\label{eq:wig_conv}
W(\mathbf{r},\boldsymbol\uptheta) = N_e \iint W_0(\mathbf{r}-\mathbf{r_e},\boldsymbol\uptheta-\boldsymbol\uptheta_\mathbf{e}) P(\mathbf{r_e}, \boldsymbol\uptheta_\mathbf{e})\,d^2 \mathbf{r_e} \,d^2 \boldsymbol\uptheta_\mathbf{e}.
\end{equation}

The effect of energy spread in electron beam can be quite significant, and most generally it is accounted by extending the integration variable $\mathcal{V}_e$ to also include the energy. E.g.~the effect of a small energy spread $\delta_e \equiv \Delta\gamma_e/\gamma_e \ll 1$, where $\gamma_e$ is the normalized electron energy (later denoted as simply $\gamma$), leads to
\begin{equation}\label{eq:wig_conv_en}
W(\mathbf{r},\boldsymbol\uptheta) = N_e \iint W_0(\mathbf{r}-\mathbf{r_e},\boldsymbol\uptheta-\boldsymbol\uptheta_\mathbf{e}; \delta_e) P(\mathbf{r_e}, \boldsymbol\uptheta_\mathbf{e}, \delta_e)\,d^2 \mathbf{r_e} \,d^2 \boldsymbol\uptheta_\mathbf{e} \,d\delta_e.
\end{equation}
Note that for a small energy change $\delta_e \sim 1/N_u$ with a large number of undulator periods $N_u \gg 1$, the effect on the radiation pattern at a given frequency $\omega_0$ is identical to that of the on-energy particle $\Delta\delta_e = 0$ while tuning the radiation frequency off the resonance by $\Delta\omega/\omega_0 = -2 \delta_e$.

The evaluation of Eq.~\ref{eq:wig_conv_en} can be quite involved in terms of computational resources required even if being straightforward in all other respects. However, if the electron distribution $P(\mathcal{V}_e)$ is separable, i.e. $P(\mathcal{V}_e) = P_x(x_e,\theta_{ex}) P_y(y_e, \theta_{ey}) P_\gamma (\delta_e)$, then the 2D projection of the WDF, $\mathcal{B}_x(x,\theta_x)$ and $\mathcal{B}_y(y, \theta_y)$ can be easily computed.

It is instructive to consider the requirements for when Eq.~\ref{eq:wig_offset} is applicable in case of a planar undulator (horizontally deflecting). As mentioned previously, two effects can change the shape of the WDF depending on the (small) electron trajectory offsets in position and angle in vertical plane. One is the $\cosh$-dependence of the vertical field, whereas the other is the natural undulator focusing.

The equation of motion for the average vertical position $y_{av}$ when $B_{0x} = 0$ for the undulator, Eqs.~\ref{eq:und_bfield}, can be written as \cite{scharlemann:2154}
\begin{equation}\label{eq:und_focusing}
\frac{d^2y_{av}}{dz^2} = -k_{\beta_y}^2 y_{av},
\end{equation}
where the vertical focusing strength is given by $k_{\beta_y} = B_{0y} e / \sqrt{2} \gamma m_e c$, or in terms of the period of oscillations due to focusing $L_{\beta_y} = 2 \pi/k_{\beta_y} = \sqrt{2} \gamma \lambda_u/K_y $, where $K_y$ is the undulator $K$-value and $\gamma = E/m_e c^2$ is the normalized energy of the electron. Typically, $L_{\beta_y} \gg L_u$, i.e.~the slow oscillation phase increment due to the focusing is $\ll 2 \pi$ in undulators. Nevertheless, in order to be able to treat vertically offset trajectories as simple copies of each other, we require that the slow sine-like oscillations due to focusing produce a change in the electron trajectory's deviation over the length of the undulator that is much smaller than the natural cone of the radiation, $\sqrt{\lambda/L_u}$ \cite{Kim89}. Integrating Eq.~\ref{eq:und_focusing} for a typical vertical size $\sigma_y$, the angle \textit{change} of the electron trajectory is of the order $\sigma_y k_{\beta_y}^2 L_u$, which leads to the following requirement
\begin{equation}\label{eq:und_focusing_req}
\sigma_y \ll \frac{1}{k_{\beta_y}^2 L_u} \sqrt\frac{\lambda}{L_u}.
\end{equation}

\begin{figure}[tbh]
\begin{center}
	\includegraphics[width=0.9\columnwidth]{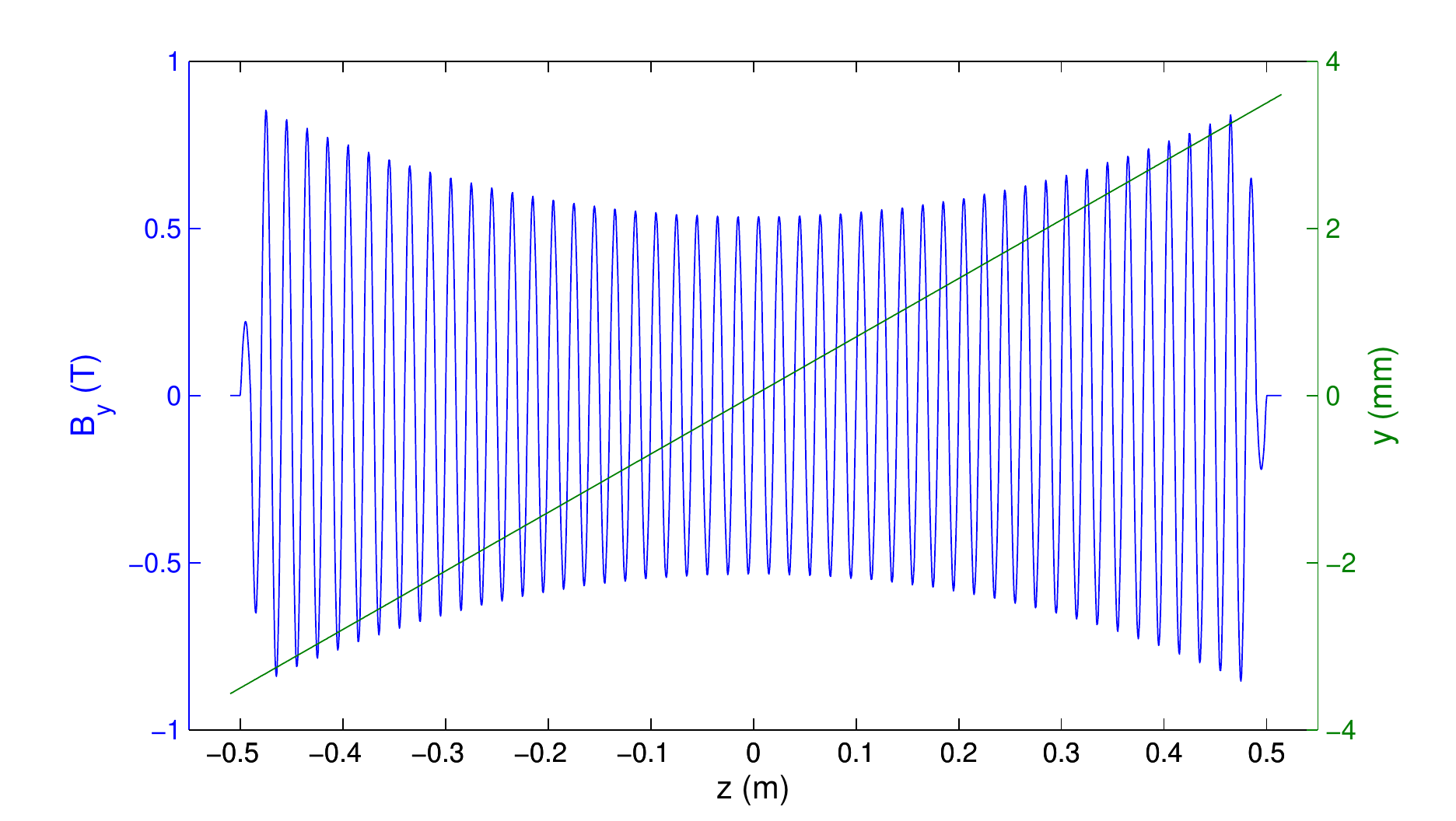}
	\caption{An example of a vertical trajectory in a planar undulator and the magnetic field as seen by the particle. The angular offset is taken to be rather large to illustrate the effect of $\cosh$-like dependence of the field on vertical postion.\label{fig:und_cosh_angle}}
\end{center}
\end{figure}

Similarly, the vertical dependence of the magnetic field in the undulator $B_y \propto \cosh(k_u y)$, leads to the vertical trajectories with an offset to effectively sample a larger $K_y$ value. Therefore, to enable the simpler treatment, we require that $\Delta K_y/ K_y \approx (k_y y)^2/2$ produces a change in the undulator wavelength $\lambda = \lambda_u/2\gamma^2(1+K_y^2/2)$ which is much smaller than the natural undulator bandwidth $\Delta\lambda/\lambda \sim 1/N_u$. This leads to another requirement for the electron beam size
\begin{equation}\label{eq:und_offset_req}
\sigma_y \ll \frac{\lambda_u}{2\pi} \sqrt{\frac{1}{N_u}}\sqrt{\frac{2+K^2}{K^2}}.
\end{equation}
Electrons coming with a vertical angle into a planar undulator generally sample a more complicated magnetic field pattern, such as shown in Fig.~\ref{fig:und_cosh_angle}. Therefore, the following requirement can be imposed on the vertical angular size $\sigma_{y'}$
\begin{align}
& L_u \sigma_{y'} \ll \frac{\lambda_u}{2\pi} \sqrt{\frac{1}{N_u}}\sqrt{\frac{2+K^2}{K^2}}, \\
& \sigma_{y'} \ll \frac{1}{2\pi N_u^{3/2}} \sqrt{\frac{2+K^2}{K^2}}.\label{eq:und_angle_req}
\end{align}

In summary, if the requirements \ref{eq:und_focusing_req}, \ref{eq:und_offset_req}, and \ref{eq:und_angle_req} are satisfied, the simple convolution of a single electron radiation pattern with that of the electron bunch phase-space distribution, Eq.~\ref{eq:wig_conv} or Eq.~\ref{eq:wig_conv_en} can be used. Otherwise, the more general integral, Eq.~\ref{eq:wig_add_gen}, needs to be evaluated. We note that the potential complications discussed here apply only to the vertical plane for a planar undulator with exact translational symmetry of fields in the $x$-direction.

\subsection{Revisiting emittance definition}
Rms emittance, Eq.~\ref{eq:sig-mat}, is widely used as a measure of beam quality in accelerator physics. While this definition is attractive due to the fact that it can be applied to a variety of different distributions, the connection of the rms emittance to phase space density or brightness available in the beam is generally distribution dependent. Whereas equilibrium processes (e.g.~radiation damping in storage rings, equilibrium beam in a focusing channel under the influence of space charge \cite{reiser2008theory}, etc.) lead to a Gaussian distribution in phase space, beams in linear accelerators are rarely in equilibrium. As a result a meaningful characterization of the phase space of electron beams or, as we shall see later, the synchrotron radiation, needs a more flexible metric than the rms emittance alone. Short of the complete knowledge of the actual phase space distribution, a useful way to reduce and represent the information is to extend the concept of the rms emittance to the so-called brightness curve or rms emittance vs.~beam fraction \cite{Lejeune1980}. As we will see, a wide class of practical phase space distributions can be effectively characterized by such a curve as the beam fraction is varied from 0 to 100\%. Three parameters, the usual rms emittance ($\epsilon = \epsilon(100\%)$ with 100\% denoting that the entire beam is included in the emittance calculation), core emittance, $\epsilon_c$, and core fraction $f_c$ can convey the information not only about the second moments of the beam distribution, but also the peak brightness and what fraction of the beam effectively contributes to this brightness. The situation is somewhat analogous to how the peak height and the full-width at half maximum complement the rms width information for arbitrary (unimodal and finite integrable in the second moment sense) pulses.

Below is one prescription for obtaining emittance vs.~fraction curve. Here we only consider the case of a two-dimensional phase space, $\mathbf{x} = (x,p)^\intercal$ where $x$ is the transverse coordinate and $p$ can represent (normalized) transverse momentum or angle. The phase space distribution function $P(x,p)$ is assumed to be normalized, $\iint P(x,p)\,dx\,dp = 1$. One can apply the following procedure:
\begin{enumerate}[1)]
\item For an ellipse of a fixed area $\pi a$, choose Twiss parameters $\mathbf{T}$ of the ellipse (c.f. Eq.~\ref{eq:sig-mat}) that maximize the beam fraction contained therein:

\begin{equation}\label{eq:frac}
f(a) = \max \Bigl[ \iint_{D(a)} P(x,p)\,dx\,dp \Bigr] \mbox{ with }D(a) = \{\mathbf{x}:~ \mathbf{x}^\intercal \mathbf{T}^{-1} \mathbf{x} \leq a \}.
\end{equation}

\item Obtain the rms emittance $\epsilon(a)$ for $\mathbf{x} \in D(a)$ of Eq.~\ref{eq:frac}:

\begin{equation}\label{eq:emit}
\epsilon(a) = \sqrt{\braket{x^2}_D\braket{p^2}_D-\braket{xp}_D^2},\mbox{ with } \braket{u}_D = \iint_{D(a)} u\frac{P(x,p)}{f(a)}dx\,dp.
\end{equation}

The parametric curve $(f(a), \epsilon(a))$ is the emittance vs.~fraction curve, $\epsilon(f)$.

\item Define the core emittance, $\epsilon_c$, and the core fraction, $f_c$, according to
\begin{align}
& \epsilon_c \equiv \left. \frac{d\epsilon(f)}{df} \right|_{f \rightarrow 0},\\
& f_c: \epsilon(f_c) = \epsilon_c.
\end{align}
\end{enumerate}
We have assumed that each individual ellipse $D(a)$ remains centered around the origin as does the corresponding centroid of the beam fraction. Generalization to when this is not the case is straightforward by allowing the clipping ellipse to shift. This procedure for obtaining emittance vs.~fraction curve is meaningful for distributions which are unimodal (i.e.~with a single hump) and finite integrable (for second moments).

It is easy to show that the core emittance is directly related to the peak phase space density or brightness $P_0 = \max\{P(x,p)\}$:
\begin{equation}
\epsilon_c = \frac{1}{4\pi P_0}.
\end{equation}
To see that one simply needs to note that a small area clipping ellipse in the limit $a \rightarrow 0$ cuts out a uniform slice containing the beam fraction $\pi a P_0$ and having the rms emittance of $a/4$. It is interesting to note that because of the Property \ref{prop:bound}, which states that a maximum Wigner distribution is $h/2$ for \textit{any} even pure state, and a corresponding 2D equivalent in optics of $\lambda/2$, the minimum core emittance (the diffraction limit) is therefore
\begin{equation}
\min(\epsilon_c) = \frac{\lambda}{8\pi},
\end{equation}
and it can only be larger for a symmetric mode when the coherence $\mu_g^2 < 1$. Thus, the core emittance (or peak brightness) is a more general than the rms emittance indicator of whether the radiation is coherent. This is because the rms emittance minimum is restricted only to a Gaussian coherent mode, whereas the minimum core emittance is realized for \textit{any symmetric} coherent mode.

\begin{figure}[htb]
\begin{center}
	\includegraphics[width=0.9\columnwidth]{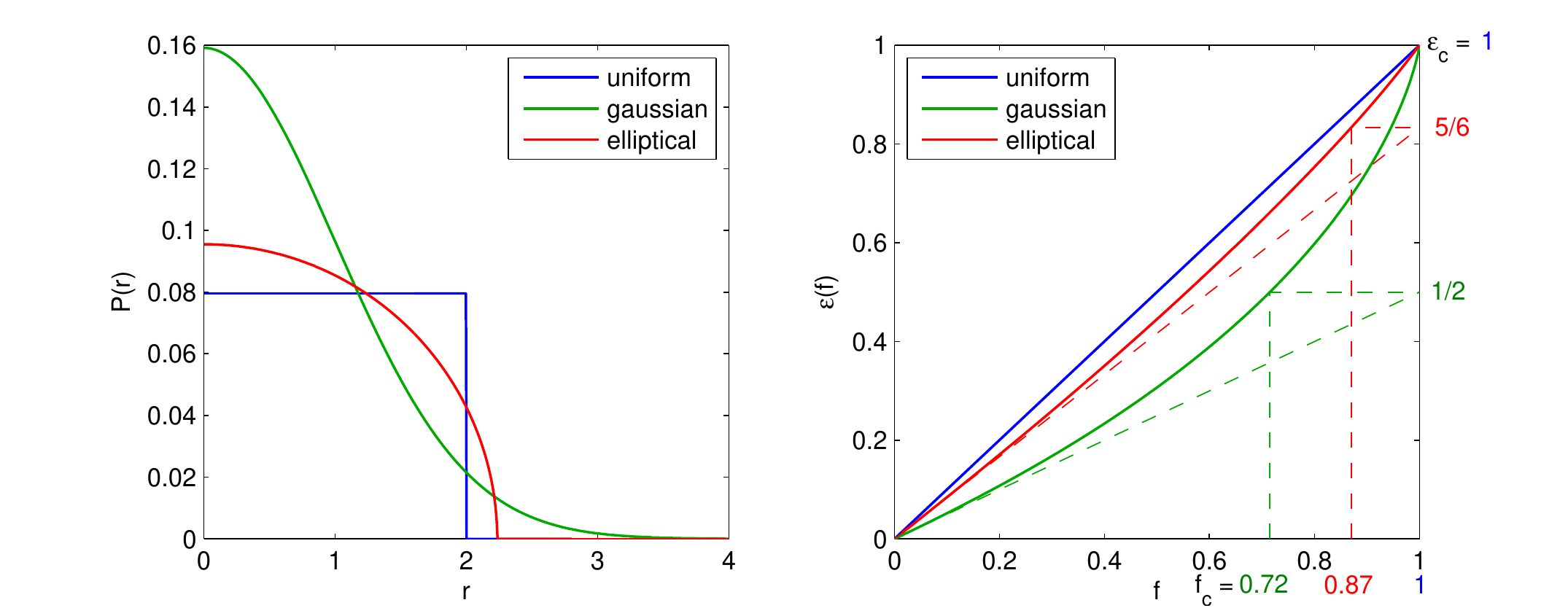}
	\caption{Radial phase-space distributions (left) and corresponding emittance vs.~fraction curves (right). All distributions are scaled to have $\epsilon = 1$. Core fraction and emittance for different distribution types are shown as well.\label{fig:three_dist}}
\end{center}
\end{figure}

Fig.~\ref{fig:three_dist} further illustrates the concept of the emittance vs.~fraction by showing the curve for 3 different distributions: uniform, gaussian, and elliptical. The correlation in $x$ and $p$ is removed and the units for $x$ and $p$ are chosen so that the distributions can be written as a radial function of $r = \sqrt{x^2 + p^2}$. Furthermore, to facilitate the comparison, each distribution is normalized to have $\epsilon = \sigma_x = \sigma_p = 1$ in these natural units. As seen, the core emittance conveniently captures the fact that the peak brightness of a Gaussian is $\times2$ larger than that of the uniform distribution of the same rms width, as well as the fact that the core fraction in the Gaussian is smaller (0.715 vs. 1 for the uniform).

\begin{figure}[htb]
\begin{center}
	\includegraphics[width=0.9\columnwidth]{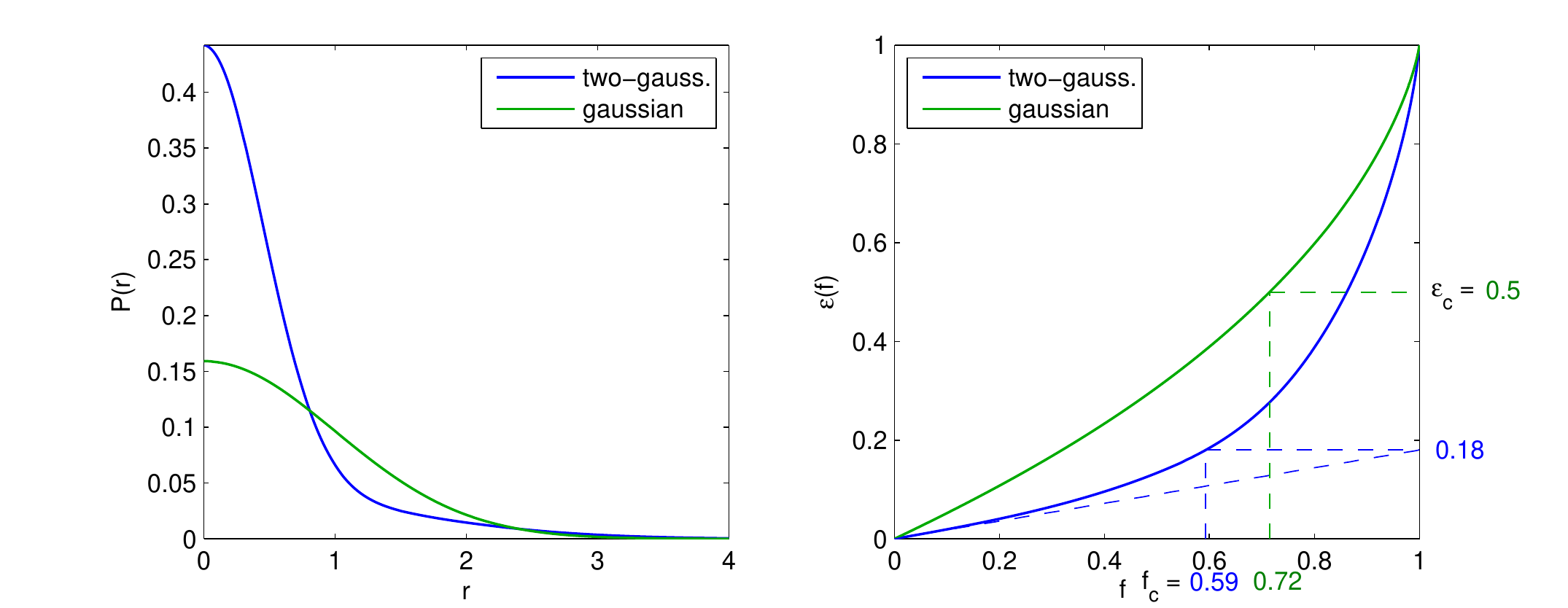}
	\caption{Radial phase-space distributions (left) and corresponding emittance vs.~fraction curves (right). Both distributions are scaled to have $\epsilon = 1$. Core fraction and emittance for the two distribution are shown as well.\label{fig:frac_two_gauss}}
\end{center}
\end{figure}

Another example, Fig.~\ref{fig:frac_two_gauss}, compares a Gaussian distribution $P(r) = \tfrac{1}{2\pi} e^{-\tfrac{r^2}{2}}$ and $P(r) = \tfrac{p}{2\pi\epsilon_1} e^{-\tfrac{r^2}{2 \epsilon_1}}+\tfrac{1-p}{2\pi\epsilon_2} e^{-\tfrac{r^2}{2 \epsilon_2}}$, with $p = 0.5$, $\epsilon_1 = \tfrac{1}{5}$, and $\epsilon_2 = \tfrac{9}{5}$. The total emittance in this case is $\epsilon = p \epsilon_1 + (1-p) \epsilon_2 = 1$. Once again, the information about the peak brightness is lost with the rms emittance quoted only, but is conveyed conveniently with the three parameters: $\{\epsilon, \epsilon_c, f_c\}$. A practical measure of beam brightness available can be defined as $f_c/\epsilon_c$, a subject that we explore further below.

\subsection{Possible definitions of brightness}
As a phase space quasi-probability, the WDF \textit{is} the generalized brightness (also known as microscopic brightness \cite{Lejeune1980}), $\mathcal{B}(\mathbf{r},\boldsymbol\uptheta) \equiv W(\mathbf{r},\boldsymbol\uptheta)$. It is convenient, however, to be able to reduce the information to a single parameter, which, for example, can facilitate comparison of various partially coherent sources. Here we revisit several of the definitions that can be useful for this purpose remembering that no single reduced parameter or a figure of merit can suit all the practical purposes.

\begin{enumerate}[1)]
\item The following definition, which we denote as classical, can be written (modulo a prefactor that generally depends on the actual distribution shape) as
\begin{equation}\label{eq:def_b_class}
\mathcal{B}_{cl} = \frac{\mathcal{F}}{\epsilon_x \epsilon_y}.
\end{equation}
$\mathcal{F}$ is the overall (spectral) flux. In the definition above we have assumed that the 4D emittance can be represented as a product of two 2D emittances. This definition, which gives a positive quantity, is easy to compute and can serve as a measure of brightness. One drawback is in the use of rms emittance, which, as discussed previously fails to capture the peak brightness available in the beam and tends to exaggerate the importance of tails when non-Gaussian distributions are encountered. A possible modification to the definition of Eq.~\ref{eq:def_b_class} can be made to write the effective brightness in terms of
\begin{equation}\label{eq:def_b_class_eq}
\mathcal{B}_{cl,alt} = \frac{\mathcal{F} f_{cx} f_{cy}}{\epsilon_{cx} \epsilon_{cy}},
\end{equation}
where in place of rms emittances as a measure of effective phase space area we use the core emittance $\epsilon_{cx,y}$ while at the same time reducing the participating flux by the product of the core fractions in each plane $f_{cx,y}$. All the necessary quantities in Eq.~\ref{eq:def_b_class_eq} can be found from the WDF as discussed previously.

These classical definitions, however, fail to capture the concept of coherence. A mode with a large dispersion (emittance) but perfectly coherent is indistinguishable from its incoherent analog of the same emittance.

\item As we have seen, the WDF contains the information about the density matrix, which, to the overall flux factor leads to the following natural definition for brightness (denoted as \textit{average} brightness)

\begin{equation}\label{eq:def_b_av}
\mathcal{B}_{av} = \frac{\iint W^2 d^2\mathbf{r}\,d^2\boldsymbol\uptheta}{\mathcal{F}}.
\end{equation}

This definition is discussed in a classical context in \cite{rhee:1674}, though its genuine justification becomes clear from the connection to quantum or wave phenomena. The brightness of Eq.~\ref{eq:def_b_av} is higher for more coherent radiation, even though the dispersion or emittance no longer comes into this definition. In particular, as pointed out previously, a pure mode, no matter how dispersed it gets, would have the same $\mathcal{B}_{av}$ provided the flux remains unchanged.

\item Another definition is simply to quote the on-axis peak brightness

\begin{equation}\label{eq:def_b_peak}
\mathcal{B}_0 = W(\mathrm{r}=0, \boldsymbol\uptheta = 0).
\end{equation}

An obvious drawback of this definition is that the WDF is not guaranteed to be positive. However, as previously discussed, the on-axis WDF is always positive for symmetric (even) modes, which are of most practical interest for synchrotron radiation. Additionally, the peak brightness due to the boundness property (Property~\ref{prop:bound}) can serve as a measure of coherence because for any pure and symmetric (even) mode $\mathcal{B}_0$ is guaranteed to be related to the total (coherent) spectral flux according to
\begin{equation}\label{eq:def_b_peak_pure}
\mathcal{B}_{0,pure,ev} = \left(\frac{2}{\lambda}\right)^{\!2} \mathcal{F}.
\end{equation}
As pointed out previously, the core emittance is inversely related to the peak brightness.
\end{enumerate}

Finally, for the purpose of the numerical examples below, it will be convenient to consider 2D projections of the WDF which are easy to visualize. The extension of the above definitions to 2D is straightforward and the equivalent meaning remains intact only when the mode is separable in $x,y$-planes. In particular, the Eq.~\ref{eq:def_b_av} in 2D becomes
\begin{equation}\label{eq:def_b_avx}
\mathcal{B}_{avx} = \frac{\iint \mathcal{B}_{x}^2 dx\,d\theta_x}{\mathcal{F}},
\end{equation}
and equations Eqs.~\ref{eq:def_b_peak} and \ref{eq:def_b_peak_pure}
\begin{align}
& \mathcal{B}_{0x} = \mathcal{B}_x(x=0, \theta_x = 0), \\
& \mathcal{B}_{0x,pure,ev} = \frac{2}{\lambda} \mathcal{F}\label{eq:def_b_peak_purex},
\end{align}
with equivalent expressions for $y$-plane. We are going to use Eq.~\ref{eq:def_b_avx} even when the mode is not separable as a measure of \textit{effective} average brightness in one plane.

\subsection{Numerical examples}

Here we demonstrate numerical examples of using the Wigner distribution function formalism for undulator radiation. We start out with a zero emittance electron case. For convenience, we scale the results to 100\,mA average current for otherwise perfect (pencil) electron beam. Throughout all the examples, the electron energy is set to 5\,GeV and the undulator period $\lambda_u = 2$\,cm.

\begin{figure}[htb]
\begin{center}
	\includegraphics[width=0.9\columnwidth]{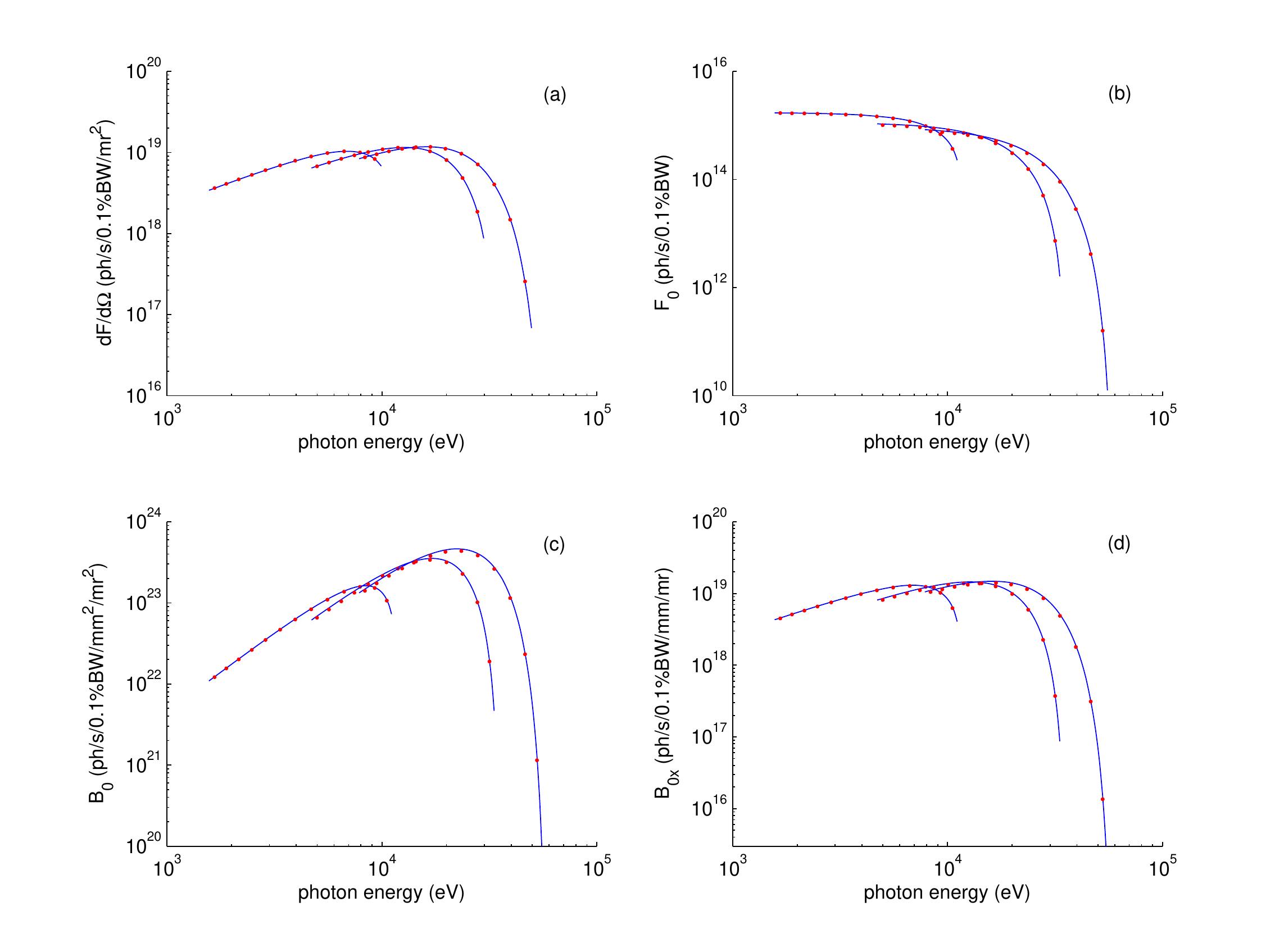}
	\caption{Comparison of calculated (dots) vs.~theoretical values (solid curves) for undulator radiation: (a) angular flux density, (b) central cone flux, (c) on-axis 4D brightness, (d) on-axis 2D brightness. The first three odd harmonics are shown. The undulator is planar with $N_u = 250$. Refer to text for other parameters.\label{fig:und_benchmarking}}
\end{center}
\end{figure}

Fig.~\ref{fig:und_benchmarking} illustrates the calculated angular flux and the central cone for an undulator with $N_u = 250$ periods. To convert the computed from radiation fields quantities to the standard units of photons/s/0.1\%BW for spectral flux and corresponding angular (per mrad$^2$ or mrad in one plane projection) and areal densities (per mm$^2$ or mm in one plane projection), we note that the spectral flux density is related to the computed fields $|\mathbf{E}(\mathbf{r})|^2$ in frequency domain according to
\begin{equation}
\frac{d^2 \mathcal{F}}{dA d\omega/\omega} = \frac{I}{e} \frac{c\epsilon_0}{\pi \hbar} |\mathbf{E}(\mathbf{r})|^2, \label{eq:connection_to_flux}
\end{equation}
where $I$ is the average beam current (non-FEL process is assumed), and $\epsilon_0$ is the vacuum permittivity.

Fig.~\ref{fig:und_benchmarking}a checks the angular flux density (or the on-axis radiation field) against the well-known expression for the planar undulator \cite{Kim89}
\begin{equation}\label{eq:angular_flux_kim}
\frac{d^2 \mathcal{F}}{d\Omega d\omega/\omega} = \frac{I}{e} \alpha N_u^2 \gamma^2 F_n(K),
\end{equation}
with fine-structure constant $\alpha$, and function $F_n(K) = K^2n^2/(1+\tfrac{K^2}{2})^2 [J\!J]$ for the harmonic number $n$ and the undulator $K$ where $[J\!J] = [J_{(n-1)/2}(\xi)-J_{(n+1)/2}(\xi)]^2$ in terms of the Bessel functions and $\xi = nK^2/(4+2 K^2)$. The area $dA$ and solid angle $d\Omega$ elements are related via $dA = R^2 d\Omega$ with $R$ being the distance to the source, allowing to cross-check the angular flux, Eq.~\ref{eq:angular_flux_kim}, in terms of the computed fields via Eq.~\ref{eq:connection_to_flux}. Fig.~\ref{fig:und_benchmarking}b compares on-resonance spectral flux with the analytical result
\begin{equation}\label{eq:flux_kim}
\frac{\mathcal{F}_0}{d\omega/\omega} = \frac{1}{2} \frac{I}{e} \pi\alpha N_u Q_n(K),
\end{equation}
where $Q_n(K) = (1+K^2/2) F_n(K)/n$. In what follows, we denote the spectral flux by simply $\mathcal{F}_0$ implicitly assuming the usual 0.1\% bandwidth scaling. To find the total flux, the detector in simulations is placed 50\,m away from the undulator center and the electric field is computed on a $1024\times1024$ 3\,mm square grid.

To check that the code correctly computes on-axis (peak) brightness for 4D and 2D WDF computed from the fields, we use the Eq.~\ref{eq:flux_kim} with Eqs.~\ref{eq:def_b_peak_pure} and \ref{eq:def_b_peak_purex}, which relate the total flux to the peak brightness of \textit{any} symmetric coherent mode according to $\mathcal{B}_0 = (2/\lambda)^2 \mathcal{F}_0$ and $\mathcal{B}_{0x} = (2/\lambda) \mathcal{F}_0$. The results of this cross-check are shown in Fig.~\ref{fig:und_benchmarking}c and \ref{fig:und_benchmarking}d.

\subsubsection{Example: helical undulator on resonance}

\begin{figure}[htb]
\begin{center}
	\includegraphics[width=0.9\columnwidth]{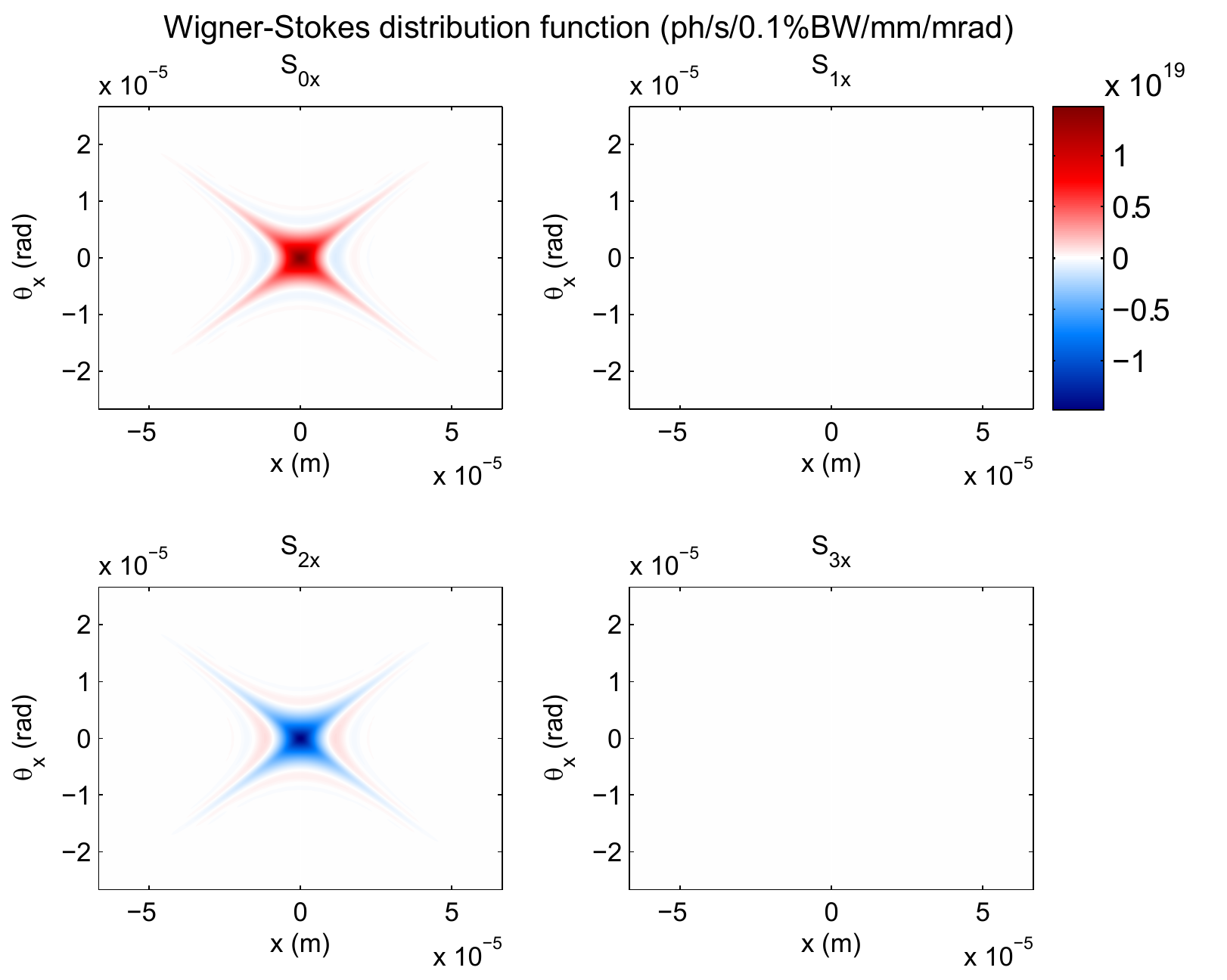}
	\caption{Wigner-Stokes density distribution functions computed for a helical undulator. The x-ray phase space is back-propagated to the undulator center. Refer to text for details.\label{fig:und_helical}}
\end{center}
\end{figure}

While the planar undulator radiation on-axis is fully horizontally polarized, Fig.~\ref{fig:und_helical} shows the WDF for a helical undulator at its first harmonic ($N_u = 250$, $K_x = K_y = 0.696$, $\hbar\omega = 8$\,keV). The WDF is obtained from the detector plane placed 50\,m away from the undulator center, and subsequent back-propagation of the radiation phase space back to the center of the undulator. As discussed previously, the case of a (nearly) pure circularly polarized wave leads to $|S_0| = |S_1|$ with other generalized Stokes parameters being approximately zero as seen in Fig.~\ref{fig:und_helical}.

The $\times$-like shape of the x-ray phase space is persistent throughout all the examples. The explanation behind it is simple --- undulator, being an extended source, has radiation emitted from its beginning and the end, which must advance different distances to reach the observer, or when (back)propagated to the undulator center. This results in the $\times$-like shape, with the two branches corresponding to the undulator ends.

In the remainder of this section, we limit our numerical examples to planar undulators investigating x-ray phase space for radiation on and off resonance, the segmented undulator with a quadrupole focusing in between, and a 25-m long undulator including electron emittance and energy spread effects.

\subsubsection{Example: planar undulator on resonance}

\begin{figure}[htb]
\begin{center}
	\includegraphics[width=0.9\columnwidth]{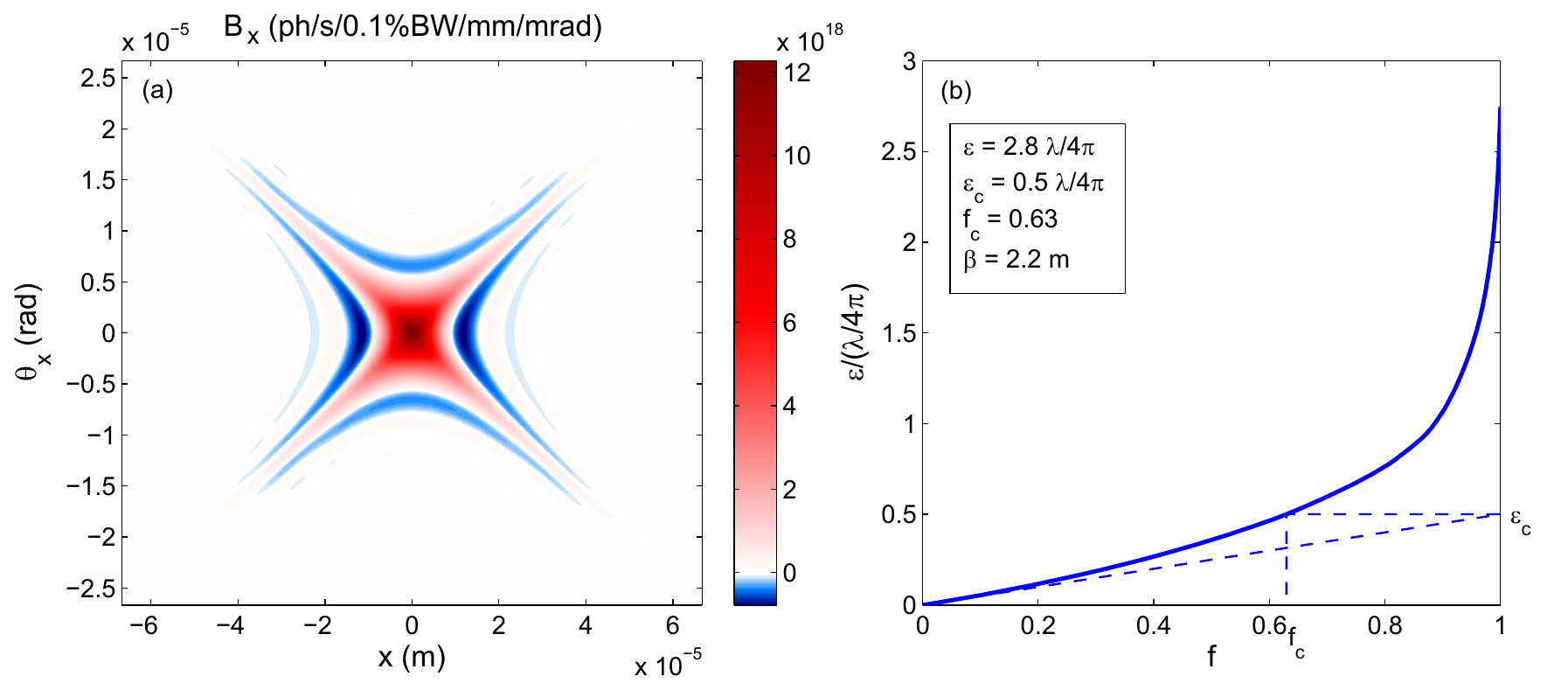}
	\caption{(a) Wigner 2D distribution for a planar undulator with $N_u = 250$ on resonance of the 1st harmonic at 8\,keV photon energy along with (b) the emittance of light vs.~fraction curve. The WDF is back-propagated to the undulator center.\label{fig:wigner_planar_resonance}}
\end{center}
\end{figure}

This example illustrates the x-ray phase space for radiation at undulator resonance, Fig.~\ref{fig:wigner_planar_resonance}, along with the emittance vs.~fraction curve. It is seen that $M^2 > 1$ or emittance is not the minimum possible for the fully coherent mode. On the other hand, the core emittance is its possible minimum as discussed previously. Also, note the value of the $\beta$-function or Rayleigh range, is somewhat different than $L_u/2$ or $L_u/2\pi$ values commonly quoted in the literature. Additionally, the full beam and its core have different $\beta$-function values. Therefore, a proper matching with the electron beam depends on whether one maximizes the peak brightness or minimizes the overall rms emittance of light.

\subsubsection{Example: segmented undulator with quad focusing}

\begin{figure}[htb]
\begin{center}
	\includegraphics[width=0.8\columnwidth]{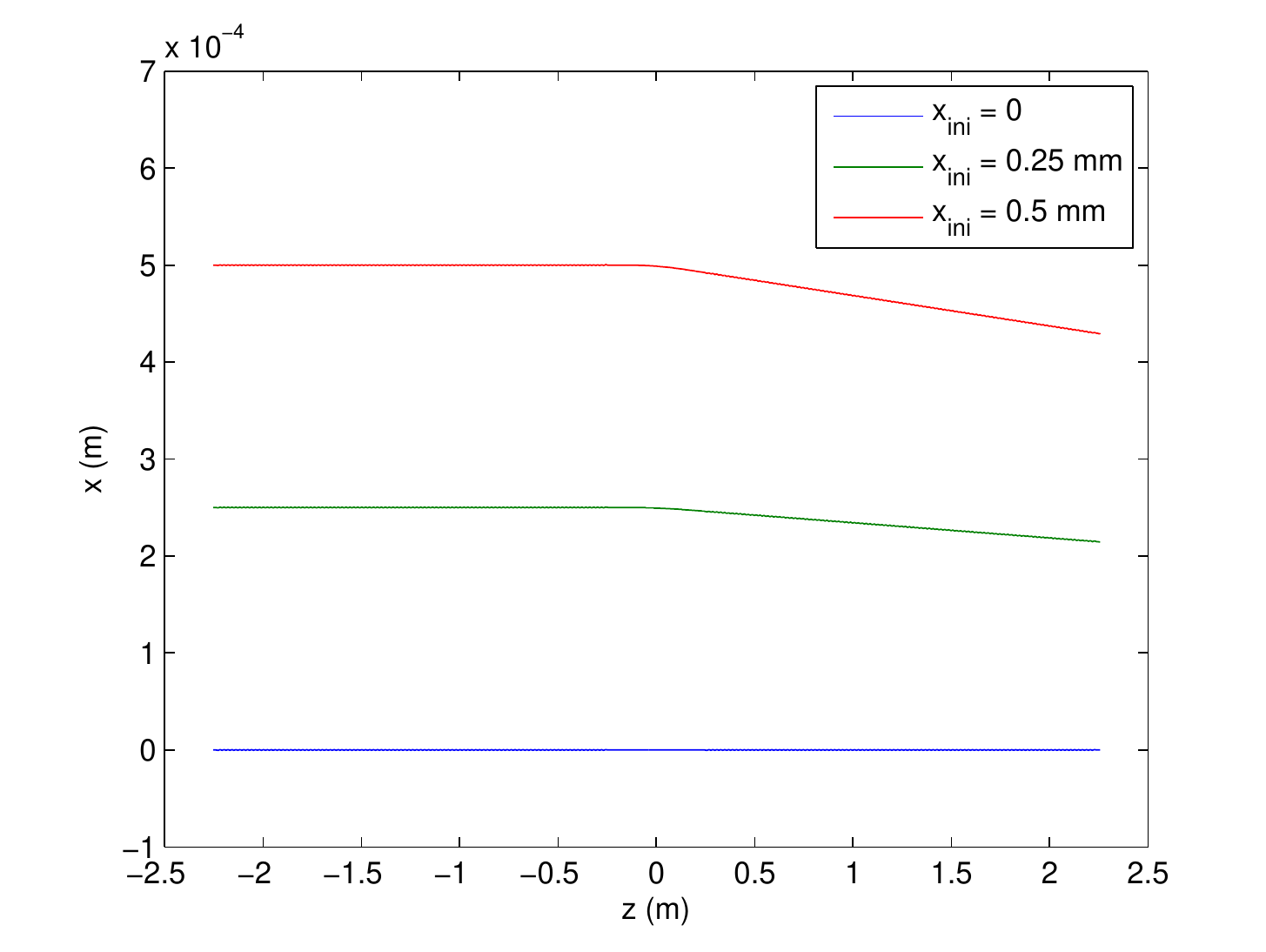}
	\caption{Trajectories inside a segmented undulator with a quadrupole focusing. The undulator has two segments each with $N_u = 100$ periods. The separation between the two is 0.486\,m. A single horizontally focusing quadrupole of length 0.3\,m is located at the center with 3.5\,T/m gradient.\label{fig:segmented_trajectory}}
\end{center}
\end{figure}

Next, we consider a segmented undulator with a quadrupole focusing in between the two segments. Fig.~\ref{fig:segmented_trajectory} shows trajectories for different horizontal offsets of electrons going into the undulator.

\begin{figure}[tbh]
\begin{center}
	\includegraphics[width=0.7\columnwidth]{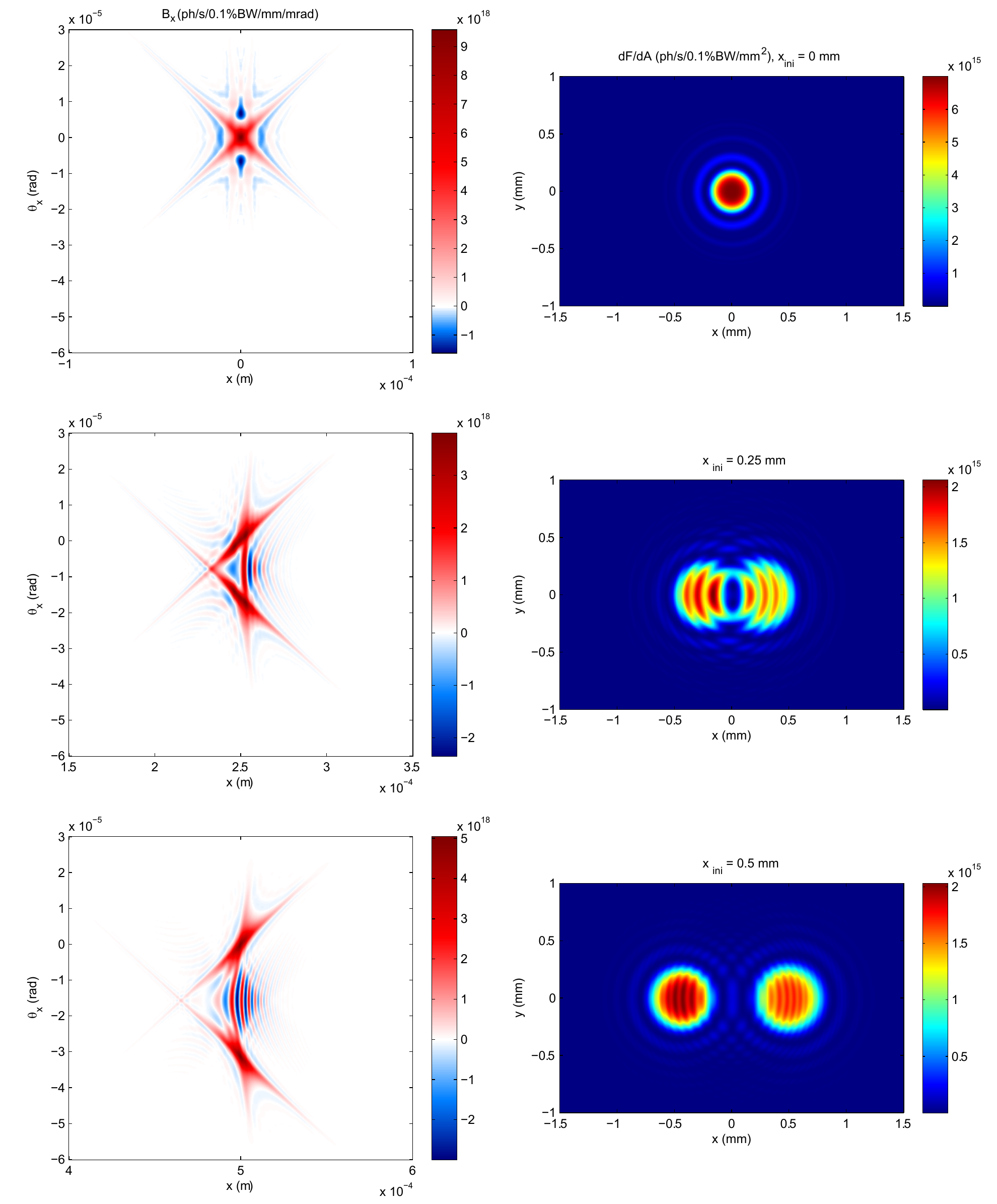}
	\caption{Radiation of the 1st harmonic on 8\,keV energy resonance from the segmented undulator with quadrupole focusing of Fig.~\ref{fig:segmented_trajectory}. Different rows correspond to different trajectory offsets, $x_{ini}$, as shown. The left column is 2D WDF back-propagated to the center of the undulator, which corresponds to the right column showing the radiation spectral flux density at the detector 30\,m away from the undulator center.\label{fig:segmented}}
\end{center}
\end{figure}

Fig.~\ref{fig:segmented} shows the flux at a detector 30\,m away from the center of the undulator and the WDF back-propagated to the undulator center. It is seen that in this case the electron interferes with \textit{itself} and the WDF clearly shows features present in a coherent superposition of two modes.

\subsubsection{Example: radiation off undulator resonance}

Here we consider the radiation off the undulator resonance. This is not only of interest for practical cases of detuning or selecting photon energy in a monochromator but also when considering off-energy electrons (electron beams with energy spread). This is because for undulators with a large number of periods, the effect of tuning off resonance is identical to keeping the radiation frequency $\omega_0$ the same but changing the electron energy according to $\Delta\omega/\omega_0 = -2 \Delta\gamma/\gamma$.

\begin{figure}[htb]
\begin{center}
	\includegraphics[width=0.5\columnwidth]{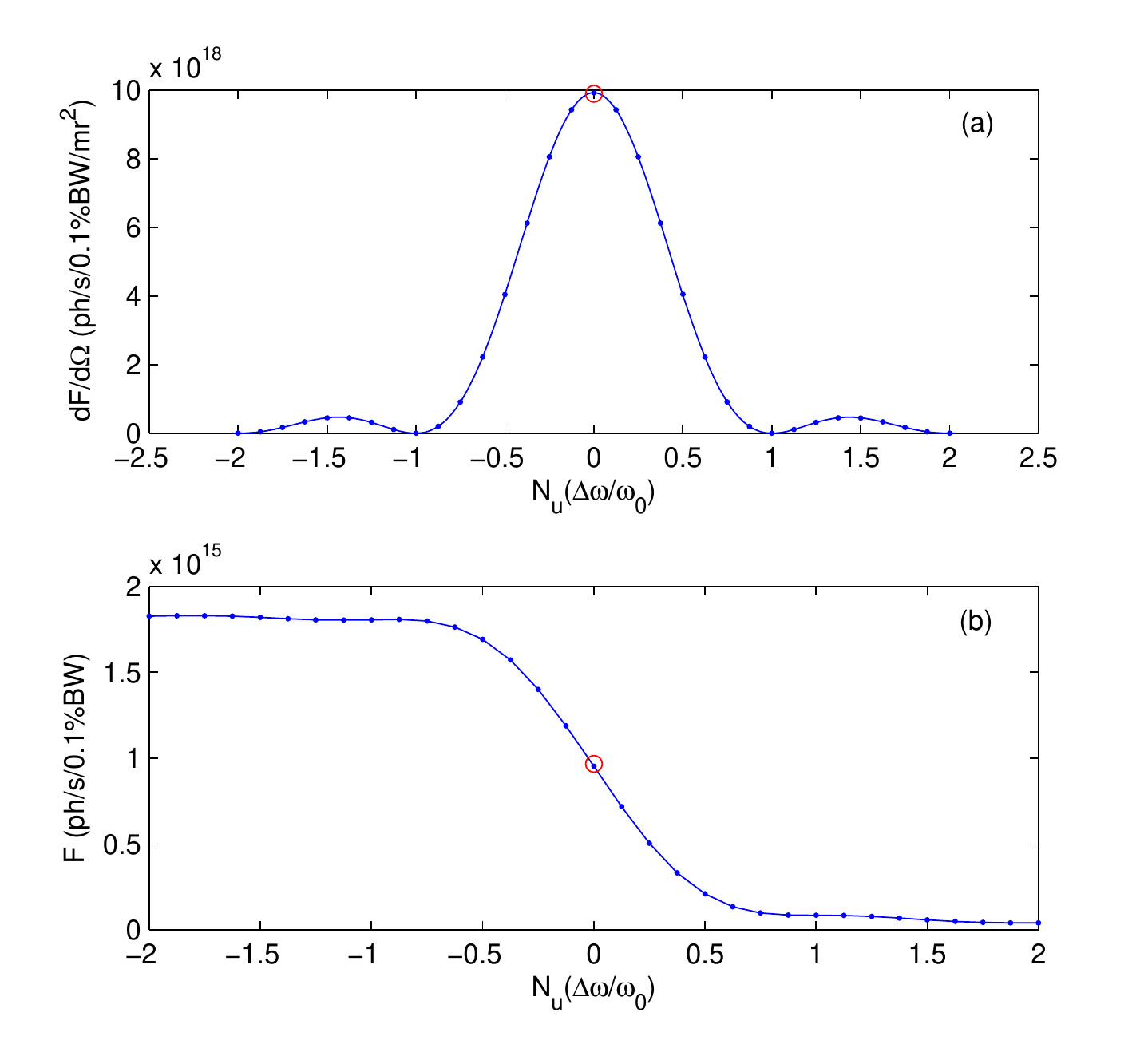}
	\caption{Scanning around the 1st harmonic resonance of 8\,keV for $N_u = 250$ period planar undulator: (a) angular flux on axis, (b) integrated (central cone) flux. Red circles denote analytical values.\label{fig:around_resonance_flux}}
\end{center}
\end{figure}

Fig.~\ref{fig:around_resonance_flux} shows the effect of the radiation frequency scanning around the resonance of the 1st harmonic on the angular (spectral) flux density on axis and integrated flux. Twice the spectral flux is available for radiation $\Delta\omega/\omega_0 \sim 1/N_u$ below the resonance.

\begin{figure}[htb]
\begin{center}
	\includegraphics[width=1.0\columnwidth]{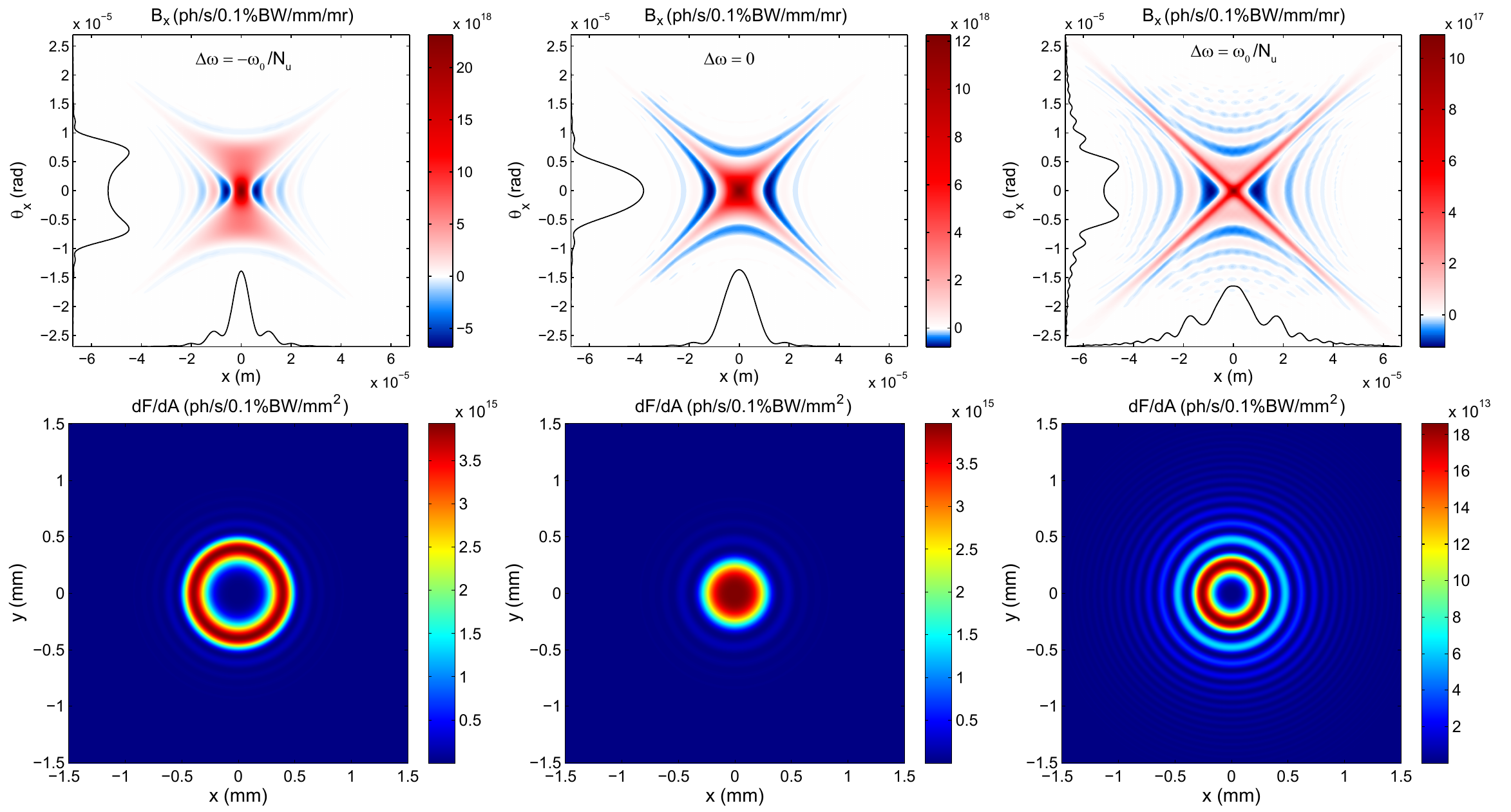}
	\caption{The WDF (top row) corresponding to Fig.~\ref{fig:around_resonance_flux} for different detuning off the resonance. The WDF is back-propagated to the undulator center. Solid lines show position and angular intensity projections. The radiation pattern used to calculate WDF (bottom row) is obtained at 50\,m from the undulator center.\label{fig:around_resonance_wigner}}
\end{center}
\end{figure}

Fig.~\ref{fig:around_resonance_wigner} demonstrates the light phase space for 3 different values of the radiation frequency detuning along with the radiation pattern (50\,m from the undulator center).

\begin{figure}[htb]
\begin{center}
	\includegraphics[width=1.0\columnwidth]{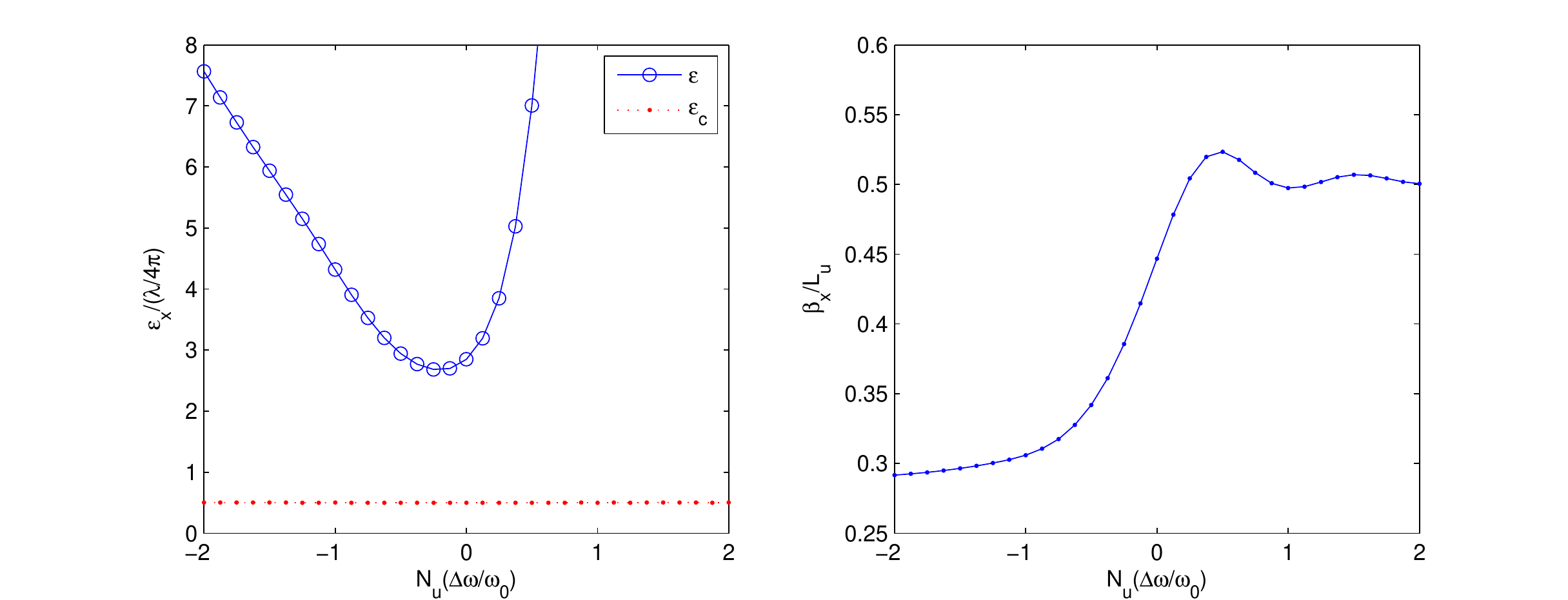}
	\caption{Light 100\% and core emittances (left) and beta function (right) at the undulator center corresponding to Fig.~\ref{fig:around_resonance_flux} for different detuning off the resonance.\label{fig:around_resonance_twiss}}
\end{center}
\end{figure}

Fig.~\ref{fig:around_resonance_twiss} shows the emittance and $\beta$-function of light for scanning the radiation frequency around the resonance. As shown previously, the core emittance is $\lambda/8\pi$ in all cases, whereas the rms emittance is minimal (though with $M^2 > 1$) around the resonance.

\begin{figure}[htb]
\begin{center}
	\includegraphics[width=1.0\columnwidth]{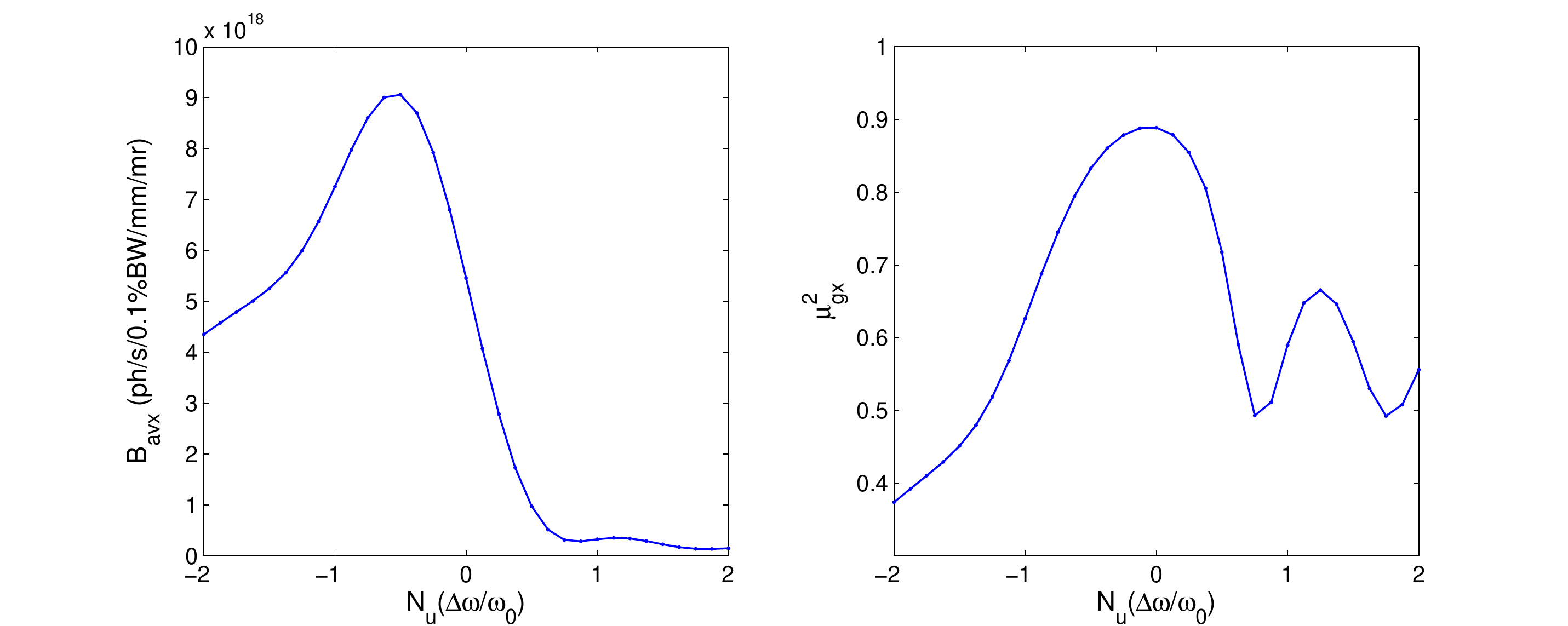}
	\caption{The average 2D brightness $\mathcal{B}_{avx}$, and $\mu_{gx}^2$ corresponding to Fig.~\ref{fig:around_resonance_flux}. \label{fig:around_resonance_brightness}}
\end{center}
\end{figure}

Finally, Fig.~\ref{fig:around_resonance_brightness} shows the effective 2D average brightness $\mathcal{B}_{avx} = \iint W_{x}^2(x,\theta_x)\,dx\,d\theta_x$ and $\mu_{gx}^2 = \lambda \iint W_x^2(x,\theta_x)\,dx\,d\theta_x$ where the normalized WDF $W_x(x,\theta_x) = \mathcal{B}_x(x,\theta_x)/\mathcal{F}$ with $\mathcal{F} = \iint \mathcal{B}_{x}(x,\theta_x)\,dx\,d\theta_x$. The deviation of $\mu_{gx}^2$ from 1 is due to the fact that the radiation mode is not separable, even though the radiation is fully transversely coherent in this case and therefore the full 4D $\mu_g^2 = 1$. The peak 2D brightness, which is not shown, simply follows the trend of Fig.~\ref{fig:around_resonance_flux} since it is related to the flux according to $\mathcal{B}_{0x} = \mathcal{F} (2/\lambda)$.

\subsubsection{Example: including emittance and energy spread of electrons}

\begin{figure}[htb]
\begin{center}
	\includegraphics[width=1.0\columnwidth]{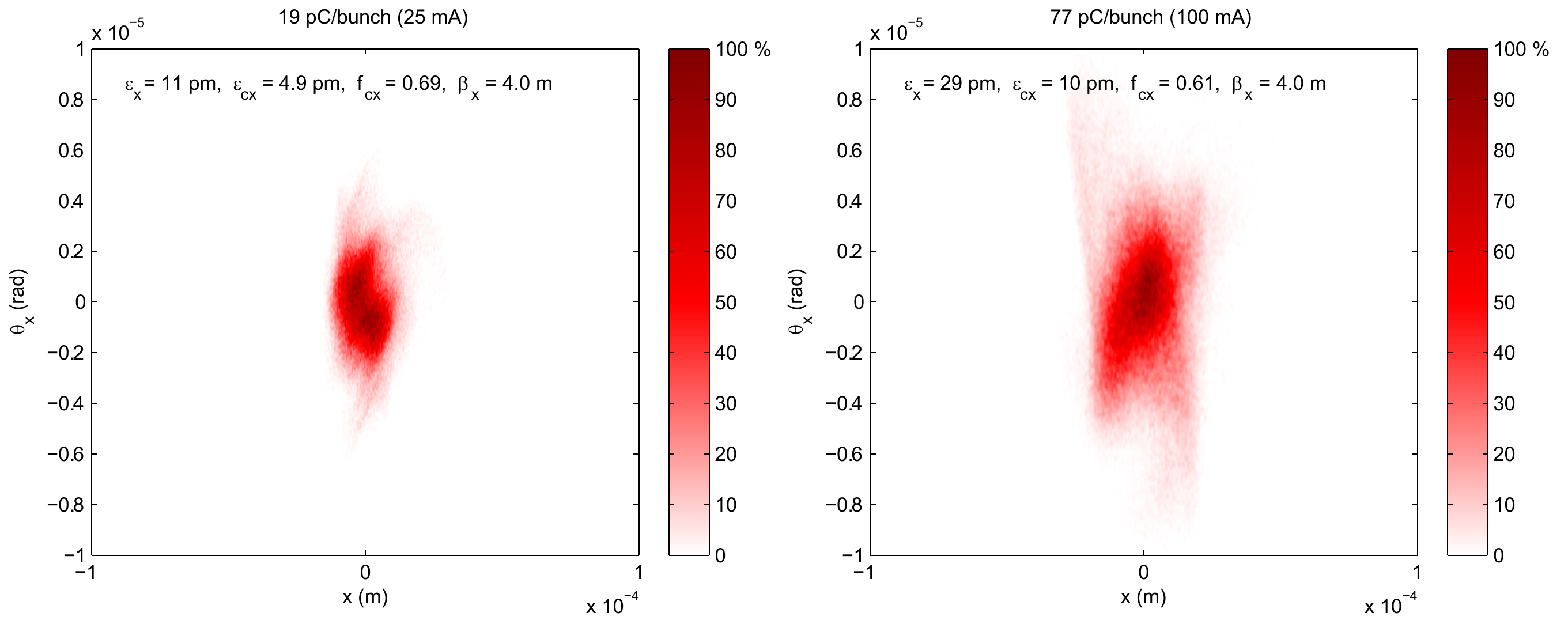}
	\caption{Electron phase space at the center of 25\,m long undulator corresponding to two different currents: 25\,mA (left) and 100\,mA (right).\label{fig:electron_phase_space}}
\end{center}
\end{figure}

Here we provide an example of including emittance and energy spread to the calculated WDF. For simplicity, we continue to limit ourselves to 2D projection of the Wigner distribution function and treat electron phase space probability distribution function as separable $P(\mathbf{r_e}, \boldsymbol\uptheta_\mathbf{e}, \delta_e) = P_x(x_e, \theta_{ex}) P_y(y_e, \theta_{ey}) P_\delta(\delta_e)$. Fig.~\ref{fig:electron_phase_space} shows the horizontal phase space at 5\,GeV obtained from the simulations of the photoinjector for 77\,pC per bunch and 1.3\,GHz repetition rate (average current of 100\,mA), including the effects of the merger and the linear accelerator \cite{pddr}. See Fig.~\ref{fig:electron_phase_space}. The energy spread of the electron beam is $\sigma_{\delta_e} = 2\times10^{-4}$. To illustrate its effect, we consider a 25-m long undulator with $N_u = 1250$ periods. Table~\ref{tab:25m_long} summarizes the parameters used in this example. As seen, the radiation is computed slightly below the resonance where the flux is roughly doubled.

\begin{table}[h!]
\caption{Parameters used in computing the radiation phase space.}
\label{tab:25m_long}
\begin{center}
\begin{tabular}{|c l c|}
\hline
& Number of periods, $N_u = 1250$ & \\
& Undulator period, $\lambda_u = 2$\,cm & \\
& Harmonic number, $n = 1$ & \\
& Resonant photon energy, $\hbar\omega = 8$\,keV & \\
& Detuning radiation frequency, $\Delta\omega = -0.75\omega/N_u$ & \\
& Beam energy, $E = 5$\,GeV & \\
& Electron energy spread, $\sigma_{\delta_e} = 2\times10^{-4}$ & \\
& Electron emittance, $\epsilon_x = 11, 29$\,pm & \\
& Average current, $I = 25, 100$\,mA & \\
& $\beta$-function, $\beta_x = 4$\,m & \\
\hline
\end{tabular}
\end{center}
\end{table}

\begin{figure}[htb]
\begin{center}
	\includegraphics[width=1.0\columnwidth]{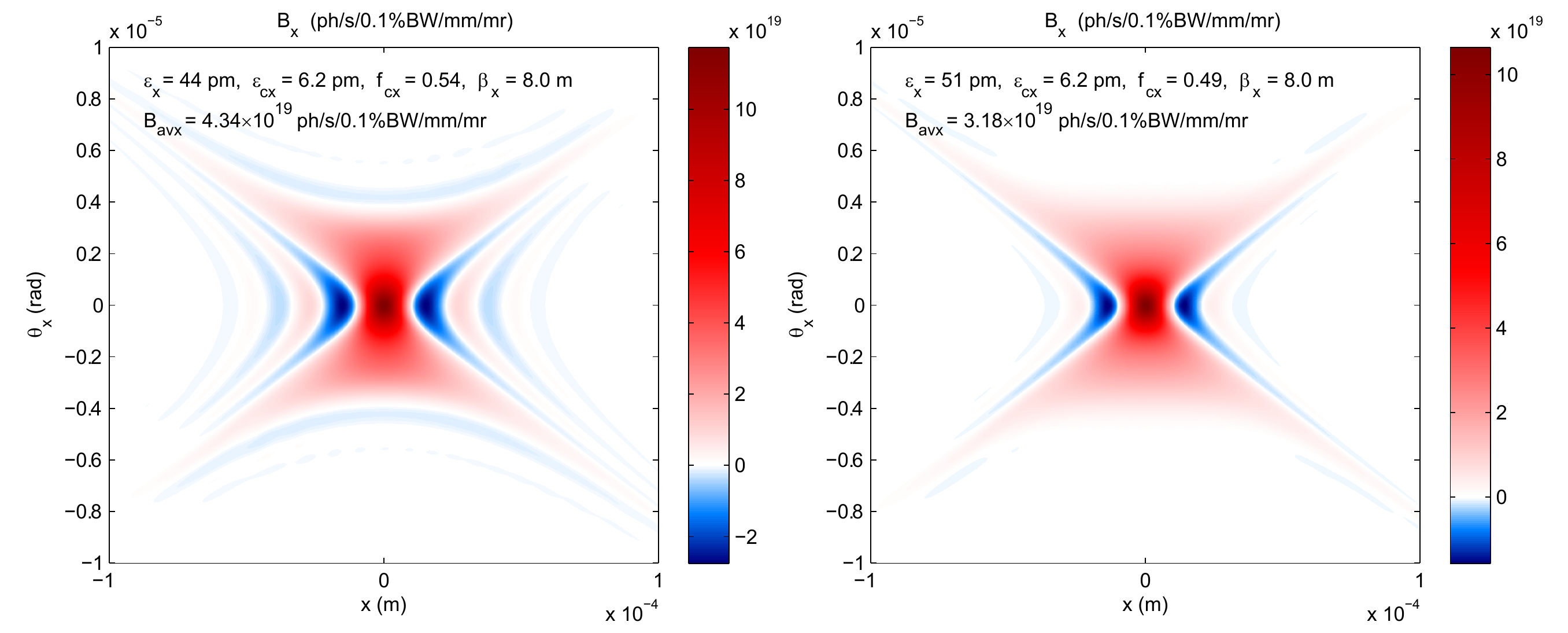}
	\caption{X-ray phase space corresponding to zero emittance and zero energy spread (left) and non-zero energy spread $\sigma_{\delta_e} = 2\times10^{-4}$ (right). The beam current is 100\,mA.\label{fig:ideal_and_energy_spread}}
\end{center}
\end{figure}

To provide more optimal matching for the core of the beam, $\beta_x$ is chosen to be $\beta_x = 4$\,m close to the Rayleigh range of the core of the radiation from a pencil (zero emittance) beam, Fig.~\ref{fig:ideal_and_energy_spread}a. Fig.~\ref{fig:ideal_and_energy_spread}b shows the effect of the energy spread for otherwise ideal (zero emittance) beam. Some degradation of the $\mathcal{B}_{avx}$ can be seen.

\begin{figure}[htb]
\begin{center}
	\includegraphics[width=1.0\columnwidth]{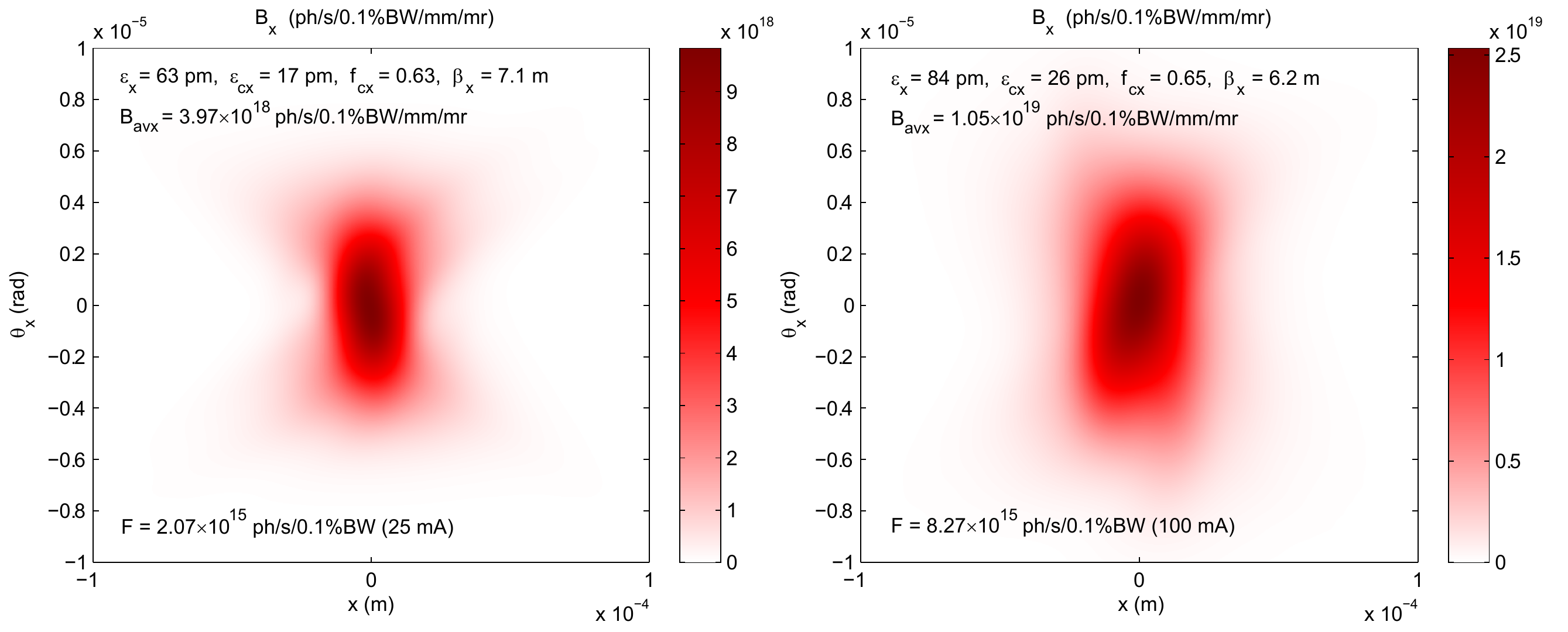}
	\caption{X-ray phase space including the effects of emittance and energy spread for the two different currents.\label{fig:xrays_convolved}}
\end{center}
\end{figure}

Fig.~\ref{fig:xrays_convolved} shows the effect of the beam emittance on the radiation phase space. It can be seen that the negative parts of the WDF are no longer present after the convolution, and the average 2D brightness is degraded by a factor of about 2 and 3 for 25 and 100\,mA cases respectively compared to zero emittance. Other relevant parameters of the radiation are shown in Fig.~\ref{fig:xrays_convolved}.

\section{Conclusions}

The Wigner distribution function approach to describe partially coherent radiation in phase space has been presented. Despite the general nature of the approach, the true power of the method to describe modern and future x-ray synchrotron sources is to employ 5 or 6D phase space (times 4 for arbitrary polarized light) complementing the 4D transverse phase space with frequency and time where the timing structure is important. Though straightforward, such a description is rather challenging from the point of computational requirements, even though a sampled approach similar to particle tracking in accelerator physics can be employed to represent the radiation in the entire 6D phase space (the microscopic brightness is allowed to take on negative values). When the x-ray optics beamline consists of drifts and perfect lenses without clipping apertues, this description is complete and allows to fully account for the light properties following geometric optics transformation rules. Introduction of apertures in the beam, however, requires the convolution of the transmissive mask's WDF with that of the beam. In this cases, it might be more efficient to consider decomposition of the partially coherent light into orthogonal mutually incoherent modes and to include the diffraction effects on each mode separately.

Nevertheless, the Wigner distribution function is demonstrated in this paper to be a rigorous and insightful way to describe the coherence and other properties of the synchrotron radiation. Its use will grow in importance as synchrotron x-ray sources with higher coherence become more prevalent.

\section{Acknowledgements}

I would like to acknowledge stimulating discussions with Keith Nugent, who pointed out his work on the Wigner distribution measurements for partially coherent x-rays and the subsequent spatial mode decomposition. Andrew Gasbarro has assisted in tests and design of various MATLAB scripts used in this work. David Sagan is acknowledged for initial discussions on the synchrotron radiation calculation approaches.

\bibliographystyle{apsrev4-1}

\begin{thebibliography}{37}%
\makeatletter
\providecommand \@ifxundefined [1]{%
 \@ifx{#1\undefined}
}%
\providecommand \@ifnum [1]{%
 \ifnum #1\expandafter \@firstoftwo
 \else \expandafter \@secondoftwo
 \fi
}%
\providecommand \@ifx [1]{%
 \ifx #1\expandafter \@firstoftwo
 \else \expandafter \@secondoftwo
 \fi
}%
\providecommand \natexlab [1]{#1}%
\providecommand \enquote  [1]{``#1''}%
\providecommand \bibnamefont  [1]{#1}%
\providecommand \bibfnamefont [1]{#1}%
\providecommand \citenamefont [1]{#1}%
\providecommand \href@noop [0]{\@secondoftwo}%
\providecommand \href [0]{\begingroup \@sanitize@url \@href}%
\providecommand \@href[1]{\@@startlink{#1}\@@href}%
\providecommand \@@href[1]{\endgroup#1\@@endlink}%
\providecommand \@sanitize@url [0]{\catcode `\\12\catcode `\$12\catcode
  `\&12\catcode `\#12\catcode `\^12\catcode `\_12\catcode `\%12\relax}%
\providecommand \@@startlink[1]{}%
\providecommand \@@endlink[0]{}%
\providecommand \url  [0]{\begingroup\@sanitize@url \@url }%
\providecommand \@url [1]{\endgroup\@href {#1}{\urlprefix }}%
\providecommand \urlprefix  [0]{URL }%
\providecommand \Eprint [0]{\href }%
\providecommand \doibase [0]{http://dx.doi.org/}%
\providecommand \selectlanguage [0]{\@gobble}%
\providecommand \bibinfo  [0]{\@secondoftwo}%
\providecommand \bibfield  [0]{\@secondoftwo}%
\providecommand \translation [1]{[#1]}%
\providecommand \BibitemOpen [0]{}%
\providecommand \bibitemStop [0]{}%
\providecommand \bibitemNoStop [0]{.\EOS\space}%
\providecommand \EOS [0]{\spacefactor3000\relax}%
\providecommand \BibitemShut  [1]{\csname bibitem#1\endcsname}%
\let\auto@bib@innerbib\@empty
\bibitem [{\citenamefont {Wigner}(1932)}]{PhysRev.40.749}%
  \BibitemOpen
  \bibfield  {author} {\bibinfo {author} {\bibfnamefont {E.}~\bibnamefont
  {Wigner}},\ }\href {\doibase 10.1103/PhysRev.40.749} {\bibfield  {journal}
  {\bibinfo  {journal} {Phys. Rev.}\ }\textbf {\bibinfo {volume} {40}},\
  \bibinfo {pages} {749} (\bibinfo {year} {1932})}\BibitemShut {NoStop}%
\bibitem [{\citenamefont {Walther}(1968)}]{WALTHER:68}%
  \BibitemOpen
  \bibfield  {author} {\bibinfo {author} {\bibfnamefont {A.}~\bibnamefont
  {Walther}},\ }\href {\doibase 10.1364/JOSA.58.001256} {\bibfield  {journal}
  {\bibinfo  {journal} {J. Opt. Soc. Am.}\ }\textbf {\bibinfo {volume} {58}},\
  \bibinfo {pages} {1256} (\bibinfo {year} {1968})}\BibitemShut {NoStop}%
\bibitem [{\citenamefont {Kim}(1986)}]{KwangJe198671}%
  \BibitemOpen
  \bibfield  {author} {\bibinfo {author} {\bibfnamefont {K.-J.}\ \bibnamefont
  {Kim}},\ }\href {\doibase 10.1016/0168-9002(86)90048-3} {\bibfield  {journal}
  {\bibinfo  {journal} {Nuclear Instruments and Methods in Physics Research
  Section A: Accelerators, Spectrometers, Detectors and Associated Equipment}\
  }\textbf {\bibinfo {volume} {246}},\ \bibinfo {pages} {71 } (\bibinfo {year}
  {1986})}\BibitemShut {NoStop}%
\bibitem [{\citenamefont {Bastiaans}(1986)}]{Bastiaans:86}%
  \BibitemOpen
  \bibfield  {author} {\bibinfo {author} {\bibfnamefont {M.~J.}\ \bibnamefont
  {Bastiaans}},\ }\href {\doibase 10.1364/JOSAA.3.001227} {\bibfield  {journal}
  {\bibinfo  {journal} {J. Opt. Soc. Am. A}\ }\textbf {\bibinfo {volume} {3}},\
  \bibinfo {pages} {1227} (\bibinfo {year} {1986})}\BibitemShut {NoStop}%
\bibitem [{\citenamefont {{Luis}}(2007)}]{2007JEOS....2E7030L}%
  \BibitemOpen
  \bibfield  {author} {\bibinfo {author} {\bibfnamefont {A.}~\bibnamefont
  {{Luis}}},\ }\href {http://dx.doi.org/10.2971/jeos.2007.07030} {\bibfield
  {journal} {\bibinfo  {journal} {Journal European Optical Society - Rapid
  Publications}\ }\textbf {\bibinfo {volume} {2}},\ \bibinfo {pages} {07030}
  (\bibinfo {year} {2007})}\BibitemShut {NoStop}%
\bibitem [{\citenamefont {Kim}(1989)}]{Kim89}%
  \BibitemOpen
  \bibfield  {author} {\bibinfo {author} {\bibfnamefont {K.~J.}\ \bibnamefont
  {Kim}},\ }\href {\doibase 10.1063/1.38046} {\bibfield  {journal} {\bibinfo
  {journal} {AIP Conference Proceedings}\ }\textbf {\bibinfo {volume} {184}},\
  \bibinfo {pages} {565} (\bibinfo {year} {1989})}\BibitemShut {NoStop}%
\bibitem [{\citenamefont {Co{\"\i}sson}\ and\ \citenamefont
  {Marchesini}(1997)}]{Coisson:mb0007}%
  \BibitemOpen
  \bibfield  {author} {\bibinfo {author} {\bibfnamefont {R.}~\bibnamefont
  {Co{\"\i}sson}}\ and\ \bibinfo {author} {\bibfnamefont {S.}~\bibnamefont
  {Marchesini}},\ }\href {\doibase 10.1107/S0909049597008169} {\bibfield
  {journal} {\bibinfo  {journal} {Journal of Synchrotron Radiation}\ }\textbf
  {\bibinfo {volume} {4}},\ \bibinfo {pages} {263} (\bibinfo {year}
  {1997})}\BibitemShut {NoStop}%
\bibitem [{\citenamefont {Geloni}\ \emph {et~al.}(2008)\citenamefont {Geloni},
  \citenamefont {Saldin}, \citenamefont {Schneidmiller},\ and\ \citenamefont
  {Yurkov}}]{Geloni2008463}%
  \BibitemOpen
  \bibfield  {author} {\bibinfo {author} {\bibfnamefont {G.}~\bibnamefont
  {Geloni}}, \bibinfo {author} {\bibfnamefont {E.}~\bibnamefont {Saldin}},
  \bibinfo {author} {\bibfnamefont {E.}~\bibnamefont {Schneidmiller}}, \ and\
  \bibinfo {author} {\bibfnamefont {M.}~\bibnamefont {Yurkov}},\ }\href
  {\doibase 10.1016/j.nima.2008.01.089} {\bibfield  {journal} {\bibinfo
  {journal} {Nuclear Instruments and Methods in Physics Research Section A:
  Accelerators, Spectrometers, Detectors and Associated Equipment}\ }\textbf
  {\bibinfo {volume} {588}},\ \bibinfo {pages} {463 } (\bibinfo {year}
  {2008})}\BibitemShut {NoStop}%
\bibitem [{\citenamefont {Tran}\ \emph {et~al.}(2007)\citenamefont {Tran},
  \citenamefont {Williams}, \citenamefont {Roberts}, \citenamefont {Flewett},
  \citenamefont {Peele}, \citenamefont {Paterson}, \citenamefont {de~Jonge},\
  and\ \citenamefont {Nugent}}]{PhysRevLett.98.224801}%
  \BibitemOpen
  \bibfield  {author} {\bibinfo {author} {\bibfnamefont {C.~Q.}\ \bibnamefont
  {Tran}}, \bibinfo {author} {\bibfnamefont {G.~J.}\ \bibnamefont {Williams}},
  \bibinfo {author} {\bibfnamefont {A.}~\bibnamefont {Roberts}}, \bibinfo
  {author} {\bibfnamefont {S.}~\bibnamefont {Flewett}}, \bibinfo {author}
  {\bibfnamefont {A.~G.}\ \bibnamefont {Peele}}, \bibinfo {author}
  {\bibfnamefont {D.}~\bibnamefont {Paterson}}, \bibinfo {author}
  {\bibfnamefont {M.~D.}\ \bibnamefont {de~Jonge}}, \ and\ \bibinfo {author}
  {\bibfnamefont {K.~A.}\ \bibnamefont {Nugent}},\ }\href {\doibase
  10.1103/PhysRevLett.98.224801} {\bibfield  {journal} {\bibinfo  {journal}
  {Phys. Rev. Lett.}\ }\textbf {\bibinfo {volume} {98}},\ \bibinfo {pages}
  {224801} (\bibinfo {year} {2007})}\BibitemShut {NoStop}%
\bibitem [{\citenamefont {Smithey}\ \emph {et~al.}(1993)\citenamefont
  {Smithey}, \citenamefont {Beck}, \citenamefont {Raymer},\ and\ \citenamefont
  {Faridani}}]{PhysRevLett.70.1244}%
  \BibitemOpen
  \bibfield  {author} {\bibinfo {author} {\bibfnamefont {D.~T.}\ \bibnamefont
  {Smithey}}, \bibinfo {author} {\bibfnamefont {M.}~\bibnamefont {Beck}},
  \bibinfo {author} {\bibfnamefont {M.~G.}\ \bibnamefont {Raymer}}, \ and\
  \bibinfo {author} {\bibfnamefont {A.}~\bibnamefont {Faridani}},\ }\href
  {\doibase 10.1103/PhysRevLett.70.1244} {\bibfield  {journal} {\bibinfo
  {journal} {Phys. Rev. Lett.}\ }\textbf {\bibinfo {volume} {70}},\ \bibinfo
  {pages} {1244} (\bibinfo {year} {1993})}\BibitemShut {NoStop}%
\bibitem [{\citenamefont {Hillery}\ \emph {et~al.}(1984)\citenamefont
  {Hillery}, \citenamefont {O'Connell}, \citenamefont {Scully},\ and\
  \citenamefont {Wigner}}]{Hillery1984121}%
  \BibitemOpen
  \bibfield  {author} {\bibinfo {author} {\bibfnamefont {M.}~\bibnamefont
  {Hillery}}, \bibinfo {author} {\bibfnamefont {R.}~\bibnamefont {O'Connell}},
  \bibinfo {author} {\bibfnamefont {M.}~\bibnamefont {Scully}}, \ and\ \bibinfo
  {author} {\bibfnamefont {E.}~\bibnamefont {Wigner}},\ }\href {\doibase
  10.1016/0370-1573(84)90160-1} {\bibfield  {journal} {\bibinfo  {journal}
  {Physics Reports}\ }\textbf {\bibinfo {volume} {106}},\ \bibinfo {pages} {121
  } (\bibinfo {year} {1984})}\BibitemShut {NoStop}%
\bibitem [{\citenamefont {Tatarski\u{i}}(1983)}]{Tatarskii83}%
  \BibitemOpen
  \bibfield  {author} {\bibinfo {author} {\bibfnamefont {V.~I.}\ \bibnamefont
  {Tatarski\u{i}}},\ }\href {http://stacks.iop.org/0038-5670/26/i=4/a=R02}
  {\bibfield  {journal} {\bibinfo  {journal} {Soviet Physics Uspekhi}\ }\textbf
  {\bibinfo {volume} {26}},\ \bibinfo {pages} {311} (\bibinfo {year}
  {1983})}\BibitemShut {NoStop}%
\bibitem [{\citenamefont {Testorf}\ \emph {et~al.}(2009)\citenamefont
  {Testorf}, \citenamefont {Hennelly},\ and\ \citenamefont
  {Ojeda-Casta{\~n}eda}}]{testorf2009phase}%
  \BibitemOpen
  \bibfield  {author} {\bibinfo {author} {\bibfnamefont {M.}~\bibnamefont
  {Testorf}}, \bibinfo {author} {\bibfnamefont {B.}~\bibnamefont {Hennelly}}, \
  and\ \bibinfo {author} {\bibfnamefont {J.}~\bibnamefont
  {Ojeda-Casta{\~n}eda}},\ }\href
  {http://books.google.com/books?id=C7PU0hz3MXQC} {\emph {\bibinfo {title}
  {Phase-space optics: fundamentals and applications}}}\ (\bibinfo  {publisher}
  {McGraw-Hill},\ \bibinfo {year} {2009})\BibitemShut {NoStop}%
\bibitem [{\citenamefont {Dragoman}()}]{Dragoman_2005}%
  \BibitemOpen
  \bibfield  {author} {\bibinfo {author} {\bibfnamefont {D.}~\bibnamefont
  {Dragoman}},\ }\href
  {http://www.hindawi.com/journals/asp/2005/264967.abs.html} {\bibfield
  {journal} {\bibinfo  {journal} {EURASIP Journal on Advances in Signal
  Processing}\ }\textbf {\bibinfo {volume} {2005}},\ \bibinfo {pages}
  {1520}}\BibitemShut {NoStop}%
\bibitem [{\citenamefont {Sheng}(2000)}]{Sheng2000}%
  \BibitemOpen
  \bibfield  {author} {\bibinfo {author} {\bibfnamefont {Y.}~\bibnamefont
  {Sheng}},\ }in\ \href {http://dx.doi.org/10.1201/9781420036756.ch10} {\emph
  {\bibinfo {booktitle} {Electrical Engineering Handbook}}}\ (\bibinfo
  {publisher} {CRC Press},\ \bibinfo {year} {2000})\ pp.\ \bibinfo {pages}
  {10--1}\BibitemShut {NoStop}%
\bibitem [{\citenamefont {Boudreaux-Bartels}(2000)}]{Faye2000}%
  \BibitemOpen
  \bibfield  {author} {\bibinfo {author} {\bibfnamefont {F.}~\bibnamefont
  {Boudreaux-Bartels}},\ }in\ \href
  {http://dx.doi.org/10.1201/9781420036756.ch12} {\emph {\bibinfo {booktitle}
  {Electrical Engineering Handbook}}}\ (\bibinfo  {publisher} {CRC Press},\
  \bibinfo {year} {2000})\ pp.\ \bibinfo {pages} {12--1}\BibitemShut {NoStop}%
\bibitem [{\citenamefont {{Rioux}}(2009)}]{2009arXiv0912.2333R}%
  \BibitemOpen
  \bibfield  {author} {\bibinfo {author} {\bibfnamefont {F.}~\bibnamefont
  {{Rioux}}},\ }\href@noop {} {\bibfield  {journal} {\bibinfo  {journal} {ArXiv
  e-prints}\ } (\bibinfo {year} {2009})},\ \Eprint
  {http://arxiv.org/abs/0912.2333} {arXiv:0912.2333 [physics.gen-ph]}
  \BibitemShut {NoStop}%
\bibitem [{\citenamefont {Leaf}(1968)}]{leaf:65}%
  \BibitemOpen
  \bibfield  {author} {\bibinfo {author} {\bibfnamefont {B.}~\bibnamefont
  {Leaf}},\ }\href {\doibase 10.1063/1.1664478} {\bibfield  {journal} {\bibinfo
   {journal} {Journal of Mathematical Physics}\ }\textbf {\bibinfo {volume}
  {9}},\ \bibinfo {pages} {65} (\bibinfo {year} {1968})}\BibitemShut {NoStop}%
\bibitem [{\citenamefont {{Bertrand}}\ \emph {et~al.}(1983)\citenamefont
  {{Bertrand}}, \citenamefont {{Doremus}}, \citenamefont {{Izrar}},
  \citenamefont {{Nguyen}},\ and\ \citenamefont
  {{Feix}}}]{1983PhLA...94..415B}%
  \BibitemOpen
  \bibfield  {author} {\bibinfo {author} {\bibfnamefont {P.}~\bibnamefont
  {{Bertrand}}}, \bibinfo {author} {\bibfnamefont {J.~P.}\ \bibnamefont
  {{Doremus}}}, \bibinfo {author} {\bibfnamefont {B.}~\bibnamefont {{Izrar}}},
  \bibinfo {author} {\bibfnamefont {V.~T.}\ \bibnamefont {{Nguyen}}}, \ and\
  \bibinfo {author} {\bibfnamefont {M.~R.}\ \bibnamefont {{Feix}}},\ }\href
  {\doibase 10.1016/0375-9601(83)90841-1} {\bibfield  {journal} {\bibinfo
  {journal} {Physics Letters A}\ }\textbf {\bibinfo {volume} {94}},\ \bibinfo
  {pages} {415} (\bibinfo {year} {1983})}\BibitemShut {NoStop}%
\bibitem [{\citenamefont {Schleich}()}]{Schleich2001}%
  \BibitemOpen
  \bibfield  {author} {\bibinfo {author} {\bibfnamefont {W.~P.}\ \bibnamefont
  {Schleich}},\ }\href {http://dx.doi.org/10.1002/3527602976} {\emph {\bibinfo
  {title} {{Quantum Optics in Phase Space}}}},\ \bibinfo {edition} {1st}\ ed.\
  (\bibinfo  {publisher} {Wiley-VCH})\BibitemShut {NoStop}%
\bibitem [{\citenamefont {Menski\u{i}}(2000)}]{menskii­2000quantum}%
  \BibitemOpen
  \bibfield  {author} {\bibinfo {author} {\bibfnamefont {M.~B.}\ \bibnamefont
  {Menski\u{i}}},\ }\href {http://books.google.com/books?id=Bo7jujlMqL8C}
  {\emph {\bibinfo {title} {Quantum measurements and decoherence: models and
  phenomenology}}},\ Fundamental theories of physics\ (\bibinfo  {publisher}
  {Kluwer Academic Publishers},\ \bibinfo {year} {2000})\BibitemShut {NoStop}%
\bibitem [{\citenamefont {Flewett}\ \emph {et~al.}(2009)\citenamefont
  {Flewett}, \citenamefont {Quiney}, \citenamefont {Tran},\ and\ \citenamefont
  {Nugent}}]{Flewett:09}%
  \BibitemOpen
  \bibfield  {author} {\bibinfo {author} {\bibfnamefont {S.}~\bibnamefont
  {Flewett}}, \bibinfo {author} {\bibfnamefont {H.~M.}\ \bibnamefont {Quiney}},
  \bibinfo {author} {\bibfnamefont {C.~Q.}\ \bibnamefont {Tran}}, \ and\
  \bibinfo {author} {\bibfnamefont {K.~A.}\ \bibnamefont {Nugent}},\ }\href
  {\doibase 10.1364/OL.34.002198} {\bibfield  {journal} {\bibinfo  {journal}
  {Opt. Lett.}\ }\textbf {\bibinfo {volume} {34}},\ \bibinfo {pages} {2198}
  (\bibinfo {year} {2009})}\BibitemShut {NoStop}%
\bibitem [{\citenamefont {Gori}(1980)}]{Gori80}%
  \BibitemOpen
  \bibfield  {author} {\bibinfo {author} {\bibfnamefont {F.}~\bibnamefont
  {Gori}},\ }\href {\doibase 10.1016/0030-4018(80)90382-X} {\bibfield
  {journal} {\bibinfo  {journal} {Optics Communications}\ }\textbf {\bibinfo
  {volume} {34}},\ \bibinfo {pages} {301 } (\bibinfo {year}
  {1980})}\BibitemShut {NoStop}%
\bibitem [{\citenamefont {Starikov}\ and\ \citenamefont
  {Wolf}(1982)}]{Starikov:82}%
  \BibitemOpen
  \bibfield  {author} {\bibinfo {author} {\bibfnamefont {A.}~\bibnamefont
  {Starikov}}\ and\ \bibinfo {author} {\bibfnamefont {E.}~\bibnamefont
  {Wolf}},\ }\href {\doibase 10.1364/JOSA.72.000923} {\bibfield  {journal}
  {\bibinfo  {journal} {J. Opt. Soc. Am.}\ }\textbf {\bibinfo {volume} {72}},\
  \bibinfo {pages} {923} (\bibinfo {year} {1982})}\BibitemShut {NoStop}%
\bibitem [{\citenamefont {Mandel}\ and\ \citenamefont
  {Wolf}(1995)}]{mandel1995optical}%
  \BibitemOpen
  \bibfield  {author} {\bibinfo {author} {\bibfnamefont {L.}~\bibnamefont
  {Mandel}}\ and\ \bibinfo {author} {\bibfnamefont {E.}~\bibnamefont {Wolf}},\
  }\href {http://books.google.com/books?id=FeBix14iM70C} {\emph {\bibinfo
  {title} {Optical coherence and quantum optics}}}\ (\bibinfo  {publisher}
  {Cambridge University Press},\ \bibinfo {year} {1995})\BibitemShut {NoStop}%
\bibitem [{\citenamefont {Papoulis}(1968)}]{papoulis1968systems}%
  \BibitemOpen
  \bibfield  {author} {\bibinfo {author} {\bibfnamefont {A.}~\bibnamefont
  {Papoulis}},\ }\href {http://books.google.com/books?id=gqoeAQAAIAAJ} {\emph
  {\bibinfo {title} {Systems and transforms with applications in optics}}},\
  McGraw-Hill series in systems science\ (\bibinfo  {publisher} {McGraw-Hill},\
  \bibinfo {year} {1968})\BibitemShut {NoStop}%
\bibitem [{\citenamefont {Luis}(2007)}]{Luis:07}%
  \BibitemOpen
  \bibfield  {author} {\bibinfo {author} {\bibfnamefont {A.}~\bibnamefont
  {Luis}},\ }\href {\doibase 10.1364/JOSAA.24.002070} {\bibfield  {journal}
  {\bibinfo  {journal} {J. Opt. Soc. Am. A}\ }\textbf {\bibinfo {volume}
  {24}},\ \bibinfo {pages} {2070} (\bibinfo {year} {2007})}\BibitemShut
  {NoStop}%
\bibitem [{\citenamefont {O'Connell}\ and\ \citenamefont
  {Wigner}(1984)}]{PhysRevA.30.2613}%
  \BibitemOpen
  \bibfield  {author} {\bibinfo {author} {\bibfnamefont {R.~F.}\ \bibnamefont
  {O'Connell}}\ and\ \bibinfo {author} {\bibfnamefont {E.~P.}\ \bibnamefont
  {Wigner}},\ }\href {\doibase 10.1103/PhysRevA.30.2613} {\bibfield  {journal}
  {\bibinfo  {journal} {Phys. Rev. A}\ }\textbf {\bibinfo {volume} {30}},\
  \bibinfo {pages} {2613} (\bibinfo {year} {1984})}\BibitemShut {NoStop}%
\bibitem [{\citenamefont {Goldstein}(2003)}]{goldstein2003polarized}%
  \BibitemOpen
  \bibfield  {author} {\bibinfo {author} {\bibfnamefont {D.~H.}\ \bibnamefont
  {Goldstein}},\ }\href {http://books.google.com/books?id=aWX5jj603PoC} {\emph
  {\bibinfo {title} {Polarized light}}}\ (\bibinfo  {publisher} {Marcel
  Dekker},\ \bibinfo {year} {2003})\BibitemShut {NoStop}%
\bibitem [{\citenamefont {Chubar}(1995)}]{chubar:1995}%
  \BibitemOpen
  \bibfield  {author} {\bibinfo {author} {\bibfnamefont {O.~V.}\ \bibnamefont
  {Chubar}},\ }\href {\doibase 10.1063/1.1145810} {\bibfield  {journal}
  {\bibinfo  {journal} {Review of Scientific Instruments}\ }\textbf {\bibinfo
  {volume} {66}},\ \bibinfo {pages} {1872} (\bibinfo {year}
  {1995})}\BibitemShut {NoStop}%
\bibitem [{\citenamefont {Landau}\ and\ \citenamefont
  {Lifshits}(1975)}]{landau1975classical}%
  \BibitemOpen
  \bibfield  {author} {\bibinfo {author} {\bibfnamefont {L.}~\bibnamefont
  {Landau}}\ and\ \bibinfo {author} {\bibfnamefont {E.}~\bibnamefont
  {Lifshits}},\ }\href {http://books.google.com/books?id=X18PF4oKyrUC} {\emph
  {\bibinfo {title} {The classical theory of fields}}},\ Course of theoretical
  physics\ (\bibinfo  {publisher} {Butterworth Heinemann},\ \bibinfo {year}
  {1975})\BibitemShut {NoStop}%
\bibitem [{\citenamefont {Geloni}\ \emph {et~al.}(2007)\citenamefont {Geloni},
  \citenamefont {Saldin}, \citenamefont {Schneidmiller},\ and\ \citenamefont
  {Yurkov}}]{Geloni2007167}%
  \BibitemOpen
  \bibfield  {author} {\bibinfo {author} {\bibfnamefont {G.}~\bibnamefont
  {Geloni}}, \bibinfo {author} {\bibfnamefont {E.}~\bibnamefont {Saldin}},
  \bibinfo {author} {\bibfnamefont {E.}~\bibnamefont {Schneidmiller}}, \ and\
  \bibinfo {author} {\bibfnamefont {M.}~\bibnamefont {Yurkov}},\ }\href
  {\doibase 10.1016/j.optcom.2007.03.051} {\bibfield  {journal} {\bibinfo
  {journal} {Optics Communications}\ }\textbf {\bibinfo {volume} {276}},\
  \bibinfo {pages} {167 } (\bibinfo {year} {2007})}\BibitemShut {NoStop}%
\bibitem [{\citenamefont {Scharlemann}(1985)}]{scharlemann:2154}%
  \BibitemOpen
  \bibfield  {author} {\bibinfo {author} {\bibfnamefont {E.}~\bibnamefont
  {Scharlemann}},\ }\href {\doibase 10.1063/1.335980} {\bibfield  {journal}
  {\bibinfo  {journal} {Journal of Applied Physics}\ }\textbf {\bibinfo
  {volume} {58}},\ \bibinfo {pages} {2154} (\bibinfo {year}
  {1985})}\BibitemShut {NoStop}%
\bibitem [{\citenamefont {Reiser}(2008)}]{reiser2008theory}%
  \BibitemOpen
  \bibfield  {author} {\bibinfo {author} {\bibfnamefont {M.}~\bibnamefont
  {Reiser}},\ }\href {http://books.google.com/books?id=eegK9Mqgpi4C} {\emph
  {\bibinfo {title} {Theory and design of charged particle beams}}},\ Wiley
  Series in Beam Physics and Accelerator Technology\ (\bibinfo  {publisher}
  {Wiley-VCH},\ \bibinfo {year} {2008})\BibitemShut {NoStop}%
\bibitem [{\citenamefont {Lejeune}\ and\ \citenamefont
  {Aubert}(1980)}]{Lejeune1980}%
  \BibitemOpen
  \bibfield  {author} {\bibinfo {author} {\bibfnamefont {C.}~\bibnamefont
  {Lejeune}}\ and\ \bibinfo {author} {\bibfnamefont {K.}~\bibnamefont
  {Aubert}},\ }in\ \href@noop {} {\emph {\bibinfo {booktitle} {Applied Charged
  Particle Optics}}},\ \bibinfo {series and number} {Advances in Electronics
  and Electron Physics, Supplement 13A},\ \bibinfo {editor} {edited by\
  \bibinfo {editor} {\bibfnamefont {A.}~\bibnamefont {Septier}}}\ (\bibinfo
  {publisher} {Academic Press, New York},\ \bibinfo {year} {1980})\BibitemShut
  {NoStop}%
\bibitem [{\citenamefont {Rhee}(1992)}]{rhee:1674}%
  \BibitemOpen
  \bibfield  {author} {\bibinfo {author} {\bibfnamefont {M.~J.}\ \bibnamefont
  {Rhee}},\ }\href {\doibase 10.1063/1.860076} {\bibfield  {journal} {\bibinfo
  {journal} {Physics of Fluids B: Plasma Physics}\ }\textbf {\bibinfo {volume}
  {4}},\ \bibinfo {pages} {1674} (\bibinfo {year} {1992})}\BibitemShut
  {NoStop}%
\bibitem [{pdd(2011)}]{pddr}%
  \BibitemOpen
  \href@noop {} {\enquote {\bibinfo {title} {Cornell energy recovery linac
  project definition design report},}\ }\bibinfo {howpublished} {unpublished}
  (\bibinfo {year} {2011})\BibitemShut {NoStop}%
\end{thebibliography}

\end{document}